\documentclass[aps,prd,showpacs,nofootinbib]{revtex4}
\usepackage{amsfonts,bbm,amsmath,amssymb}
\usepackage{graphicx,calc,epsfig}

\usepackage{psfrag}
\usepackage{latexsym}

\usepackage{graphicx}

\newcommand{\n}{\nonumber}

\newcommand{\be}{\nopagebreak[3]\begin{equation}}
\newcommand{\ee}{\end{equation}}
\newcommand{\ba}{\nopagebreak[3]\begin{eqnarray}}
\newcommand{\ea}{\end{eqnarray}}

\newcommand{\C}{\mathbb{C}}

\newcommand{\N}{\mathbb{N}}
\newcommand{\va}{\scriptscriptstyle}
\newcommand{\vani}{\scriptstyle}
\newcommand{\R}{\mathbb{R}}

\DeclareFontFamily{U}{rsfs}{}         
\DeclareFontShape{U}{rsfs}{m}{n}{<5> rsfs5 <6><7> rsfs7          %
  <8><9><10><10.95><12><14.4><17.28><20.74><24.88> rsfs10}{}     %
\DeclareMathAlphabet{\mathfs}{U}{rsfs}{m}{n}                     %
\newcommand{\mfs}[1]{\mathfs {#1}}                               %
\newcommand{\van}{\scriptstyle}

\newcommand{\sK}{{\mfs K}}

\newcommand{\sY}{{\mfs Y}}

\newcommand{\sH}{{\mfs H}}
\newcommand{\sL}{{\mfs L}}

\newcommand{\sO}{{\mfs O}}

\def\pb#1{\rlap{\lower1.5ex\hbox{$\longleftarrow$}}{#1}}
\def\dpb#1{\rlap{\lower1.5ex\hbox{$\Longleftarrow$}}{#1}}
\def\spb#1{\rlap{\lower1.5ex\hbox{$\leftarrow$}}{#1}}
\def\sdpb#1{\rlap{\lower1.5ex\hbox{$\Leftarrow$}}{#1}}

\newcommand{\bn}{\mathbf{n}}
\newcommand{\bm}{\mathbf{m}}

\newcommand{\la}{\langle}
\newcommand{\ra}{\rangle}

\newcommand{\SU}{\mathrm{SU}}

\newcommand{\Spin}{\mathrm{Spin}}

\newcommand{\ftj}{\mbox{15}j}   

\newcommand{\nb}{\mathbf{n}} 

\newcommand{\rd}{\mathrm{d}}

\def\lsim{\
  \lower-2.0pt\vbox{\hbox{\rlap{$<$}\lower5.5pt\vbox{\hbox{$\sim$}}}}\ }
\def\gsim{\
  \lower-2.0pt\vbox{\hbox{\rlap{$>$}\lower5.5pt\vbox{\hbox{$\sim$}}}}\ }

\begin{document}
\title{ The new spin foam models and quantum gravity}

\date{\today}

\author{Alejandro Perez}

\affiliation{Centre de Physique Th\'eorique\footnote{Unit\'e
Mixte de Recherche (UMR 6207) du CNRS et des Universit\'es
Aix-Marseille I, Aix-Marseille II, et du Sud Toulon-Var; laboratoire
afili\'e \`a la FRUMAM (FR 2291)}, Campus de Luminy, 13288
Marseille, France.}

\begin{abstract}
In this article we give a systematic definition of the recently introduced 
spin foam models for four dimensional quantum gravity reviewing the main results 
on their semiclassical limit on fixed discretizations. \end{abstract}
\maketitle

\section{Introduction}

The quantization of the gravitational interaction is a major open challenge 
in theoretical physics. This review presents the status of the {\em spin foam approach}
to the problem. Spin foam models are definitions of the path integral formulation 
of quantum general relativity and are expected to be the covariant counterpart of the
background independent canonical quantization of general relativity known as 
{\em loop quantum gravity} \cite{book, bookt, ash10}. 

This article concentrates on the definition of the recently introduced
 Engle-Pereira-Rovelli-Livine (EPRL) model\cite{Engle:2007uq, Engle:2007wy} and the closely related Freidel-Krasnov (FK) model \cite{Freidel:2007py}.
 An important original feature of the present 
paper is the explicit derivation of both the Riemannian and the Lorentzian models 
in terms of a notation that exhibits the close relationship between the two at the algebraic level that might 
signal a possible deeper relationship at the level of transition amplitudes. 
   
We will take Plebanski's perspective where general relativity is formulated as a constrained BF theory (for a review introducing the new models from 
a bottom-up perspective see \cite{Rovelli:2011eq}; for an extended version of the present review including a wide collection of related work see \cite{mimi}).
For that reason it will be convenient to start this review by introducing the exact spin foam quantization of BF in the following section.
In Section \ref{eprl-r} we present the EPRL model in both its Riemannian and Lorentzian versions.
A unified treatment of the representation theory of the relevant gauge groups is presented in that section.
In Section \ref{fk} we introduce the FK model and discuss its relationship with the EPRL model.
In Section \ref{boundarya} we describe the structure of the boundary states of these 
model and emphasize the relationship with the kinematical Hilbert space of loop quantum gravity.
In Section \ref{further} we give a compendium of important issues (and associated references) 
that have been left out but which are important for future developpement.  Finally, in section \ref{semiclas} we present the resent 
encouraging results on the nature of the semiclassical limit of the new models.

\section{Spinfoam quantization of BF theory}\label{BF}

We will start by briefly reviewing  the spin foam quantization of BF theory. This section will be the basic building block for the construction of the models of quantum 
gravity that this article is about. The key idea is that the quantum transition amplitudes (computes in the path integral representation) of gravity can be obtained by suitably restricting the histories that are summed over in the
spin foam representation of exactly solvable BF theory. We describe the nature of these constraints at the end of this section.

Here one follow the perspective of \cite{baez5}.
Let $G$ be a compact group whose Lie algebra $\frak g$ has an invariant 
inner product here denoted $\langle \rangle$, and $\cal M$ a ${\rm d}$-dimensional manifold.
Classical BF theory  is defined by the action
\begin{equation}\label{BFT}
S[{\rm B},\omega]=\int \limits_{\cal M}\langle {\rm B}\wedge {\rm F}(\omega) \rangle,
\end{equation}
where ${\rm B}$ is a $\frak g$ valued $({\rm d}-2)$-form,
$\omega$ is a connection on a $G$ principal bundle over
$\cal M$. The theory has no local excitations: all solutions of the equations of motion are locally related by gauge transformations. 
More precisely, the  gauge symmetries of the action are
the local $G$ gauge transformations
\begin{equation}\label{gauge1g}
\delta {\rm B} = \left[{\rm B},\alpha \right], \ \ \ \ \ \ \ \ \ \delta \omega
= {\rm d}_{\omega} \alpha,
\end{equation}
where $\alpha$ is a $\frak g$-valued $0$-form, and the
`topological' gauge transformation
\begin{equation}\label{gauge2g}
\delta {\rm B} = {\rm d}_{\omega} \eta, \ \ \ \ \ \ \ \ \ \delta \omega =
0,
\end{equation}
where ${\rm d}_{\omega}$ denotes the covariant exterior derivative
and $\eta$ is a ${\frak g}$-valued $0$-form. The first
invariance is manifest from the form of the action, while the
second is a consequence of the Bianchi identity, ${\rm d}_{
\omega}F(\omega)=0$. The gauge symmetries are so vast that all
the solutions to the equations of motion are locally pure gauge.
The theory has only global or topological degrees of freedom.
%

For the moment we assume ${\cal M}$ to be a compact and
orientable.
The partition
function, ${\cal Z}$, is formally given by
\begin{equation}\label{zbfg}
{\cal Z}=\int  {\cal D}[{\rm B}] {\cal D}[\omega]\ \ \exp(i \int_{\va \cal M}
\langle {\rm B}\wedge F(\omega)\rangle).
\end{equation}
Formally integrating over the ${\rm B}$ field in (\ref{zbfg}) we
obtain
\begin{equation}\label{VAg}
{\cal Z}=\int {\cal D}[\omega] \ \ \delta \left(F(\omega)\right).
\end{equation}
The partition function ${\cal Z}$ corresponds to the `volume' of
the space of flat connections on $\cal M$.

\begin{figure}[h]
 \centerline{\hspace{0.5cm} \(
\begin{array}{c}
\includegraphics[width=5cm]{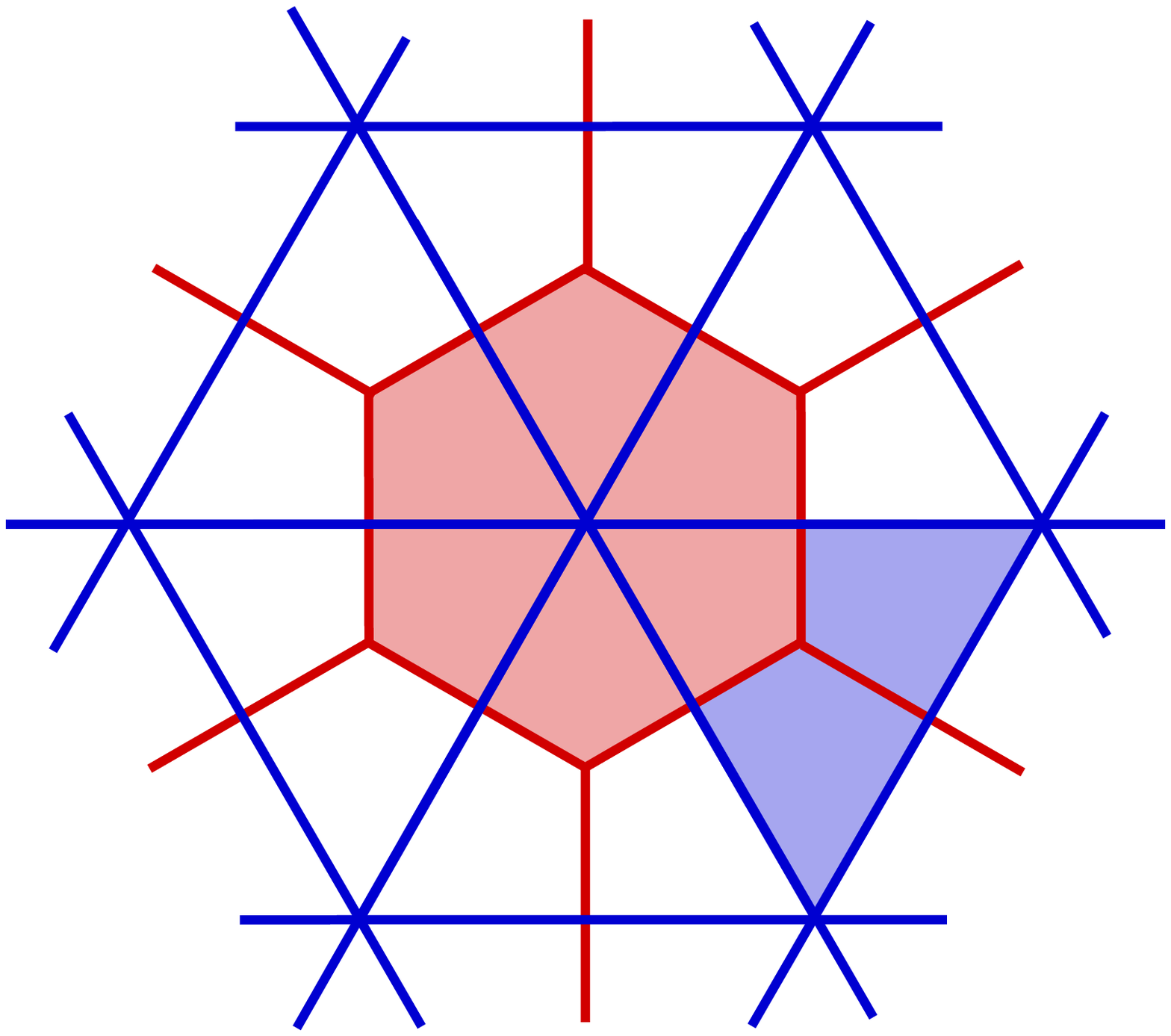}
\end{array}\ \ \ \ \ \ \ \ \  \ \ \ \ \ \ \ \ \ \ \ 
\begin{array}{c}
\includegraphics[width=4.5cm]{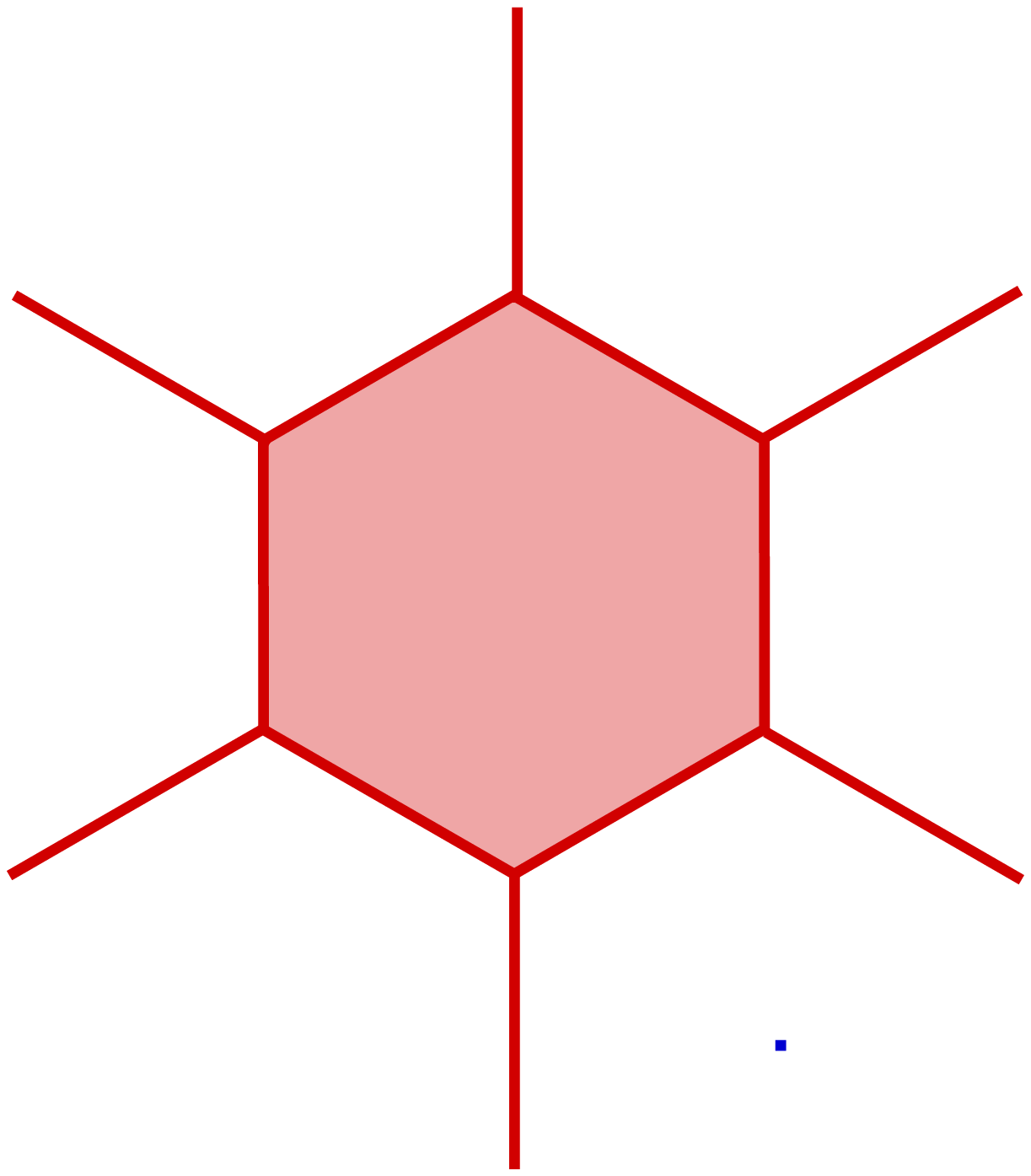}
\end{array}
\)}
\caption{On the left: a triangulation and its dual in two dimensions. On the right: the dual two complex; faces (shaded polygone) are dual to $0$-simpleces in 2d.} \label{cell2}
\end{figure}

\begin{figure}[h]
 \centerline{\hspace{0.5cm} \(
\begin{array}{c}
\includegraphics[width=5cm]{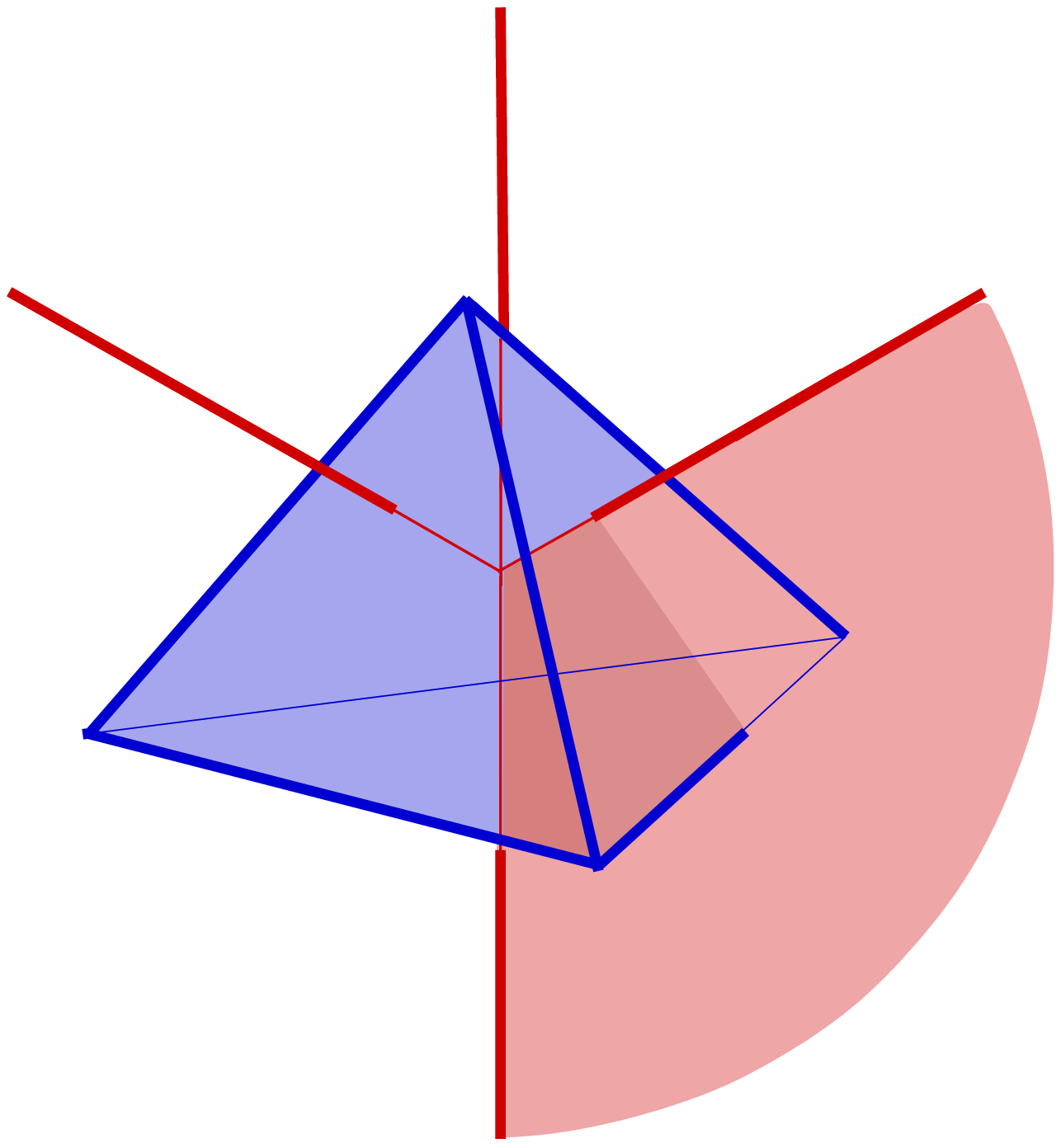}
\end{array}\ \ \ \ \ \ \ \ \  \ \ \ \ \ \ \ \ \ \ \ 
\begin{array}{c}
\includegraphics[width=4cm]{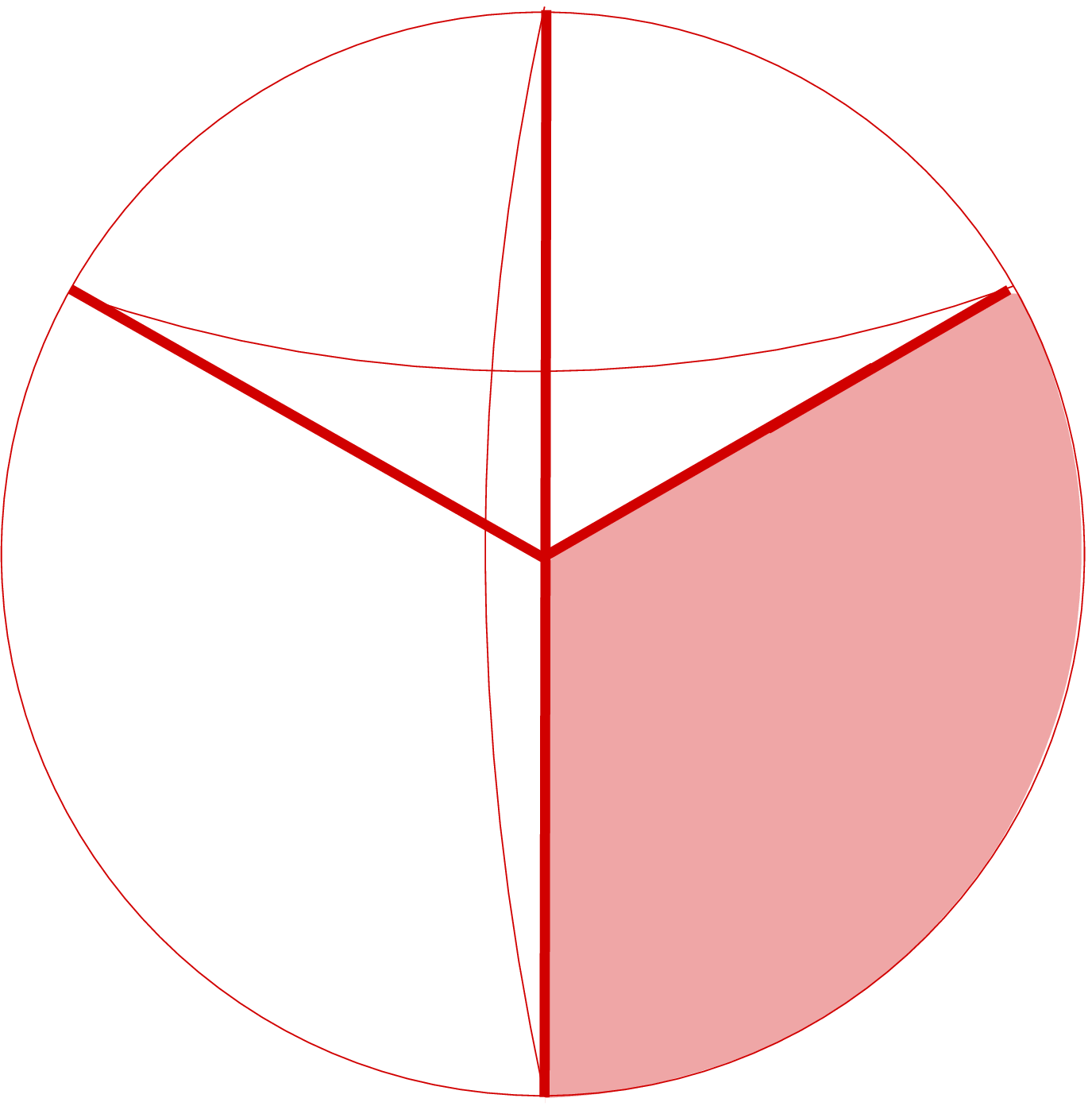}
\end{array}\) }
\caption{On the left: a triangulation and its dual in three dimensions. On the right: the dual two complex; faces (shaded wedge) are dual to $1$-simpleces in 3d.} \label{cell3}
\end{figure}

\begin{figure}[h]
 \centerline{\hspace{0.5cm} \(
\begin{array}{c}
\includegraphics[width=4.5cm]{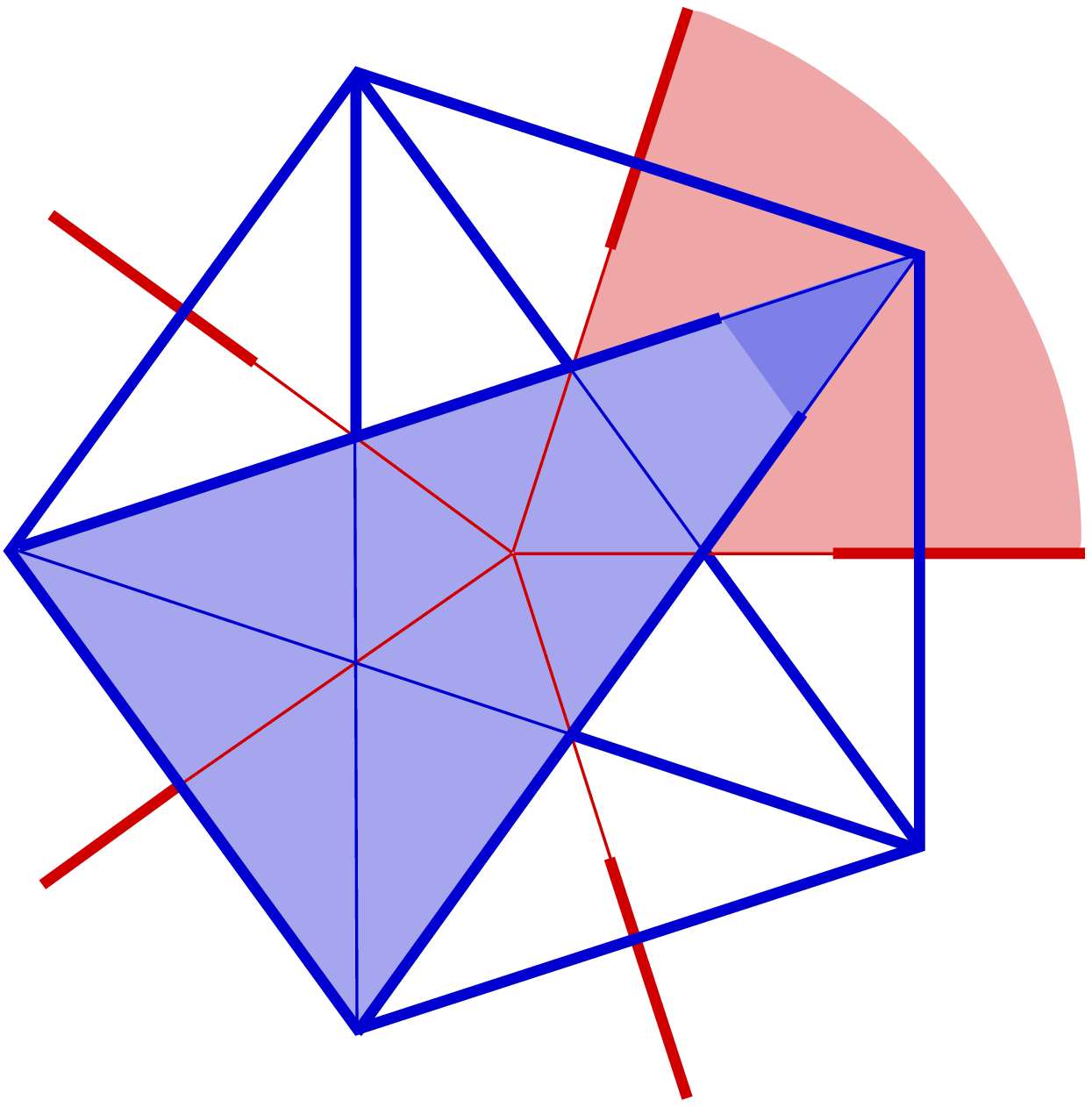}
\end{array}\ \ \ \ \ \ \ \ \  \ \ \ \ \ \ \ \ \ \ \ 
\begin{array}{c}
\includegraphics[width=4cm]{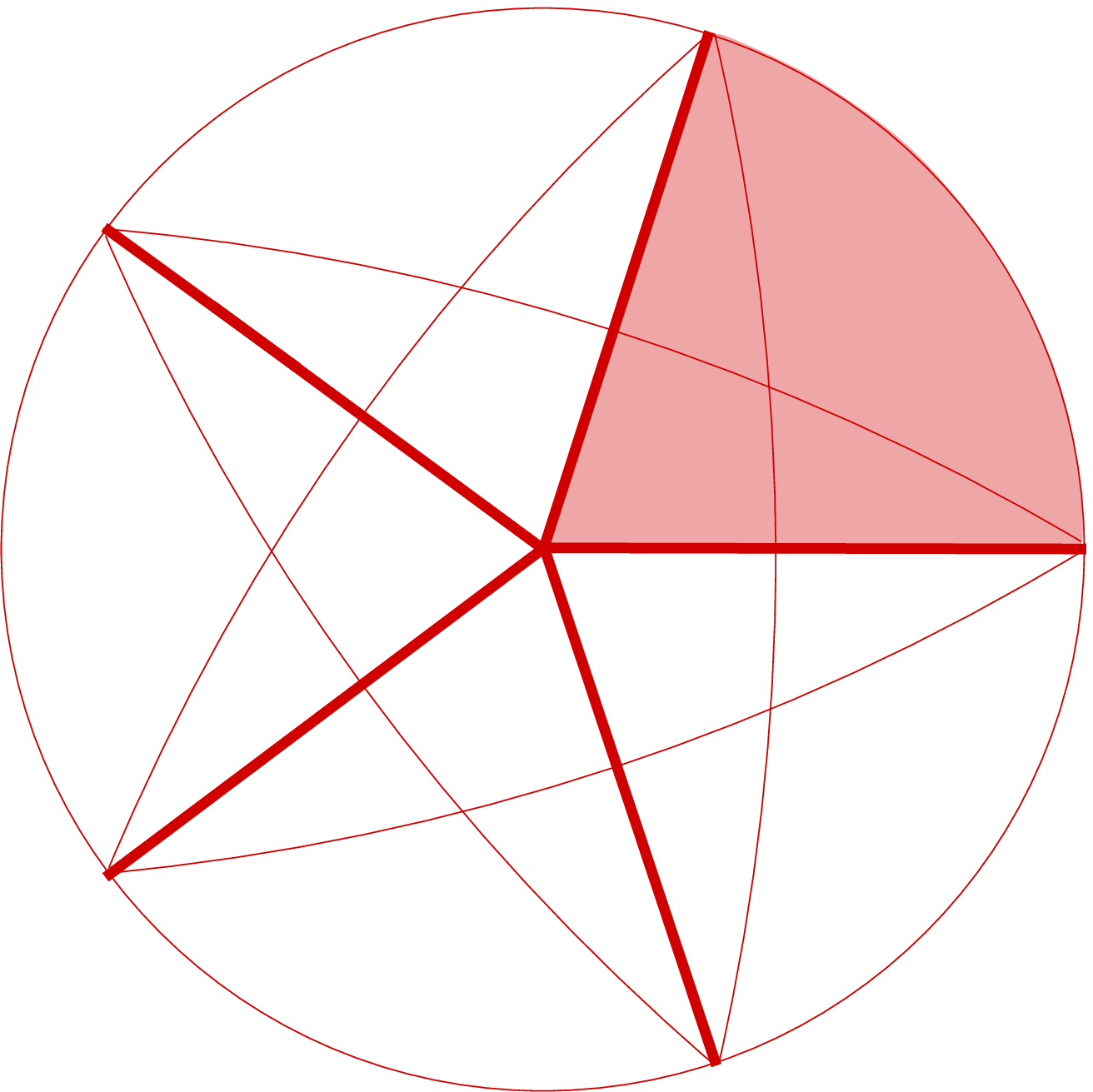}
\end{array}
\) }
\caption{On the left: a triangulation and its dual in four dimensions. On the right: the dual two complex; faces (shaded wedge) are dual to triangles in 4d. The shaded triangle dual to the shaded face is exhibited.} \label{cell4}
\end{figure}

In order to give a meaning to the formal expressions above, we
replace the ${\rm d}$-dimensional manifold ${\cal M}$ with an arbitrary
cellular decomposition $\Delta$. We also need the notion of the
associated dual 2-complex of $\Delta$ denoted by 
$\Delta^{\star}$. The dual 2-complex $\Delta^{\star}$ is a
combinatorial object defined by a set of vertices 
$v\in \Delta^{\star}$ (dual to d-cells in $\Delta$) edges 
$e\in \Delta^{\star}$ (dual to (d$-1$)-cells in $\Delta$) and faces $f\in \Delta^{\star}$ (dual to (d$-2$)-cells in $\Delta$).
In the case where $\Delta$ is a simplicial decomposition of $\cal M$ the structure of both $\Delta$ and $\Delta^{\star}$ is illustrated in Figures \ref{cell2}, \ref{cell3}, and \ref{cell4}
in two, three, and four dimensions  respe1ctively.

For simplicity we concentrate in the 
case when $\Delta$ is a triangulation. The field ${\rm B}$ is associated
with Lie algebra elements $B_f$ assigned to faces $f\in \Delta^{\star}$. We can think of it as
the integral of the (d$-2$)-form ${\rm B}$ on the (d$-2$)-cell dual to the face $f\in \Delta^{\star}$, namely \be\label{Bdisc}
B_{f}=\int\limits_{({\rm d}-2){\rm -cell}}\!\!\! {\rm B}.
\ee In other words
$B_f$ can be interpreted as the `smearing' of the continuous (d$-2$)-form ${\rm B}$ on
the (d$-2$)-cells in  $\Delta$. We use the one-to-one
correspondence between faces $f\in {\Delta}^{\star}$ and (d$-2$)-cells in  $\Delta$ to label the discretization of the ${\rm B}$ field $B_f$.
The connection $\omega$ is discretized by the assignment of group elements $g_e \in G$
to edges $e\in \Delta^{\star}$. One can think of the group elements $g_e$ as the holonomy of $\omega$ along 
$e\in \Delta^{\star}$, namely
\be
g_e={\rm  P} \exp (-\int_e \omega),
\ee
where the symbol $``{\rm  P} \exp"$ denotes the path-order-exponential that reminds us of the relationship of the holonomy with
the connection along the path $e\in \Delta^{\star}$.

With all this the discretized version of  the path integral (\ref{zbfg}) is
\begin{equation}\label{papart}Z(\Delta)= \int \prod_{e \in \Delta^{\star}}
dg_e \prod_{f \in \Delta^{\star}} dB_f \ e^{iB_f U_f} = \int \prod_{e \in \Delta^{\star}}
dg_e \prod_{f \in \Delta^{\star}} \delta(g_{e_1} \cdots g_{e_n}),
\end{equation}
where $U_f=g_{e_1} \cdots g_{e_n}$ denotes the holonomy around faces, and 
the second equation is the result of the ${\rm B}$ integration: it can be thus regarded as the analog of (\ref{VAg}). 
The integration measure $dB_f$ is the standard Lebesgue measure while the integration 
in the group variables is done in terms of the invariant measure in $G$ (which is the unique Haar measure when $G$ is compact).
For given $h\in G$ and test function $F(g)$ the invariance property reads as follows
\be\label{invariance}
\int dg F(g)=\int dg F(g^{-1})=\int dg F(gh)=\int dg F(hg) 
\ee
The 
Peter-Weyl's theorem provides a useful formula or the Dirac delta distribution appearing in (\ref{papart}), namely
\be\label{deltaPW}
\delta(g)=\sum_{\rho} d_{\rho} {\rm Tr}[\rho(g)],\ee where $\rho$ are irreducible unitary representations of $G$. From the previous expression one obtains
\begin{equation}\label{coloring4}
{\cal Z}(\Delta)=\sum \limits_{{\cal C}:\{\rho\} \rightarrow \{
f\}} \int \ \prod_{e \in \Delta^{\star}} dg_e \ \prod_{f
\in \Delta^{\star}} {\rm d}_{\rho_f} \ {\rm
Tr}\left[\rho_f(g^1_e\dots g^{\va N}_e)\right].
\end{equation}
Integration over the
connection can be performed as follows. 
In a triangulation $\Delta$, the edges $e\in \Delta^{\star}$ bound
precisely $\rm d$ different faces; therefore, the $g_e$'s in (\ref{coloring4})
appear in $\rm d$ different traces. The relevant
formula is
\begin{equation}\label{4dp}
P^{e}_{inv}(\rho_1,\cdots, \rho_{\rm d}):= \int dg_{e}\ {\rho_1(g_{e})}\otimes \rho_2(g_{e}) \otimes \cdots \otimes \rho_{\rm d}(g_{e}).
\end{equation}
For compact $G$ it is easy to prove using the invariance (and normalization) of the the integration measure (\ref{invariance}) that $P^{e}_{inv}=(P^{e}_{inv})^2$ is the projector onto ${\rm Inv}[\rho_1\otimes \rho_2 \otimes \cdots
\otimes \rho_{\rm d}]$. In this way the spin foam amplitudes of $SO(4)$ BF theory reduce to 
\begin{eqnarray}\label{bf4} Z_{BF}(\Delta)=\sum \limits_{ {\cal
C}_f:\{f\} \rightarrow \rho_f }  \ \prod_{f \in \Delta^{\star}} {\rm d}_{\rho_f}
\prod_{e \in {\Delta^{\star}}} P^{e}_{inv}(\rho_1,\cdots, \rho_{\rm d}).
\end{eqnarray}
In other words, the $BF$ amplitude associated to a two complex $\Delta^{\star}$ is simply given by 
the sum over all possible assignments of irreducible representations of $G$ to faces of the number obtained by the natural contraction of the network of projectors $P^e_{inv}$ according to the 
pattern provided defined by the two-complex 
$\Delta^{\star}$. 

There is a nice graphical representation of the partition function of BF theory that will be very useful for some calculations.
On the one hand, using this graphical notation  one can easily prove the discretization independence of the BF amplitudes. On the other hand this graphical notation will simplify the presentation of the new spin foam models of quantum gravity that will be considered in the following sections. This useful notation was introduced by Oeckl \cite{Oeckl:2005rh,  Oeckl:2000hs} and used in \cite{Girelli:2001wr} to give a general prove of the discretization independence of the BF partition function and the Turaev-Viro invariants for their definition on general cellular decompositions.

Let us try to present this notation in more detail:
The idea is to represent each representation matrix appearing in (\ref{coloring4}) by a line (called a wire) labeled by an irreducible representation, and integrations on the group by a box (called a cable). The traces in equation (\ref{coloring4}) imply that there is a wire, labelled by the representation $\rho_f$, winding around each face  $f\in \Delta^{\star}$. In addition, there is a cable (integration on the group) associated with each edge $e\in \Delta^{\star}$.  As in (\ref{bf4}), there is a projector $P^{e}_{inv}$ is the projector into ${\rm Inv}[\rho_1\otimes \rho_2 \otimes \cdots \otimes \rho_{\rm d}]$ associated to each edge. This will be represented by a cable with $\rm d$ wires as shown in (\ref{cabled}). Such graphical representation allows for a simple diagrammatic expression of the BF quantum amplitudes. 
\be P^{e}_{inv}(\rho_1,\rho_2, \rho_3,\cdots, \rho_{\rm d })\ \equiv \!\!\!\!\!\!\!\!
\psfrag{a}{\!\!\!\!\!\!\!\!\!\!\!\!\!\!\!\!\!\!\!\!\!\!\!\!\!\!\!\!\!\!\!$$}
\psfrag{x}{\!\!$\rho_1$}\psfrag{y}{\!\!$\rho_2$}\psfrag{z}{\!\!$\rho_3$}\psfrag{w}{$\cdots$}\psfrag{u}{\!\!$\rho_{\rm d}$}
\begin{array}{c}
\includegraphics[width=3cm]{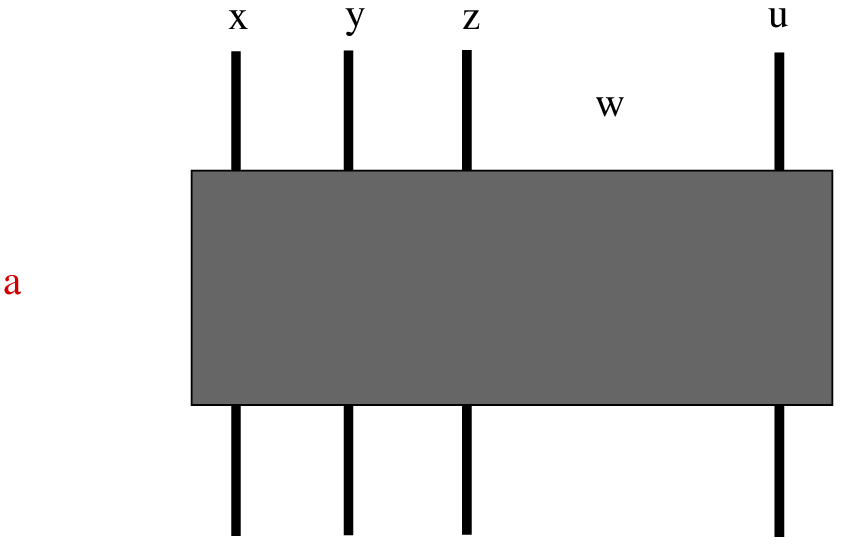}
\end{array}
\label{cabled}
\ee

The case of physical interest is $d=4$.  In such case edges are shared by four faces; each cable has now four wires. The
cable wire diagram giving the BF amplitude is dictated by the combinatorics of the dual two complex $\Delta^{\star}$. From Figure \ref{cell4} one gets 
\be
Z_{BF}(\Delta)=\sum \limits_{ {\cal
C}_f:\{f\} \rightarrow \rho_f }  \ \prod\limits_{f \in \Delta^{\star}} {\rm d}_{\rho}
\begin{array}{c}\psfrag{a}{$\rho_{1}$}
\psfrag{b}{$\rho_{2}$}
\psfrag{c}{$\rho_{3}$}
\psfrag{d}{$\rho_{4}$}
\psfrag{f}{$\rho_{5}$}
\psfrag{g}{$\rho_{6}$}
\psfrag{h}{$\rho_{7}$}
\psfrag{i}{$\rho_{8}$}
\psfrag{j}{$\rho_{9}$}
\psfrag{k}{$\rho_{10}$}
\includegraphics[width=5cm]{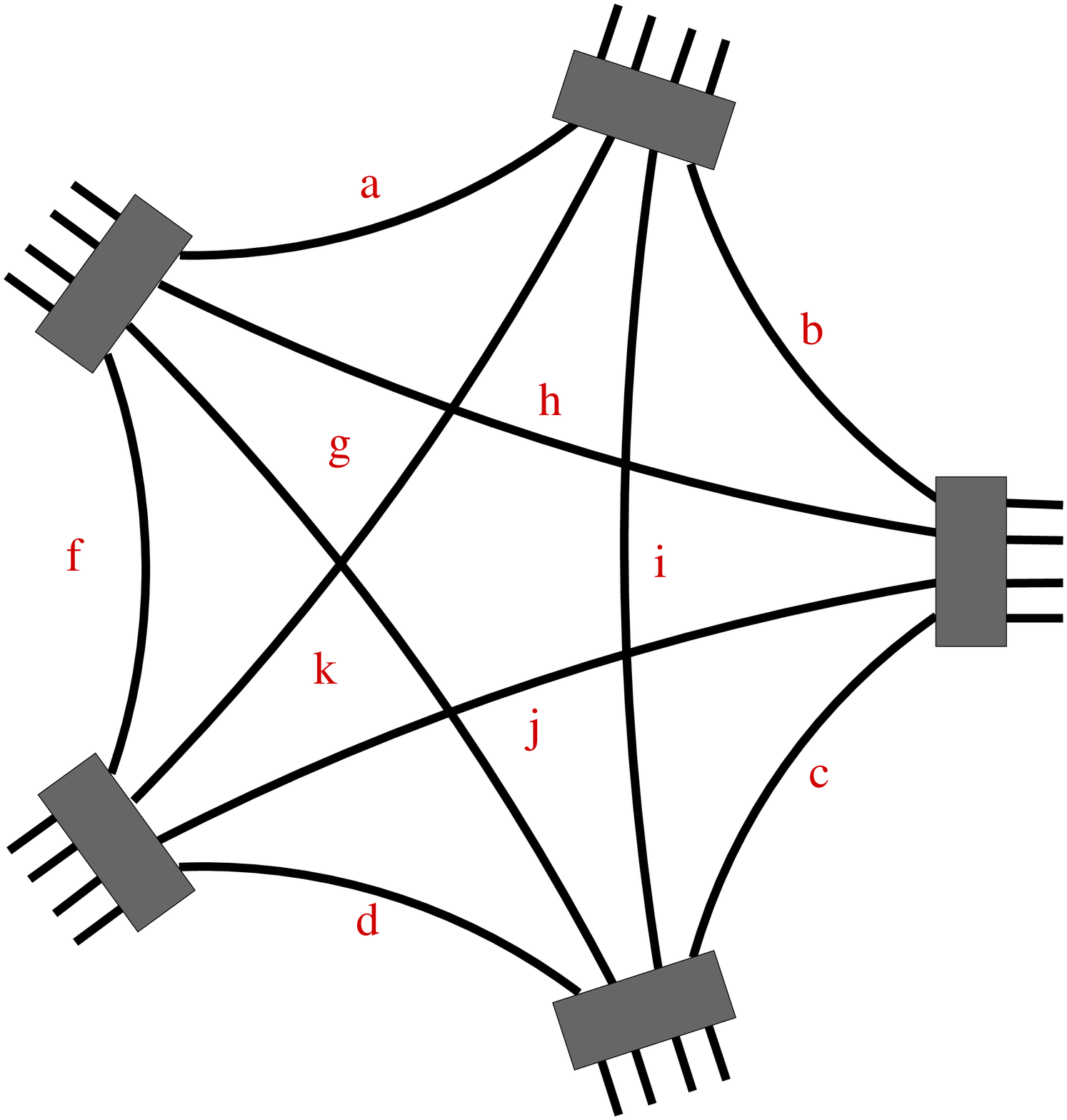}
\end{array}
\label{cabled-4d}.
\ee
 The $10$ wires corresponding to the $10$ faces $f\in \Delta^{\star}$ sharing a vertex $v\in \Delta^{\star}$ are connected to the neighbouring  vertices through the $5$ cables (representing the projectors in (\ref{bf4}) and Figure \ref{cabled}) associated to the $5$ edges  $e\in \Delta^{\star}$ sharing the vertex $v\in \Delta^{\star}$.

\subsubsection{$SU(2)\times SU(2)$ BF theory: a starting point for 4d Riemannian gravity.}

We now present the BF quantum amplitudes in the case $G=SU(2)\times SU(2)$. This special case is of fundamental importance in the construction of the gravity models presented in the following  sections. The product form of the structure group implies the simple relationship $Z_{BF}(SU(2)\times SU(2))=Z_{BF}(SU(2))^2$. Nevertheless, it is important for us to present this example in explicit form as it will provide the graphical notation that is needed to introduce the gravity models in a simple manner. The spin foam representation of the BF partition function follows from expressing the projectors in (\ref{cabled-4d}) in the orthonormal basis of intertwiners, i.e. invariant vectors in ${\rm Inv}[\rho_1\otimes\cdots\otimes\rho_4]$. From the product form of the structure group one has  
\be
\begin{array}{c}
\psfrag{x}{$\rho_1$}
\psfrag{y}{$\rho_2$}
\psfrag{z}{$\rho_3$}
\psfrag{w}{$\rho_4$}
\psfrag{h}{ $=$}
\psfrag{a}{$j^{-}_1$}
\psfrag{b}{$j^{-}_2$}
\psfrag{c}{$j^{-}_3$}
\psfrag{d}{$j^{-}_4$}
\psfrag{A}{$j^{+}_1$}
\psfrag{B}{$j^{+}_2$}
\psfrag{C}{$j^{+}_3$}
\psfrag{D}{$j^{+}_4$}
\psfrag{im}{$\iota^{-}$}
\psfrag{ip}{$\iota^{+}$}
\psfrag{g}{$={\LARGE \sum\limits_{\iota^{-}\iota^{+}}}$}
\includegraphics[height=2cm]{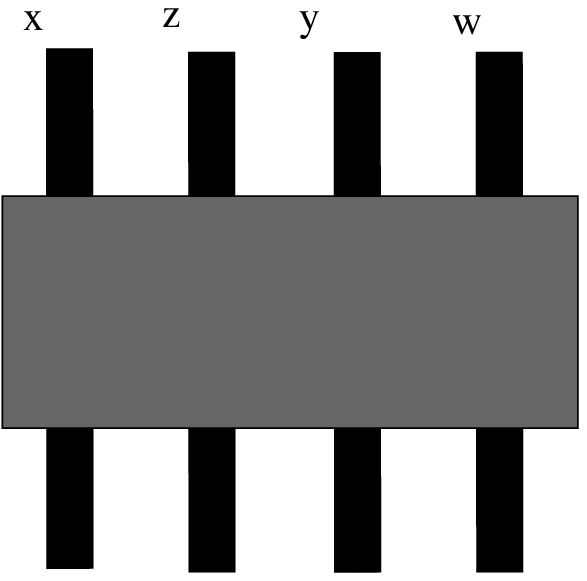}
\end{array}=
\begin{array}{c}
\psfrag{x}{$\rho_1$}
\psfrag{y}{$\rho_2$}
\psfrag{z}{$\rho_3$}
\psfrag{w}{$\rho_4$}
\psfrag{h}{ $=$}
\psfrag{a}{$j^{-}_1$}
\psfrag{b}{$j^{-}_2$}
\psfrag{c}{$j^{-}_3$}
\psfrag{d}{$j^{-}_4$}
\psfrag{A}{$j^{+}_1$}
\psfrag{B}{$j^{+}_2$}
\psfrag{C}{$j^{+}_3$}
\psfrag{D}{$j^{+}_4$}
\psfrag{im}{$\iota^{-}$}
\psfrag{ip}{$\iota^{+}$}
\psfrag{g}{$={\LARGE \sum\limits_{\iota^{-}\iota^{+}}}$}
\includegraphics[height=2cm]{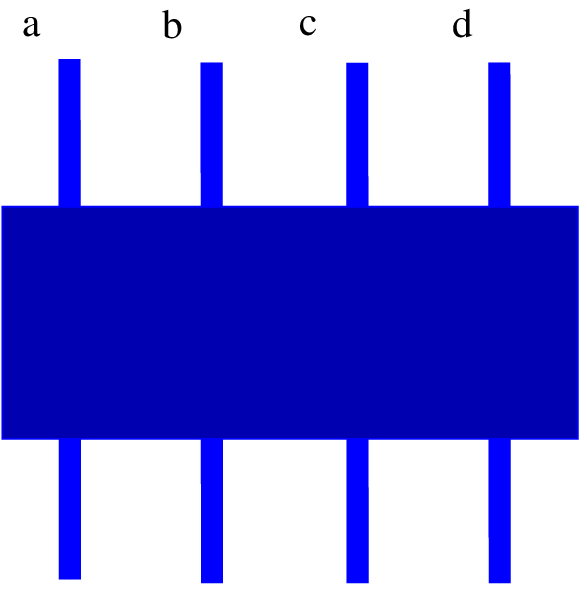}
\end{array}
\begin{array}{c}
\psfrag{x}{$\rho_1$}
\psfrag{y}{$\rho_2$}
\psfrag{z}{$\rho_3$}
\psfrag{w}{$\rho_4$}
\psfrag{h}{ $=$}
\psfrag{a}{$j^{-}_1$}
\psfrag{b}{$j^{-}_2$}
\psfrag{c}{$j^{-}_3$}
\psfrag{d}{$j^{-}_4$}
\psfrag{A}{$j^{+}_1$}
\psfrag{B}{$j^{+}_2$}
\psfrag{C}{$j^{+}_3$}
\psfrag{D}{$j^{+}_4$}
\psfrag{im}{$\iota^{-}$}
\psfrag{ip}{$\iota^{+}$}
\psfrag{g}{$={\LARGE \sum\limits_{\iota^{-}\iota^{+}}}$}
\includegraphics[height=2cm]{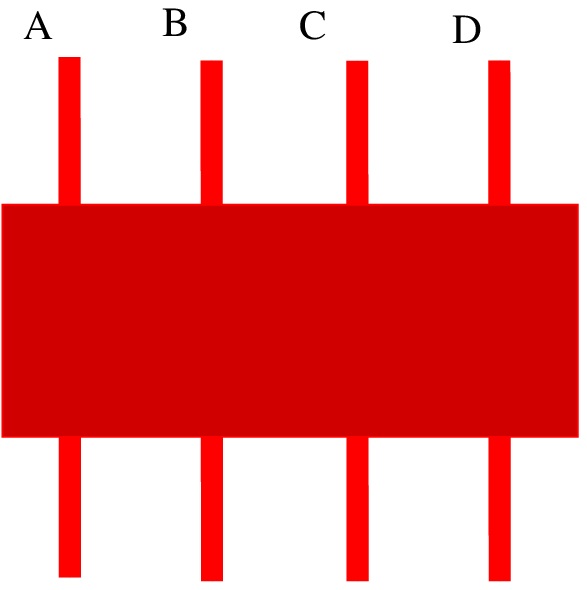}
\end{array}=\sum\limits_{\iota^{-}\iota^{+}}
\begin{array}{c}
\psfrag{x}{$\rho_1$}
\psfrag{y}{$\rho_2$}
\psfrag{z}{$\rho_3$}
\psfrag{w}{$\rho_4$}
\psfrag{h}{ $=$}
\psfrag{a}{$j^{-}_1$}
\psfrag{b}{$j^{-}_2$}
\psfrag{c}{$j^{-}_3$}
\psfrag{d}{$j^{-}_4$}
\psfrag{A}{$j^{+}_1$}
\psfrag{B}{$j^{+}_2$}
\psfrag{C}{$j^{+}_3$}
\psfrag{D}{$j^{+}_4$}
\psfrag{im}{$\iota^{-}$}
\psfrag{ip}{$\iota^{+}$}
\psfrag{g}{$={\LARGE }$}
\includegraphics[height=2cm]{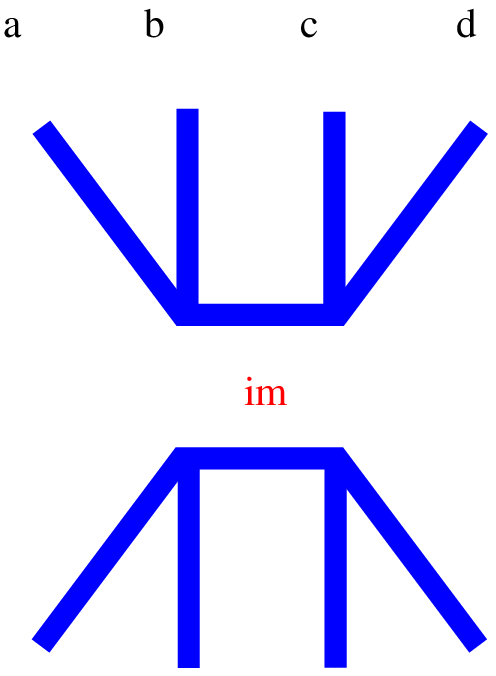}
\end{array}\ \ \ \ 
\begin{array}{c}
\psfrag{x}{$\rho_1$}
\psfrag{y}{$\rho_2$}
\psfrag{z}{$\rho_3$}
\psfrag{w}{$\rho_4$}
\psfrag{h}{ $=$}
\psfrag{a}{$j^{-}_1$}
\psfrag{b}{$j^{-}_2$}
\psfrag{c}{$j^{-}_3$}
\psfrag{d}{$j^{-}_4$}
\psfrag{A}{$j^{+}_1$}
\psfrag{B}{$j^{+}_2$}
\psfrag{C}{$j^{+}_3$}
\psfrag{D}{$j^{+}_4$}
\psfrag{im}{$\iota^{-}$}
\psfrag{ip}{$\iota^{+}$}
\psfrag{g}{$={\LARGE \sum\limits_{\iota^{-}\iota^{+}}}$}
\includegraphics[height=2cm]{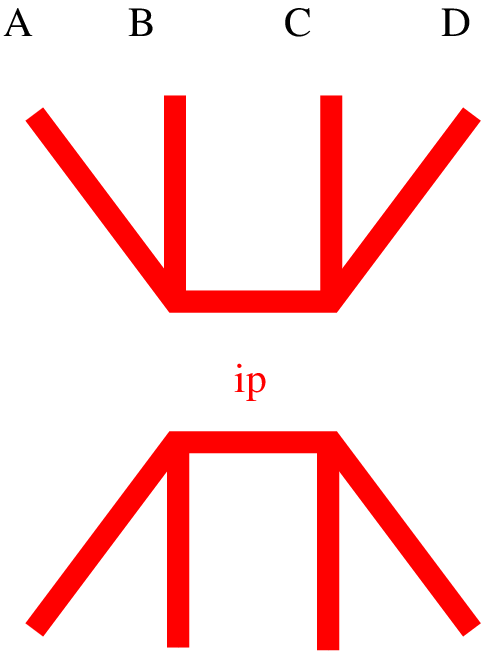},
\end{array}
\label{cab-44d}
\ee 
 where $\rho_f=j_f^{-}\otimes j_f^+$,  and $j_f^{\pm}$ and $\iota^{\pm}$ are half integers labelling left and right representations of $SU(2)$ that 
 defined the irreducible unitary representations of $G=SU(2)\times SU(2)$, and we have used the expression of the right and left $SU(2)$ projectors in a basis of intertwiners, namely
 \be
\begin{array}{c}\psfrag{x}{$\rho_1$}
\psfrag{y}{$\rho_2$}
\psfrag{z}{$\rho_3$}
\psfrag{w}{$\rho_4$}
\psfrag{h}{ $ $}
\psfrag{a}{$j^{}_1$}
\psfrag{b}{$j^{}_2$}
\psfrag{c}{$j^{}_3$}
\psfrag{d}{$j^{}_4$}
\psfrag{A}{$j^{+}_1$}
\psfrag{B}{$j^{+}_2$}
\psfrag{C}{$j^{+}_3$}
\psfrag{D}{$j^{+}_4$}
\psfrag{im}{$\iota^{}$}
\psfrag{ip}{$\iota^{+}$}
\includegraphics[height=2cm]{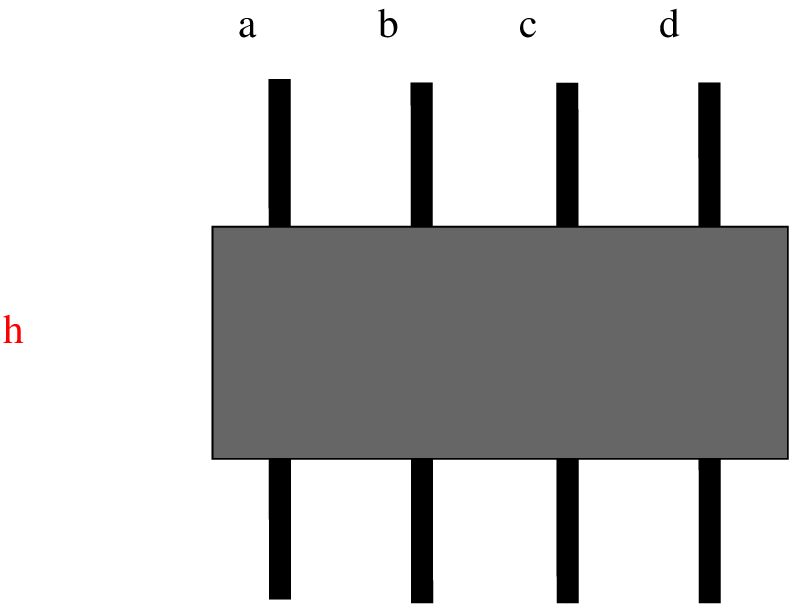}
\end{array}
={\sum\limits_{\iota^{}}}
\begin{array}{c}\psfrag{x}{$\rho_1$}
\psfrag{y}{$\rho_2$}
\psfrag{z}{$\rho_3$}
\psfrag{w}{$\rho_4$}
\psfrag{h}{ $ $}
\psfrag{a}{$j^{}_1$}
\psfrag{b}{$j^{}_2$}
\psfrag{c}{$j^{}_3$}
\psfrag{d}{$j^{}_4$}
\psfrag{A}{$j^{+}_1$}
\psfrag{B}{$j^{+}_2$}
\psfrag{C}{$j^{+}_3$}
\psfrag{D}{$j^{+}_4$}
\psfrag{im}{$\iota^{}$}
\psfrag{ip}{$\iota^{+}$}
\psfrag{g}{$$}
\includegraphics[height=2cm]{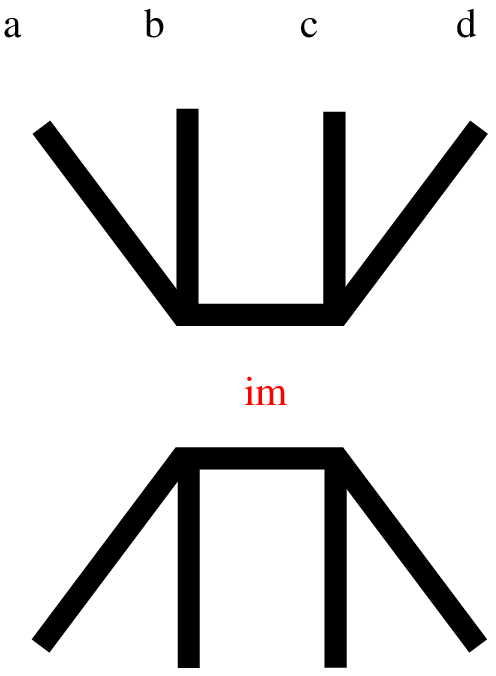}
\end{array},
\label{cab-3d}
\ee
where the four-leg objects on the right hand side denote the invariant vectors spanning a basis of ${\rm Inv}[j_1\otimes\cdots\otimes j_4]$, and $\iota$
is a half integer labelling those elements.
Accordingly, when replacing the previous expression in (\ref{cabled-4d}) one gets 
\be\label{bf-so4}
Z_{BF}(\Delta)=\sum \limits_{ {\cal
C}_f:\{f\} \rightarrow \rho_f }  \ \prod\limits_{f \in \Delta^{\star}} {\rm d}_{j_f^{-}}{\rm d}_{j_f^{+}}
\begin{array}{c}
\includegraphics[width=5cm]{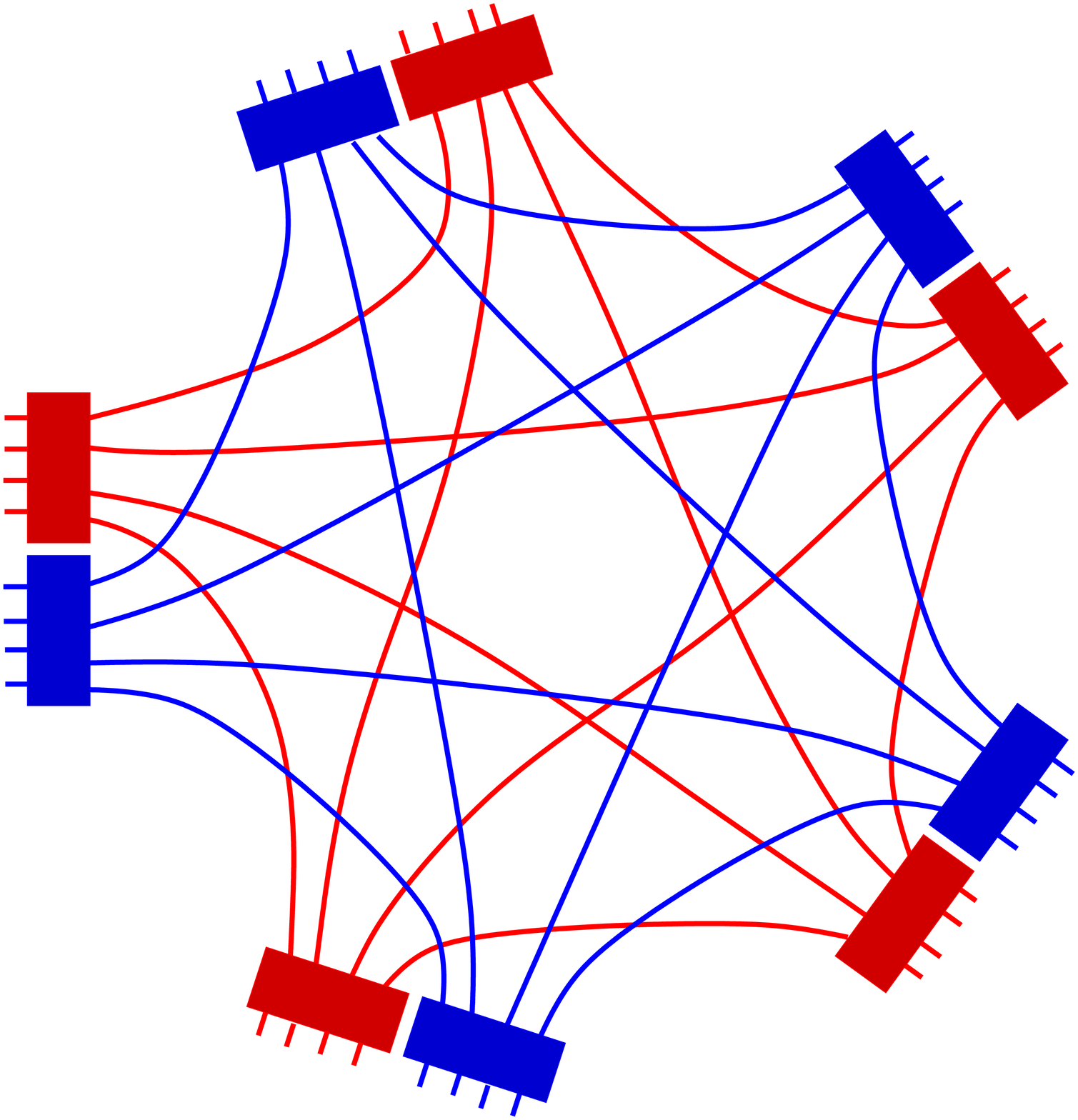}
\end{array},
\ee and equivalently
\be
Z_{BF}(\Delta)=\sum \limits_{ {\cal
C}_f:\{f\} \rightarrow \rho_f }  \ \prod\limits_{f \in \Delta^{\star}} {\rm d}_{j_f^{-}}{\rm d}_{j_f^{+}} \sum \limits_{ {\cal
C}_e:\{e\} \rightarrow \iota_e } 
\begin{array}{c}
\includegraphics[width=5cm]{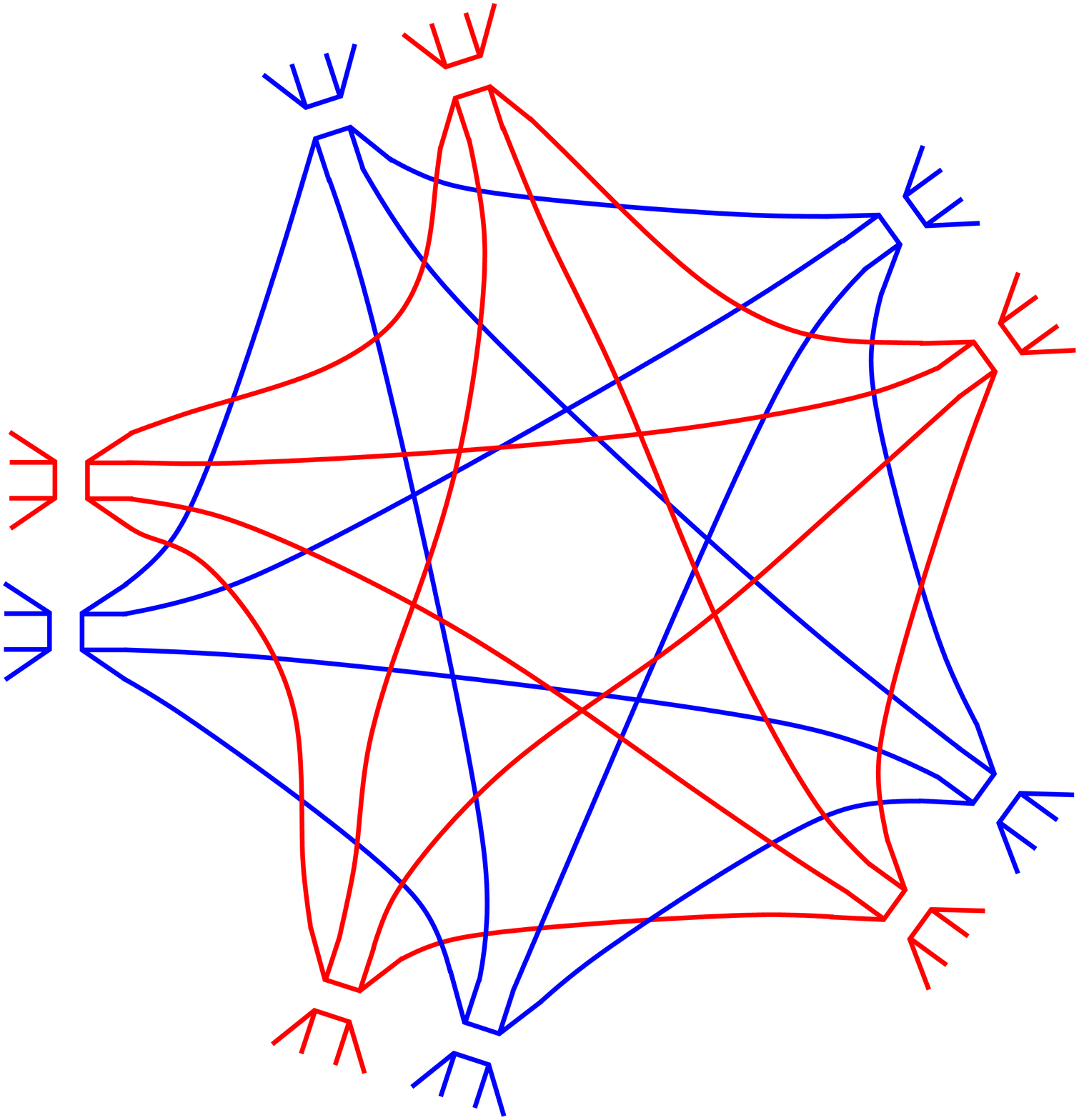}
\end{array}
\ee
from which we finally obtain the spin foam representation of the $SU(2)\times SU(2)$ partition function as a product of two $SU(2)$ amplitudes, namely
\ba &&
Z_{BF}(\Delta)=\sum \limits_{ {\cal
C}_f:\{f\} \rightarrow \rho_f }  \ \prod\limits_{f \in \Delta^{\star}} {\rm d}_{j_f^{-}}{\rm d}_{j_f^{+}} \sum \limits_{ {\cal
C}_e:\{e\} \rightarrow \iota_e } \ \prod\limits_{v\in \Delta^{\star}} \n \\ && \ \ \ \ \ \ \ \ \  \ \ \ \ \ \ \  \ \ \ \ \ \
\begin{array}{c}
\psfrag{a}{$\iota^{-}_1$}
\psfrag{b}{$\iota^{-}_2$}
\psfrag{c}{$\iota^{-}_3$}
\psfrag{d}{$\iota^{-}_4$}
\psfrag{e}{$\iota^{-}_5$}
\psfrag{ap}{$\iota^{+}_1$}
\psfrag{bp}{$\iota^{+}_2$}
\psfrag{cp}{$\iota^{+}_3$}
\psfrag{dp}{$\iota^{+}_4$}
\psfrag{ep}{$\iota^{+}_5$}
\psfrag{A}{$j^{-}_1$}
\psfrag{B}{$j^{-}_2$}
\psfrag{C}{$j^{-}_3$}
\psfrag{D}{$j^{-}_4$}
\psfrag{E}{$j^{-}_5$}
\psfrag{F}{$j^{-}_6$}
\psfrag{G}{$j^{-}_7$}
\psfrag{H}{$j^{-}_8$}
\psfrag{I}{$j^{-}_9$}
\psfrag{J}{$j^{-}_{10}$}
\psfrag{Ap}{$j^{+}_1$}
\psfrag{Bp}{$j^{+}_2$}
\psfrag{Cp}{$j^{+}_3$}
\psfrag{Dp}{$j^{+}_4$}
\psfrag{Ep}{$j^{+}_5$}
\psfrag{Fp}{$j^{+}_6$}
\psfrag{Gp}{$j^{+}_7$}
\psfrag{Hp}{$j^{+}_8$}
\psfrag{Ip}{$j^{+}_9$}
\psfrag{Jp}{$j^{+}_{10}$}
\includegraphics[height=4cm]{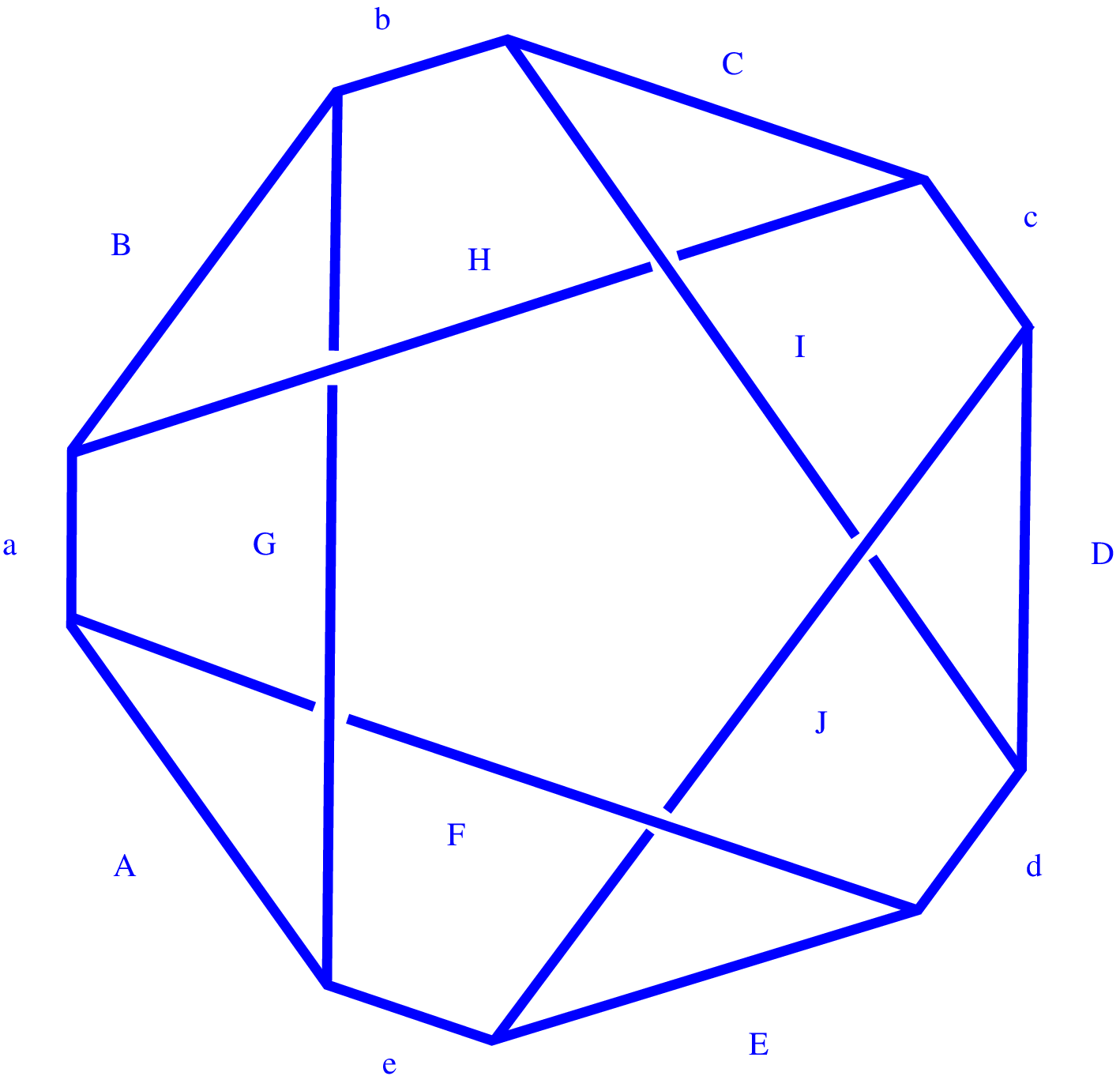}
\end{array}\ \  \ \
\begin{array}{c}
\psfrag{Q}{\!\!\!\!\!\!\!\!\!\!\!\!\!\!\!\!\!\!\!\!\!\!\!\!\!\!\!\!\!\!\!\!\!\!\!\!\!\!\!\!\!\!\!\!\!\!\!\!\!\!\!\!\!\!\!\!\!\!\!\!\!\!\!\!\!\!\!\!\!\!\!\!\!\!\!\!\!\!\!\!\!\!\!\!\!\!\!\!\!\!\!\!\!\!\!\!\!\!\!\!\!\!$Z_{BF}(\Delta)=\sum \limits_{ {\cal
C}_f:\{f\} \rightarrow \rho_f }  \ \prod\limits_{f \in \Delta^{\star}} {\rm d}_{j_f^{-}}{\rm d}_{j_f^{+}} \sum \limits_{ {\cal
C}_e:\{e\} \rightarrow \iota_e } \ \prod\limits_{v\in \Delta^{\star}} $}
\psfrag{a}{$\iota^{-}_1$}
\psfrag{b}{$\iota^{-}_2$}
\psfrag{c}{$\iota^{-}_3$}
\psfrag{d}{$\iota^{-}_4$}
\psfrag{e}{$\iota^{-}_5$}
\psfrag{ap}{$\iota^{+}_1$}
\psfrag{bp}{$\iota^{+}_2$}
\psfrag{cp}{$\iota^{+}_3$}
\psfrag{dp}{$\iota^{+}_4$}
\psfrag{ep}{$\iota^{+}_5$}
\psfrag{A}{$j^{-}_1$}
\psfrag{B}{$j^{-}_2$}
\psfrag{C}{$j^{-}_3$}
\psfrag{D}{$j^{-}_4$}
\psfrag{E}{$j^{-}_5$}
\psfrag{F}{$j^{-}_6$}
\psfrag{G}{$j^{-}_7$}
\psfrag{H}{$j^{-}_8$}
\psfrag{I}{$j^{-}_9$}
\psfrag{J}{$j^{-}_{10}$}
\psfrag{Ap}{$j^{+}_1$}
\psfrag{Bp}{$j^{+}_2$}
\psfrag{Cp}{$j^{+}_3$}
\psfrag{Dp}{$j^{+}_4$}
\psfrag{Ep}{$j^{+}_5$}
\psfrag{Fp}{$j^{+}_6$}
\psfrag{Gp}{$j^{+}_7$}
\psfrag{Hp}{$j^{+}_8$}
\psfrag{Ip}{$j^{+}_9$}
\psfrag{Jp}{$j^{+}_{10}$}
\includegraphics[height=4cm]{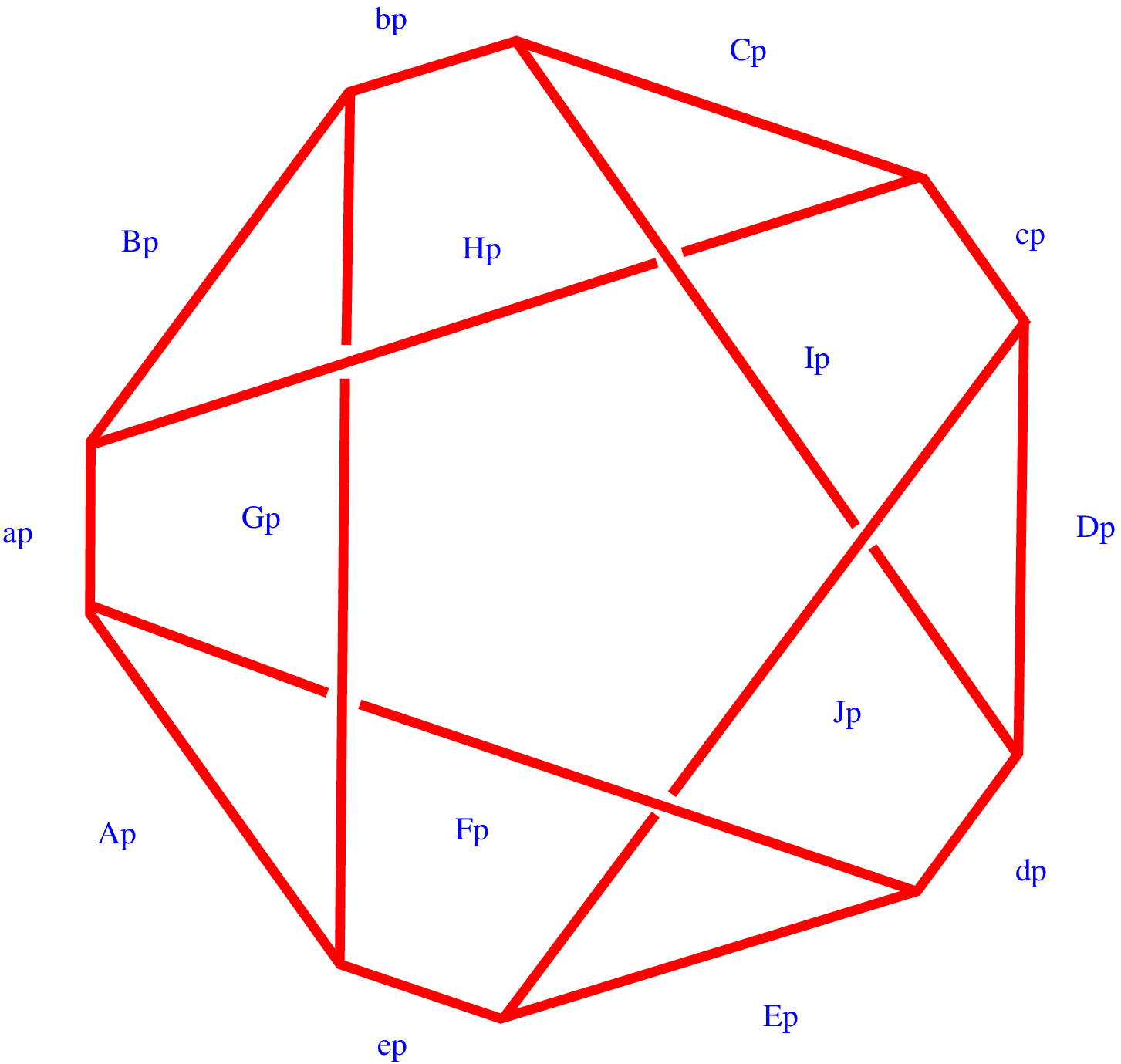}
\end{array}
\label{BF4V-b}
\ea

\subsubsection*{Extra remarks on four dimensional BF theory}

The state sum (\ref{coloring4}) is generically divergent (due to the
gauge freedom analogous to (\ref{gauge2g})). A regularized version
defined in terms of $SU_q(2)\times SU_q(2)$ was introduced by
Crane and Yetter  \cite{crane0, crane00}. As in three dimensions, if an appropriate regularization of bubble divergences is provided,
(\ref{coloring4}) is topologically invariant and the spin foam path
integral is discretization independent.

As in the three dimensional case BF theory in any dimensions can be coupled to topological defects \cite{Baez:2006sa}. In the four dimensional case
defects are string-like \cite{Fairbairn:2007fb} and can carry extra degrees of freedom such as topological Yang-Mills fields \cite{Montesinos:2007dc}.
The possibility that quantum gravity could be defined directly form these simple kind of topological theories has also been considered outside spin foams \cite{'tHooft:2008kk} 
 (for which the UV problem described in the introduction is absent) 
is attractive and should, in my view, be considered further. 

It is also possible to introduce one dimensional particles in four dimensional BF theory and gravity as shown in \cite{Freidel:2006hv}.


Two dimensional BF theory has been used as the basic theory in an attempt to define a manifold independent 
model of QFT in \cite{Livine:2003kn}. It is also related to gravity in two dimensions  in two ways: on the one hand it is  equivalent to the so-called 
Jackiw-Teitelboim model \cite{JTMODEL1, JTMODEL2}, on the other hand it is related to usual 2d gravity via constraints in a way similar to the one exploited in four dimensions
(see next section).  The first relationship has been used in the canonical quantization of the Jackiw-Teitelboim model  in \cite{Constantinidis:2008ty}.
The second relationship has been explored in \cite{Oriti:2004qk}

Three dimensional BF theory and the spin foam quantization presented above is intimately related to classical and 
quantum gravity in three dimensions (for a classic reference see \cite{carlip}). 
The state sum as presented above matches the quantum amplitudes first proposed by Ponzano and Regge in the 60's  based on their 
discovery of the asymptotic expressions of the 6j symbols \cite{ponza} and is 
often referred to as the Ponzano-Regge model. Divergences in the above formal expression require regularization.
Natural regularizations are available and that the model is well defined \cite{Barrett:2008wh, Noui:2004iy, frei9}.
For a detailed study of the divergence structure of the model see \cite{Bonzom:2010ar, Bonzom:2010zh, Bonzom:2011br}. 
The quantum deformed version of the above amplitudes  lead to the so called
Turaev-Viro model \cite{TV}  which is expected to correspond to the quantization of three dimensional Riemannian gravity in the 
presence of a non vanishing positive cosmological constant. For the definition of observables in the latter context as well as in the analog four dimensional analog see \cite{Barrett:2004im}.

The topological character of BF theory can be preserved by the coupling of the theory with topological defects that
play the role of point particles. In the spin foam literature this has been considered form the canonical perspective in \cite{Noui:2004jb, a21}
and from the covariant perspective extensively by Freidel and Louapre \cite{Freidel:2004vi}. These theories have been shown by Freidel and Livine to be 
dual, in a suitable sense, to certain non-commutative fields theories in three dimensions \cite{Freidel:2005bb, Freidel:2005me}.

Concerning coupling BF theory with non topological matter see \cite{Fairbairn:2006dn, Dowdall:2010ej} for the case of fermionic matter, and \cite{Speziale:2007mt} for gauge fields.
A more radical perspective for the definition of matter in 3d gravity is taken in \cite{Fairbairn:2007sv}. For three dimensional supersymmetric  BF theory models see \cite{Livine:2003hn, Baccetti:2010xd}

Recursion relations
for the 6j vertex amplitudes have been investigated in  \cite{Bonzom:2011jh, Dupuis:2009qw}. They provide a tool for studying dynamics in spin foams of 3d gravity and might be useful in higher dimensions \cite{Bonzom:2009zd}.

\subsection{The coherent states representation}\label{cohecohe}

In this section we introduce the coherent state representation of  the $SU(2)$ and $Spin(4)$  path integral of BF theory. 
This will be particularly important for the definition of the models defined by 
Freidel and Krasnov in \cite{Freidel:2007py} that we will address in Section \ref{fk} as well as in the semiclassical analysis of the new models reported in Section \ref{semiclas}.
The relevance of such representation for spin foams was first emphasized  by 
Livine and Speziale in \cite{Livine:2007vk}.

\subsubsection{Coherent states}

Coherent states associated to the representation theory of a compact group have been studied by Thiemann and collaborators \cite{Thiemann:2000zf, Thiemann:2000bw, Sahlmann:2001nv, Thiemann:2000bw, Thiemann:2000ca, Thiemann:2000bx, Thiemann:2000by, Thiemann:2002vj, Bahr:2007xa, Bahr:2007xn, Flori:2008nw}  see also \cite{Bianchi:2009ky}.
Their importance for the new spin foam models was put forward by Livine and Speziale in \cite{Livine:2007vk} where the emphasis is put on coherent states of intertwiners or the so-called quantum tetrahedron (see also \cite{Conrady:2009px}). Here we follow the presentation of \cite{Freidel:2007py}.

In order to built coherent states for $\Spin(4)$ we start by introducing them in the case of $SU(2)$. Starting from the representation space $\sH_j$ of dimension $\rd_j\equiv 2j+1$ one can write
the resolution of the identity in tems of the canonical orthonormal basis $|j,m\rangle$ as
\be\label{ident-m}
1_j = \sum_m |j,m \rangle \langle j,m|,
\ee
where $-j\le m\le j$. There exists an over complete basis $|j,g \rangle \in \sH_{j}$ labelled by $g\in SU(2)$ such that \be\label{ident-coherent}
1_j=
\rd_j \int\limits_{{\rm SU}(2)} dg \, |j,g \rangle \langle j,g|,
\ee
The states  $|j,g \rangle \in \sH_{j}$ are $SU(2)$ coherent states defined by the action of the group on maximum weight
states $|j,j\rangle$ (themselves coherent), namely 
\be\label{def}
|j,g \rangle \equiv g |j,j\rangle = \sum_m  |j,m \rangle D^j_{mj}(g),
\ee
where $D^j_{mj}(g)$ are the matrix elements of the unitary representations in the $|j,m\rangle$ (Wigner matrices).
Equation (\ref{ident-coherent}) follows from the orthonormality of unitary representation matrix elements, namely
\be
 \rd_j \int_{{\rm SU}(2)} dg \, |j,g \rangle \langle j,g|,= \rd_j \sum_{mm'} |j,m \rangle \langle j,m'|
\int_{{\rm SU}(2)} dg \, D^j_{mj}(g) \overline{D^j_{m'j}(g)}= \sum_m |j,m \rangle \langle j,m| ,
\ee
where in the last equality we have used the orthonormality of the matrix elements. The decomposition of the identity (\ref{ident-coherent}) can be expressed as an integral on the two-sphere of directions $S^2=SU(2)/U(1)$
 by noticing that $D^j_{mj}(g)$ and $D^j_{mj}(gh)$ differ only by a phase for 
any group element $h$ from a suitable $U(1)\subset SU(2)$. Thus one has
\be\label{patacu}
1_j = \rd_j \int_{S^2}  dn \, |j,n \rangle \langle j,n|,
\ee
where  $n\in S^2$ is integrated with the invariant measure of the sphere.
The states $|j,n\rangle$ form (an over-complete) basis in $\sH_{j}$. $SU(2)$ coherent states 
have the usual semiclassical properties. Indeed if one considers the generators $J^i$ of $su(2)$ one has
\be \label{geo}
\langle j,n|\hat{J}^{i}|j,n \rangle  = 
j \,n^{i},
\ee
where $n^i$ is the corresponding three dimensional unit vector for $n\in S^2$. 
The fluctuations of  $\hat{J}^{2}$ are also minimal with $\Delta J^{2} =\hbar^2 j$, where we have restored $\hbar$ for clarity.
The fluctuations go to zero in the limit $\hbar\to 0$ and $j\to\infty$ while $\hbar j$ is kept constant. This kind of limit will be used often as a notion of semiclassical limit in spin foams. 
The state  $|j,n\rangle$ is a semiclassical state describing a vector in $\mathbb{R}^{3}$ of length $j$ and of direction 
$n$. It will convenient to introduce the following graphical notation for equation (\ref{patacu})
\be\label{patacul}
\begin{array}{c}\psfrag{a}{$\ \ j$}
\includegraphics[height=2cm]{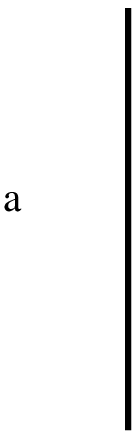}
\end{array}={\rm d}_{j}\,\int\limits_{S^2} dn \
\begin{array}{c}\psfrag{a}{$j$}\psfrag{b}{$\!n$}
\includegraphics[height=2cm]{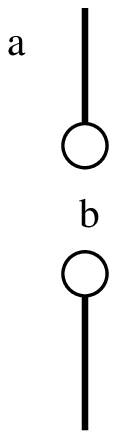}
\end{array}
\ee
Finally, an important property of $SU(2)$ coherent states stemming from the fact that \[|j,j\rangle=|{\van \frac{1}{2}},{\van \frac{1}{2}}\rangle|{\van \frac{1}{2}},{\van \frac{1}{2}}\rangle\cdots |{\van \frac{1}{2}},{\van \frac{1}{2}}\rangle\equiv |{\van \frac{1}{2}},{\van \frac{1}{2}}\rangle^{\otimes 2j}\]
is that 
\be\label{exp}
|j,n\rangle=|{\van \frac{1}{2}},n \rangle^{\otimes 2j}.
\ee
The above property will be of key importance in constructing 
effective discrete actions for spin foam models. In particular, it will play a central role in the study of the semiclassical limit of 
the EPRLand FK modesl studied in Sections \ref{eprl-r}, and \ref{fk}. In the following subsection we provide an example for $Spin(4)$ BF theory.

\subsubsection{$Spin(4)$ BF theory: amplitudes in the coherent state basis}

Here we study the coherent states representation of the path integral for $Spin(4)$ BF theory. The construction presented here can be extended to more general cases. The present case is however of particular importance for the
study of gravity models presented in  Sections \ref{eprl-r}, and \ref{fk}. With the introductions of coherent states one achieved the most difficult part of the work. In order to express the 
$Spin(4)$ BF amplitude in the coherent state representation one simply inserts a resolution of the identity in the form (\ref{patacu}) on each and every wire connecting neighbouring vertices in the expression (\ref{bf-so4})
for the BF amplitudes. The result is 
\ba\label{bf-cohe}
&& Z_{BF}(\Delta)=\sum \limits_{ {\cal
C}_f:\{f\} \rightarrow \rho_f }  \ \prod\limits_{f \in \Delta^{\star}} {\rm d}_{j_f^{-}}{\rm d}_{j_f^{+}} \n \\ && \int \prod_{e\in  \in \Delta^{\star}} {\rm d}_{j_{ef}^{-}}{\rm d}_{j_{ef}^{+}} dn^{-}_{ef}dn^{+}_{ef}
\begin{array}{c}\psfrag{w}{$$}
\psfrag{a}{$\va n^{\va -}_{ 1}$}
\psfrag{A}{$\va n^+_1$}
\psfrag{b}{$\va n^-_2$}
\psfrag{B}{$\va n^+_2$}
\psfrag{c}{$\va n^-_3$}
\psfrag{C}{$\va n^+_3$}
\psfrag{d}{$\va n^-_4$}
\psfrag{D}{$\va n^+_4$}
\includegraphics[width=7cm]{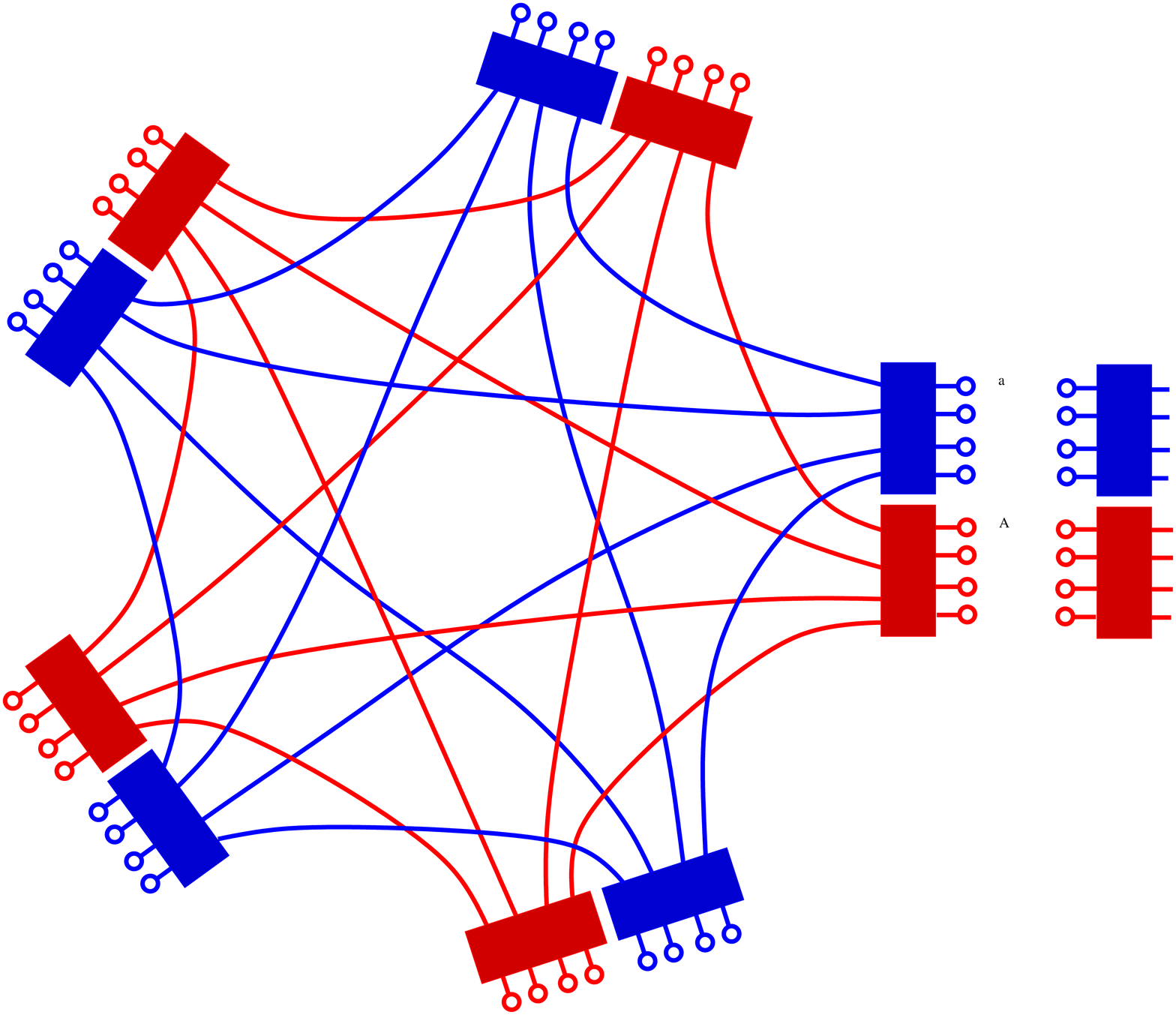}
\end{array},
\ea
where we have explicitly written the $n_{\pm}\in S^2$ integration variables only on a single cable. One observes that there is one $n_{\pm}\in S^2$ per each wire coming out at an edge $e\in\Delta^{\star}$; as 
wires are in one-to-one correspondence with faces $f\in \Delta^{\star}$ the integration variables  $n^{\pm}_{ef}\in S^2$ are labelled by an edge and face subindex. 
In order to get an expression of the BF path integral in terms of an affective action we restore at this stage the explicit group integrations represented by the 
boxes in the previous equation.  One gets,
\ba&&\!\!\!\!\!\!\!\!\!\!\!\!\!\!\!\!\!\!\!\!\!\!\!\!\!\!\!
Z_{BF}(\Delta)=\sum \limits_{ {\cal
C}_f:\{f\} \rightarrow \rho_f }  \ \prod\limits_{f \in \Delta^{\star}} {\rm d}_{j_f^{-}}{\rm d}_{j_f^{+}} \int \prod_{e\in   \Delta^{\star}} {\rm d}_{j_{ef}^{-}}{\rm d}_{j_{ef}^{+}} dn^{-}_{ef}dn^{+}_{ef} \n \\  \ \ \ \ \ \ \ \ \  \ \ \ \ \ \ 
&& \prod_{v\in \Delta^{\star}} \prod_{e,e'\in v} dg^{-}_{ef}dg^{+}_{ef} \ (\langle n^{-}_{ef} |(g^{-})^{-1}_{ef} g^-_{e'f}| n^-_{e'f}\rangle)^{2j^{-}_{f}}(\langle n^{+}_{ef} |(g^{+})^{-1}_{ef} g^+_{e'f}| n^+_{e'f}\rangle)^{2j^{+}_{f}},
\ea
where we have used the coherent states property (\ref{exp}),  and   $| n^{\pm}\rangle$ is a simplified notation for $|{\van \frac{1}{2}}, n^{\pm}\rangle$. The previous equation can be finally written as  
\ba
\label{cspig}Z_{BF}(\Delta)=\sum \limits_{ {\cal
C}_f:\{f\} \rightarrow \rho_f }  \ \prod\limits_{f \in \Delta^{\star}} {\rm d}_{j_f^{-}}{\rm d}_{j_f^{+}} \int \prod_{e\in  \Delta^{\star}} {\rm d}_{j_{ef}^{-}}{\rm d}_{j_{ef}^{+}} dn^{-}_{ef}dn^{+}_{ef} dg^{-}_{ef}dg^{+}_{ef}
\ \exp{(S^{d}_{j^{\pm},\bn^{\pm}}[g^{\pm}])}, \ea
where the discrete action \be\label{discrete-action-g}
S^{d}_{j^{\pm},\bn^{\pm}}[g^{\pm}]=\sum_{v\in\Delta^{\star}} S^v_{j_v,\bn_v}[g^{\pm}]\ee with 
\be
\label{v-action}
S^v_{j,\bn}[g] = \sum\limits^{5}_{a < b=1} 2j_{ab} \ln \, \la n_{ab}| g^{-1}_a g_b| \, n_{ba} \ra,
\ee
and the indices $a,b$ label the five edges of a given vertex. The previous expression is exactly equal to the form (\ref{coloring4}) of the BF amplitude. In the case of the gravity models studied in what follows,
the coherent state path integral representation  will be the basic tool for the study of the semiclassical limit of the models and the relationship with Regge discrete formulation of general relativity. 

\subsection{The relationship between gravity and BF theory}

The field theory described in the present section has no local degrees of freedom. It represents the simplest example of a topological field theory in four dimensions.
The interest of this theory for gravity model stems from the fact that an action for the gravitational degrees of freedom (basically equivalent to general relativity in the first order formulation) can be obtained by supplementing 
a 4d BF theory action with internal gauge group $SL(2,\C)$ (Lorentzian) or $Spin(4)$ (Riemannian) with the following set of quadratic constraints on the $B$-field 
\be \epsilon_{IJKL} B^{IJ}_{\mu\nu}B^{KL}_{\rho\sigma}-e \
 \epsilon_{\mu\nu\rho\sigma}\approx 0, \label{dual}\ee
 where $e\equiv \sigma^2 (1/4!)\epsilon_{IJKL} B^{IJ}_{\mu\nu}B^{KL}_{\rho\sigma}\epsilon^{\mu\nu\rho\sigma}$ where $\sigma^2=\pm 1$ according to we are in the Riemannian of Lorentzian case. More generally,  a one parameter family of gravity actions can be obtained obtained from the imposition of the previous 
 constraints on the following modified BF action 
 \be
 S_{\gamma}(B,\omega)=\int_{\cal M} \langle({}^{\star} B+\frac{1}{\gamma}B)\wedge F(\omega)\rangle,
 \ee
 where $\gamma$ is the Immirzi parameter.
 The strategy behind the definition of the new spin foam 
 models for quantum gravity consists of imposing these constraints on the path integral of BF theory on the momenta $J={}^{\star}B+\frac{1}{\gamma}B$ conjugate to $\omega$. In order to impose the Plebanski constraints above  it will be convenient to express the $B$ field in terms of the momenta $J$, namely
 \be B=\frac{\gamma}{1-\sigma^2 \gamma^2} (J-\gamma {}^{\star} J).\label{defidefi}\ee
 The imposition of the constraints (\ref{dual}) on the BF path integral on a fixed discretization can be done in two different ways: by directly 
 restricting the spin foam configurations (this is the EPRL approach described in the following section), or by restricting the semiclassical values of the $B$ field in the coherent state representation of the BF path integral
 (this is the FK strategy described in Section \ref{fk}).

 \section{The Engle-Pereira-Rovelli-Livine
 (EPRL) model} \label{eprl-r}
 
 In this section we introduce the Engle-Pereira-Rovelli-Livine (EPRL) model \cite{Engle:2007uq, Engle:2007wy}. 
 The section is organized as follows. 
 The relevant representation theory is introduced in Section \ref{rept}. In Section \ref{7-2} we present and discuss the linear simplicity constraints
 ---classically equivalent to the Plebanski constraints---and discuss their implementation in the quantum theory.
 In Section \ref{path-pre} we introduce the EPRL model of Riemannian gravity. In Section \ref{cuadratiqui} we prove the validity of the quadratic Plebanski constraints---reducing BF theory to general relativity---directly in the spin foam representation.
  In Section \ref{7-6} we present the coherent state representation of the Riemannian EPRL model. In Section \ref{alala} we describe the Lorentzian model. 
 The material of this section will also allow us to describe the construction of 
 the closely related (although derived from a different logic) Riemannian FK constructed in \cite{Freidel:2007py}.
 The idea that linear simplicity constraints are more convenient for dealing with the constraints that reduce BF theory to gravity 
 was pointed out by Freidel and Krasnov in this last reference.

\subsection{Representation theory of $Spin(4)$ and $SL(2,\C)$ and the canonical basis}\label{rept}

In this section we present the representation theory of the groups $Spin(4)$ and $SL(2,\C)$ that is neccesarry for the definition of the new spin foam  models for Riemannian and Lorentzian gravity respectively.
To emphasize the highly symmetric structure of the two we present them in a unified notation where a parameter $\sigma=1$ for the Riemannian sector and $\sigma=i$ for the Lorentzian one. The simple relationship between the two might be a hint of a possible relationship 
between model amplitudes in a spirit similar to the interesting link between Euclidean and Lorentzian QFT provided by Wick rotations \footnote{Such explicit relationship between gravity amplitudes in the Euclidean and Lorentzian sectors can be established by analytic continuation in 3d \cite{Buffenoir:1997ih}.}.
Unitary irreducible representations $\sH_{p,k}$ of $Spin(4)$ and $SL(2,\C)$ are labelled by two parameters $p$ and $k$.
In the case of $Spin(4)=SU(2)\times SU(2)$ the unitary irreducible representations are finite dimensional and the labels $p$ and $k$ can be expressed in terms of the half integers labelling the right and left $SU(2)$ unitary representations $j^{\pm}$ as follows
\be p= j^++j^-+1 \ \ \ \ k=|j^+-j^-|.\ee
In the $SL(2,\C)$ case the unitary irreducible representations are infinite dimensional and one has\be p \in \R^+\ \ \ \ k\in \N/2.\ee
The  two Casimirs are $C_1=\frac{1}{2}J_{IJ}J^{IJ}=L^2+ \sigma^2 K^2$ and $C_2=\frac{1}{2}{}^\star \!J_{IJ}J^{IJ}=K\cdot L$ where $L^i$ are the generators of an arbitrary rotation subgroup and
$K^i$ are the generators of the corresponding boosts. 
The Casimirs act on $ |p,k\rangle\in \sH_{p,k}$ as follows 
\ba\label{casiL}
&& C_1 |p,k\rangle=\frac{1}{2}(k^2+\sigma^2 p^2- 1)\, |p,k\rangle\n\\
&& C_2  |p,k\rangle=pk\,  |p,k\rangle. 
\ea
For detail on the representation theory of $SL(2,\C)$ see \cite{ruhl, gelfand, gelfand2}.
The definition of the EPRL model  requires the introduction of an (arbitrary)
 subgroup $SU(2)\subset Spin(4)$ or  $SU(2)\subset SL(2,\C)$ according to whether one is working in the Riemannian or in the
 Lorentzian sector. This subgroup corresponds to the internal gauge group 
 of the gravitational phase space in connection variables in the time gauge (see \cite{mimi} for details).  In the quantum theory, the representation theory of this $SU(2)$ subgroup will be hence important. This importance will soon emerge as apparent from  the imposition of the constraints that define the EPRL model.  The link between the unitary representations of $SL(2,\C)$ and those of $SU(2)$ is given by  the decomposition 
\be\label{spin4su2g}
 \sH_{p,k}=\bigoplus\limits_{j=k}^{p-1} \sH_{j}=\bigoplus\limits_{j=|j^{+} - j^{-}|}^{j^{+} + j^{-}} \sH_{j}, \ee
for the Riemannian sector, and
 \be\label{spin4su2}
 \sH_{p,k}=\bigoplus \limits_{j=k}^{\infty} \sH_{j}, \ee
 for the Lorentzian sector.
 As the unitary irreducible representations of the subgroup $SU(2)\in Spin(4)$ and   $SU(2)\in SL(2,\C)$ are essential in understanding the link of the EPRL model and the operator canonical formulation of LQG it will be convenient to express the action of the generators of the Lie algebra of the corresponding group 
 in a basis adapted to the above equation. In order to do this we first notice that the Lie algebra $spin(4)$ and $sl(2,\C)$  can be characterized in terms of the generators of a rotation subgroup $L^i$ and the remaining boost generators $K^i$ as follows
\ba&& \n
[L_3,L_{\pm}]=\pm \ L_{\pm} \ \ \ \ \ [L_+,L_{-}]=2\ L_{3} \\
&& \n
[L_+,K_{+}]= [L_-,K_{-}]=[L_3,K_{3}]=0 \\
&& \n
[K_3,L_{\pm}]=\pm \ K_{\pm} \ \ \ \ \ [L_\pm,K_{\mp}]=\pm 2\ K_{3}\ \ \ \ \ [L_3,K_{\pm}]=\pm \ K_{\pm} \\
&& [K_3,K_{\pm}]=\pm  \sigma^2 L_{\pm} \ \ \ \ \ [K_+,K_{-}]= 2 \sigma^2 L_{3},\label{liesl2c}
\ea
where $K_{\pm}=K^1 \pm i K^2$ and $L_{\pm}=L^1 {\pm}i L^2$ respectively.
The action of the previous generators in the basis $ |p,k; j ,m\rangle$ can be shown to be
\ba
&& L^3 |p,k;j,m\rangle = m |p,k;j,m\rangle, \nonumber \\
&& L^+ |p,k;j,m\rangle = \sqrt{(j+m+1)(j-m)} |p,k;j,m+1\rangle, \nonumber \\
&&  L^- |p,k;j,m\rangle = \sqrt{(j+m)(j-m+1)} |p,k;j,m-1\rangle, \nonumber \\
&&  K^3 |p,k;j,m\rangle =  \alpha_j\sqrt{j^2-m^2} |p,k;j-1,m\rangle+ \gamma_j m |p,k;j,m\rangle 
-\alpha_{j+1}\sqrt{(j+1)^2-m^2} |p,k;j+1,m\rangle,\nonumber \\
&&  K^+ |p,k;j,m\rangle = \alpha_j\sqrt{(j-m)(j-m-1)}
|p,k;j-1,m+1\rangle \nonumber \\
&& + \gamma_j\sqrt{(j-m)(j+m+1)}|p,k;j,m+1\rangle \nonumber
\\ && +\alpha_{j+1}\sqrt{(j+m+1)(j+m+2)} |p,k;j+1,m+1\rangle,\nonumber \\
&& K^- |p,k;j,m\rangle = -\alpha_j\sqrt{(j+m)(j+m-1)}
|p,k;j-1,m-1\rangle
\nonumber \\ && + \gamma_j\sqrt{(j+m)(j-m+1)} |p,k;j,m-1\rangle \nonumber \\
&& -\alpha_{j+1}\sqrt{(j-m+1)(j-m+2)}|p,k;j+1,m-1\rangle,
\label{operadores eigenedos}
\ea
where 
\be
\gamma_{j}=\frac{kp}{j(j+1)}
\ \ \ \ \ \ \ \ \ \ 
\alpha_{j}=\sigma \sqrt{ \frac{(j^2-k^2)(j^2+p^2)}{j^2(4j^2-1)}}
\ee
%
The previous equations will be important in what follows: they will allow for the characterisation of the solutions of the quantum simplicity constraints in both the Riemannian and Lorentzian models in a direct manner. 
This concludes the review of the representation theory that is necessary for the definition of the EPRL model.

\subsection{The linear simplicity constraints}\label{7-2}

As first shown in \cite{Freidel:2007py}, the  quadratic Plebanski simplicity constraints---and more precisely in their dual version presented below (\ref{dual})---are equivalent in the discrete setting to 
the linear constraint on each face of a given tetrahedron
\be\label{constrainty-e}
D_{f}^i=L_{f}^i-\frac{1}{\gamma} K_{f}^i\approx 0,
\ee
where the label $f$ makes reference to a face $f\in \Delta^{\star}$, and where (very importantly) the subgroup $SU(2)\subset Spin(4)$ or $SL(2,\C)$ that is necessary for the definition of the above 
constraints is chosen arbitrarily at each tetrahedron, equivalent on each edge $e\in \Delta^{\star}$. Such choice of the rotation subgroup is the precise analog of the time gauge in the 
canonical analysis of  general relativity. The EPRL model is defined by imposing the previous constraints as operator equations on the Hilbert spaces defined by the unitary irreducible 
representations of the internal gauge group  that take part in the state-sum of BF theory.  We will show in Section \ref{cuadratiqui} that the models constructed on the requirement of
a suitable imposition of the linear constraints (\ref{constrainty-e}) satisfy the usual quadratic Plebanski constraints---that reduce BF theory to general relativity---in the path integral formulation (up to quantum corrections which are neglected in the usual semiclassical limit). 

%

From the commutation relations (\ref{liesl2c}) of previous section we can easily compute the
commutator of the previous tetrahedron constraints and conclude that in fact it does not close, namely
\ba\label{algebry}
[D_{f}^i,D_{f'}^j]&=&\delta_{f f'}  \epsilon^{ij}_{\ \, k} \left[(1+\frac{\sigma^2}{\gamma^2}) L_{f}^k-\frac{2}{\gamma} K_e^k\right]=\n \\
&=&2 \delta_{e e'}  \epsilon^{ij}_{\ \, k} D^k+\delta_{e e'} \frac{\sigma^2-\gamma^2}{\gamma^2} \epsilon^{ij}_{\ \, k}  L_{f}^k.
\ea
The previous commutation relations imply  that the constraint algebra is not closed and cannot therefore be imposed as operator equations on the states summed over in the BF partition function in general. There are two interesting exceptions to the previous statement: 
\begin{enumerate}
\item The first one is to take $\gamma=\pm \sigma$. This corresponds to the description of the model in terms of self-dual or anti-self-dual variables.
Unfortunately, the construction of the new models is not well defined in this case for the Lorentzian theory and leads to a trivial result in the Riemannian sector: $SU(2)$ BF theory.

\item The second possibility is to work in the sector where $L^i_f=0$. This choice leads  to the 
Barret-Crane model \cite{BC2} where the degrees of freedom of BF theory seem over constrained: 
boundary states satisfying the BC constraints are a very small subset of the allowed boundary states in LQG.
This is believed to be problematic if gravity is to be recovered at low energies.
\end{enumerate}
The EPRL model is obtained by restricting the representations appearing in the expression of the BF partition 
function so that at each tetrahedron the linear constraints (\ref{constrainty-e}) the strongest possible way that is compatible with the uncertainties 
relations stemming from 
(\ref{algebry}). In addition one would  add  the requirement that the state-space of tetrahedra is compatible with the state-space 
of the analogous excitation in the canonical context of LQG so that arbitrary states in the kinematical state of LQG have non trivial 
amplitudes in the model.



%
Due to the fact that the constraints $D^{i}_f$ do not form a closed (first class) algebra in the generic case one needs to 
devise a weaker sense in which they are to be imposed. One possibility is to consider the Gupta-Bleuler criterion 
consisting in selecting a suitable class of states for which the matrix elements on $D_{f}^i$ vanish. One notices from (\ref{operadores eigenedos}) that
if we chose the subspace  $\sH_{j}\subset\sH_{p,k}$ one has
\ba && \n
  \langle p,k ,j,q|D^{3}_f |p,k ,j,m\rangle=  \delta_{q,m}m (1-\frac{\gamma_{j}}{\gamma})\\ && \n
  \langle p,k ,j,q|D^{\pm}_f |p,k ,j,m\rangle=  \delta_{q\pm 1,m}\sqrt{(j\pm m+1)(j\mp m)}(1-\frac{\gamma_{j}}{\gamma}).
\ea
The matrix elements of the linear constraints vanish in this subclass if one can chose
\be\gamma_{j}=\frac{pk}{j(j+1)}=\gamma\label{coky}
\ee
There are two cases:
\begin{enumerate}
\item {\bf Case $\gamma<1$: } Following \cite{Ding:2009jq}, in this case one restricts the representations to \be \mbox{{\bf Riemannian:} $p=j+1$, $k=\gamma j$.\ \ \ \ \ \ \ \ \ \ \ \ \ \ \ \ \ \ {\bf Lorentzian:}  $p=\gamma (j+1)$, $k= j$. } \ee which amounts to choosing the {\em maximum weight} component $j=p-1$ 
in the expansion (\ref{spin4su2}). In the Riemannian case the above choice translates into $j^{\pm}=(1\pm \gamma)j/2$ for the SU(2) right and lect representations. Notice that the solutions to the simplicity constraints in the Riemannian and Lorentzian sectors look very different for $\gamma<1$. 
Simple algebra shows that condition (\ref{coky}) is met. There are indeed other solutions \cite{Ding:2010fw} of the Gupta-Bleuler criterion in this case. 
 \item {\bf Case $\gamma>1$: } In this case \cite{Alexandrov:2010pg} one restricts the  representations to  \be \mbox{{\bf Riemannian:} $p=\gamma(j+1)$, $k=j$.\ \ \ \ \ \ \ \ \ \ \ \ \ \ \ \ \ \ {\bf Lorentzian:}  $p=\gamma (j+1)$, $k= j$. } \ee which amounts to choosing the {\em minimum weight} component $j=k$ 
 in the expansion (\ref{spin4su2}) and . For the Riemannian case we can write the solutions in terms of $j^{\pm}=( \gamma\pm 1)\frac{j}{2} +\frac{\gamma-1}{2}$. Notice that for $\gamma>1$ there is complete symmetry between the solutions of the Riemannian and Lorentzian sectors. In my opinion, this 
 symmetry deserves further investigation as it might be an indication of a deeper connexion between the Riemannian and Lorentzian models (again such relationship is a fact in 3d gravity \cite{Buffenoir:1997ih}). 
\end{enumerate}
Another criterion for weak imposition can be developed by studying the spectrum of the Master constraint $M_f=D_f\cdot D_f$. Strong imposition of the constraints $D_f^i$ would amount to
looking for the kernel of the master constraint $M_f$. However, generically the positive operator associated with the master constraint does not contain the zero eigenvalue in the spectrum due to the open nature of the constraint algebra (\ref{algebry}).  
It is convenient \cite{Rovelli:2010ed} to express the master constraint in a in a manifestly invariant way. In order to get a gauge invariant constraint one starts from the  master constraint and  uses the $D^i_f=0$ classically to write it in terms of  Casimirs, namely 
\ba\n
M_f =(1+\sigma^2 \gamma^2) C_2 - 2 C_1 \gamma,
\ea
where $C_1$ and $C_2$ are the Casimirs given in equation (\ref{casiL}). 
The minimum eigenvalue condition is
  \be \mbox{{\bf Riemannian:} $p=j$, $k=\gamma j$.\ \ \ \ \ \ \ \ \ \ \ \ \ \ \ \ \ \ {\bf Lorentzian:}  $p=\gamma j$, $k= j$. } \ee The minimum eigenvalue is $m_{\va min}=\hbar^2 \gamma j(\gamma^2-1)$ for the Riemannian case and $m_{\va min}=\gamma$ for the Lorentzian case. 
The master constraint criterion works better in the Lorentzian case as pointed out in \cite{Rovelli:2010ed}.  More recently, it has been shown that
the constraint solution $p=\gamma j$ and $k=j$ also follows naturally from a spinor formulation of the simplicity constraints \cite{Wieland:2011ru, Dupuis:2011wy, Livine:2011vk}.
The above criterion is used in the definition of the EPRL model. 

It is important to point out that 
the Riemannian case imposes strong restrictions on the allowed values of the Immirzi parameter if one  wants the spin  $j\in \N/2$ to be arbitrary (in order to have all possible boundary states allowed in LQG).
In this case the only possibilities are $\gamma=\N$ or $\gamma=1$.  This restriction is not natural from the viewpoint of LQG.  Its relevance if any remains misterious at this stage.  

Summarising,  in the Lorentzian (Riemannian) EPRL  model one restricts the  $SL(2,\C)$ ($Spin(4)$) representations of BF theory to those satisfying \be p=\gamma j \ \ \ k=j\ee
for $j\in \N/2$. From now on we denote the subset of admissible representation \be \sK_{\gamma}\subset {\rm Irrep}(SL(2,\C)) ({\rm Irrep}(Spin(4))) \ee
 The admissible quantum states $\Psi$ are elements of the subspace $\sH_{j}\subset \sH_{\gamma j,j}$ (i.e., minimum weight states) satisfy the constraints (\ref{constrainty-e}) in the following  emiclassical sense:
\be
( K^i_f-\gamma  L_f^i) \Psi=\sO_{sc},
\ee 
where the symbol $\sO_{sc}$ (order semiclassical) 
denotes a quantity that vanishes in limit $\hbar\to 0$, $j\to \infty$ with $\hbar j=$constant. 
In the Riemannian case the previous equation can be written as
\be\label{needed}
[(1-\gamma) J_+^i-(1+\gamma) J_-^i] \Psi=\sO_{sc},
\ee which in turn has a simple graphical representation in terms of spin-network grasping operators, namely
\be
\begin{array}{c}
\psfrag{x}{$\!\!\!\!\!\!\!\!\!\!-(1+\gamma)$}
\psfrag{y}{$\!\!\!\!\!\!\!\!\!\!+(1-\gamma)$}
\psfrag{z}{$=\sO_{sc}$}
\psfrag{j}{$k$}
\psfrag{jp}{$j_{-}$}
\psfrag{jm}{$j_{+}$} 
\includegraphics[height=3cm]{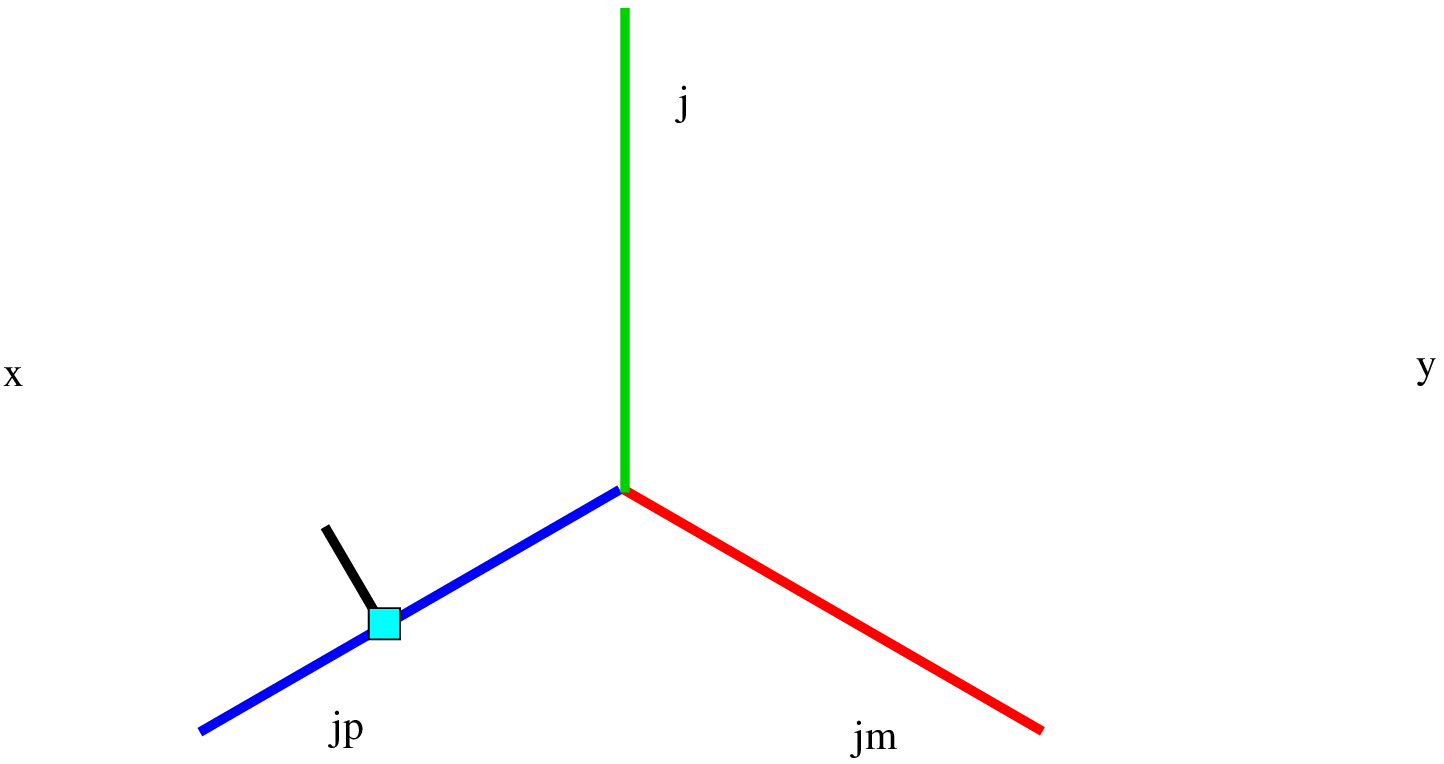} \includegraphics[height=3cm]{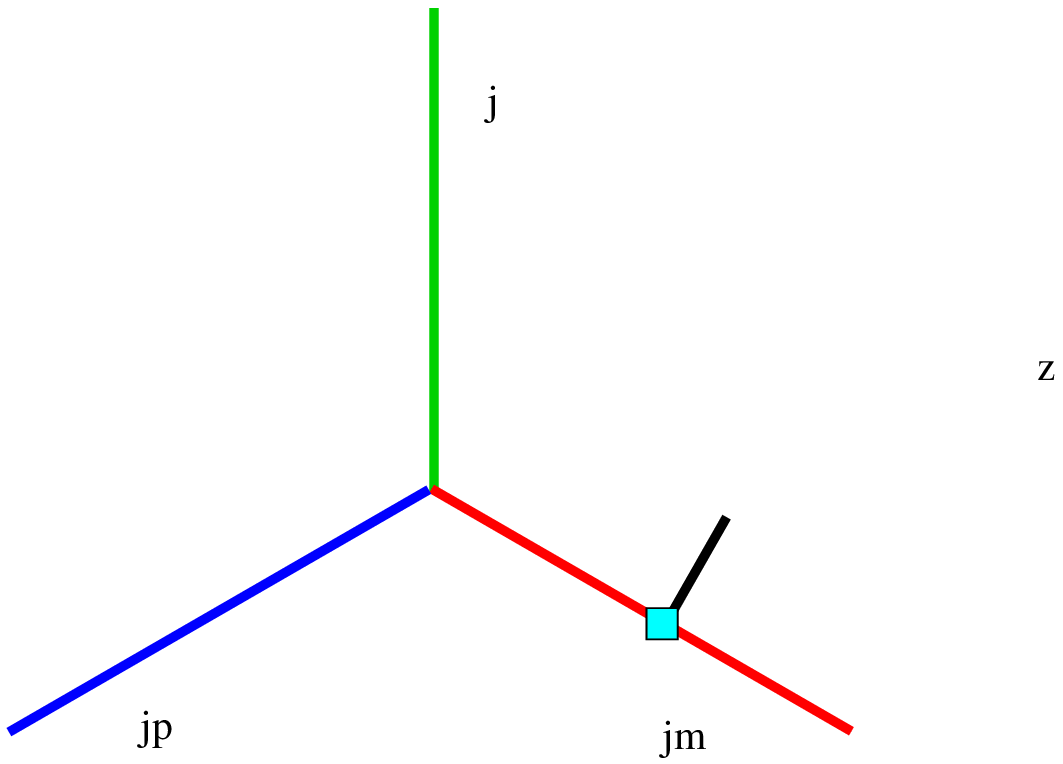}
\end{array}
\label{trival}\ee
The previous equation will be of great importance in the graphical calculus that will allow us to show that the 
linear constraint imposed here at the level of states imply the vanishing of the quadratic Plebanski constraints (\ref{dual}) and 
their fluctuations, computed in the path integral sense,  in the appropriate large spin semiclassical limit.

\subsection{Presentation of the Riemannian EPRL amplitude}\label{path-pre}

Here we complete the definition of the EPRL models by imposing the linear constraints on the BF amplitudes constructed
in Section (\ref{BF}). We will also show that the path-integral expectation value of the Plebanski constraints  (\ref{dual}), as well as their fluctuations, 
vanish in a suitable semiclassical sense.  This shows that the EPRL model can be considered as a lattice definition of the a quantum gravity theory.

We start with the Riemannian model for which a straightforward graphical notation is available. The first step is the translation of equation (\ref{spin4su2g})---for $p$ and $k$ satisfying the simplicity constraints---in terms of the graphical 
notation introduced in Section (\ref{BF}). Concretely, for $\gamma<1$ one has $j^{\pm}=(1\pm\gamma) j/2\in \sK_{\gamma}$ becomes
\vskip.5cm
\be
\begin{array}{c}
\psfrag{j}{$\alpha$}
\psfrag{jp}{$\!\!\!\!\!\!\!\!\!\!\!\!\!\!\!\!\!\!(1-\gamma) \frac{j}{2}$}
\psfrag{jm}{$(1+\gamma) \frac{j}{2}$} 
\includegraphics[height=2.5cm]{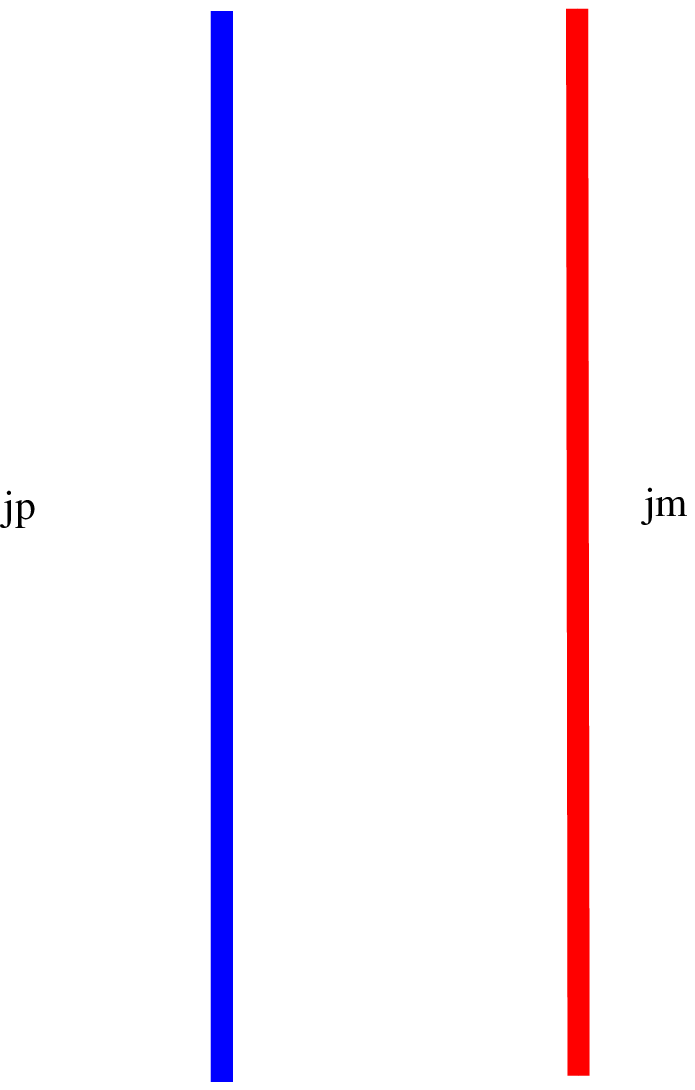}\end{array} \ \ \ \ \ \ \ \ \ \ \ \ =\bigoplus \limits_{\alpha=\gamma j}^{j} \ \ \ \ \ \ \begin{array}{c} \psfrag{j}{$\alpha$}
\psfrag{jp}{$\!\!\!\!\!\!(1-\gamma) \frac{j}{2}$}
\psfrag{jm}{$(1+\gamma) \frac{j}{2}$}  \includegraphics[height=2.5cm]{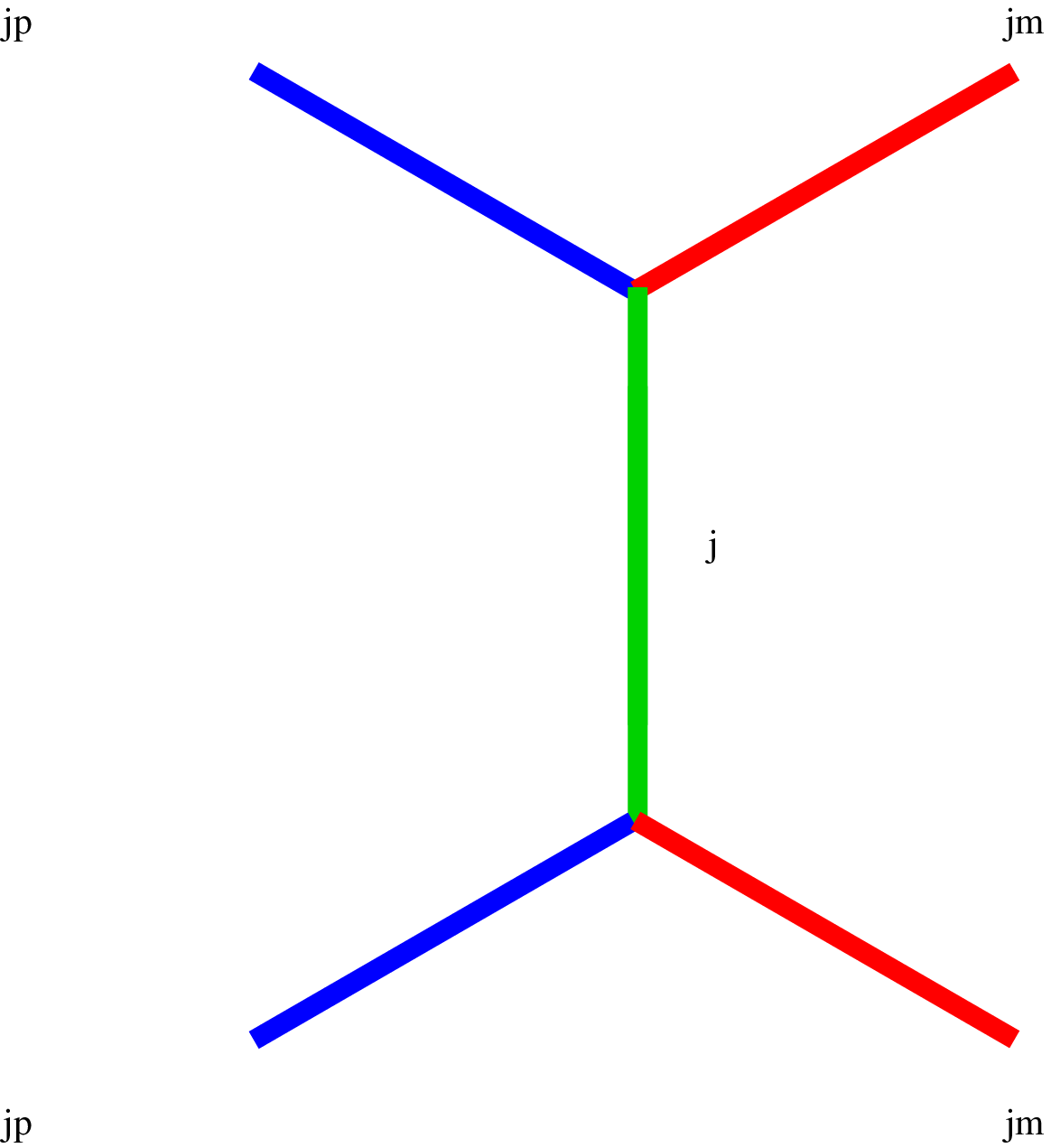}
\end{array}
\label{trivalis}\ee
For $\gamma>1$ we have 
\vskip.5cm
\be
\begin{array}{c}
\psfrag{j}{$\alpha$}
\psfrag{jp}{$\!\!\!\!\!\!\!\!\!\!\!\!\!\!\!\!\!\!(\gamma-1) \frac{j}{2}$}
\psfrag{jm}{$(1+\gamma) \frac{j}{2}$} 
\includegraphics[height=2.5cm]{travalisto.eps}\end{array} \ \ \ \ \ \ \ \ \ \ \ \ =\bigoplus \limits_{\alpha= j}^{\gamma j} \ \ \ \ \ \ \begin{array}{c} \psfrag{j}{$\alpha$}
\psfrag{jp}{$\!\!\!\!\!\!(\gamma-1) \frac{j}{2}$}
\psfrag{jm}{$(1+\gamma) \frac{j}{2}$}  \includegraphics[height=2.5cm]{travalis.eps}
\end{array}
\label{trivalisto}\ee
\vskip.5cm
The implementation of the linear constraints of Section (\ref{7-2}) consist in restricting the representations $\rho_f$ of $Spin(4)$ appearing in the state sum amplitudes of BF theory as written in Equation (\ref{bf-so4})
to the subclass $\rho_f\in \sK_{\gamma} \subset {\rm Irrep}(Spin (4))$, defined above,  while projecting to the highest weight term in (\ref{trivalis}) for $\gamma<1$. For $\gamma>1$ one must take the minimum weight term in  (\ref{trivalisto}) .
The action of this projection will be denoted $\sY_{j}:\sH_{(1+\gamma)j/2,|(1-\gamma)|j/2}\to \sH_{j}$, graphically
\be\label{Y}
\sY_{j}\left[\ \ \ \ \ \ \ \ 
\begin{array}{c}\psfrag{j}{$\alpha$}
\psfrag{jp}{$\!\!\!\!\!\!\!\!\!\!\!\!\!\!\!\!\!\! |\gamma-1| \frac{j}{2}$}
\psfrag{jm}{$(1+\gamma) \frac{j}{2}$} 
\includegraphics[width=1cm]{travalisto.eps}
\end{array}\ \ \ \ \ \ \ \ \ \right]=
\begin{array}{c} \psfrag{j}{$j$}
\psfrag{jp}{$ $}
\psfrag{jm}{$ $}  \includegraphics[height=2cm]{travalis.eps}
\end{array}.
\ee 
Explicitly, one takes the expression of the BF partition function (\ref{bf4}) and modifies it by replacing the projector
 $P^e_{inv}(\rho_1,\cdots, \rho_4)$ with $\rho_1,\cdots \rho_4\in \sK_{\gamma}$ by a new object 
 \be\label{regalo} P_{eprl}^e(j_1,\cdots, j_4) \equiv P^e_{inv} (\rho_{1}\cdots \rho_4)( \sY_{j_1}\otimes\cdots\otimes\sY_{j_4} )P^e_{inv} (\rho_{1}\cdots \rho_4)\ee with $j_1, \cdots j_4\in \N/2$
implementing the linear constraints described in the previous section. Graphically the modification of BF theory that produces the EPRL model corresponds to the replacement  \be\label{eprl-projection}
P_{inv}^e(\rho_1\cdots\rho_4)=
\begin{array}{c}\psfrag{a}{$ $}\psfrag{b}{$ $}\psfrag{c}{$ $}\psfrag{d}{$ $}
\includegraphics[height=1cm]{cable-4d-1.eps}
\end{array}\begin{array}{c}\psfrag{A}{$ $}\psfrag{B}{$ $}\psfrag{C}{$ $}\psfrag{D}{$ $}
\includegraphics[height=1cm]{cable-4d-2.eps}
\end{array} \ \ \ \ \ \begin{array}{c} \includegraphics[width=1cm]{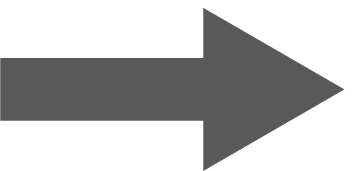}\end{array}
\ \ \ \ \  P_{eprl}^{e}(j_1\cdots j_4)=
\begin{array}{c}
\includegraphics[width=1.4cm]{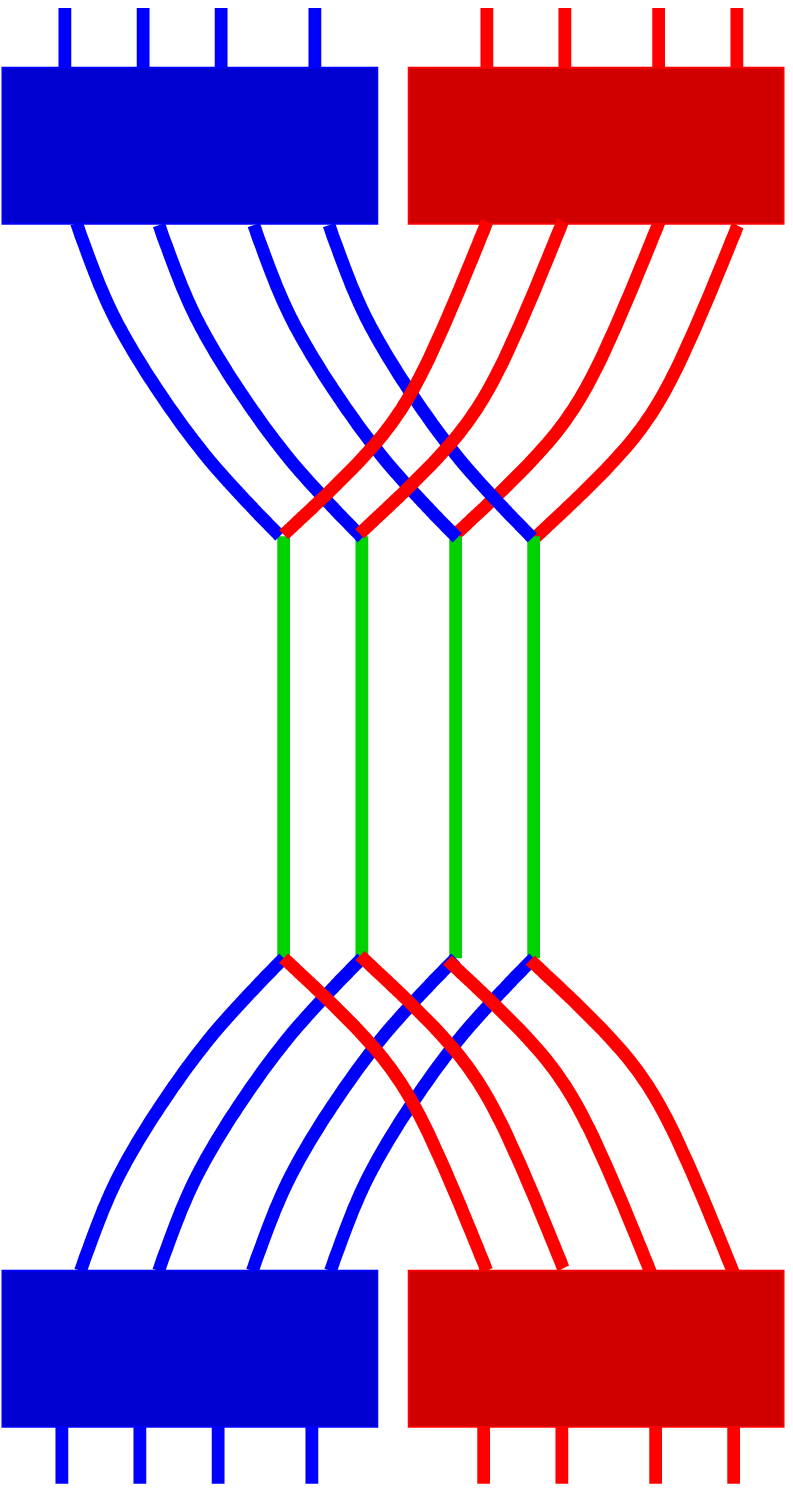}
\end{array}
\ee 
on the expression (\ref{bf-so4}), 
where we have dropped the representation labels from the figure for simplicity.  We have done the operation (\ref{Y}) on each an every  of the four pairs of representations. The $Spin(4)$ integrations represented by the two boxes at the top and bottom of the previous graphical expression restore the full $Spin(4)$ invariance as the projection (\ref{Y}) breaks this latter symmetry for being based on the selection of a special subgroup $SU(2)\subset Spin(4)$ in its definition (see section \ref{anofree} for an important implication). One should simply keep in mind that green wires in the previous 
two equations and in what follows are labeled by arbitrary spins $j$ (which are being summed over in the expression of the amplitude (\ref{eprl-so4})), while 
red and blue wires are labelled by $j^{+}=(1+\gamma)j/2$ and $j^{-}=|1-\gamma|j/2$ respectively. With this (\ref{bf-so4}) is modified to
\ba\label{eprl-so4}\n
Z^{E}_{eprl}(\Delta)&=&\sum \limits_{ \rho_f \in \sK}  \ \prod\limits_{f \in \Delta^{\star}} {\rm d}_{|1-\gamma|\frac{j}{2}}{\rm d}_{(1+\gamma)\frac{j}{2}}
\prod_{e} P^{e}_{eprl}(j_1,\cdots,j_4)=\\
&=&\sum \limits_{ \rho_f \in \sK}  \ \prod\limits_{f \in \Delta^{\star}} {\rm d}_{|1-\gamma|\frac{j}{2}}{\rm d}_{(1+\gamma)\frac{j}{2}}
\begin{array}{c}
\includegraphics[width=5cm]{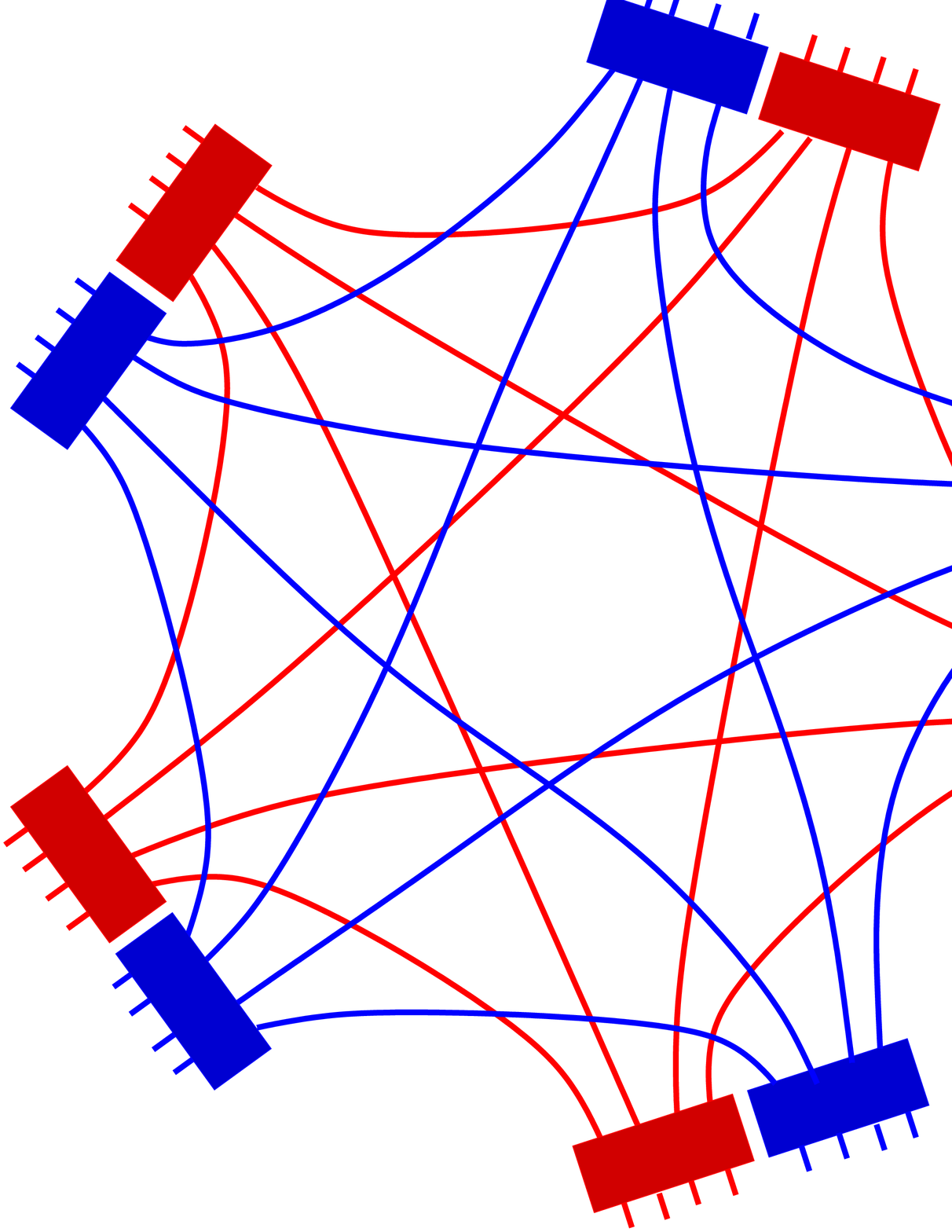}
\end{array},
\ea
The previous expression is defines the EPRL model amplitude. 

\subsubsection{The spin foam representation of the EPRL amplitude}

Now we will work out the spin foam representation of the EPRL amplitude which  at this stage will take no much more effort than the
derivation of the spin foam representation for $Spin(4)$ BF theory as we went from equation (\ref{bf-so4}) to (\ref{BF4V-b}) 
in Section \ref{BF}. The first step is given in the following equation
\ba  \begin{array}{c}
\psfrag{w}{$$}
\includegraphics[width=6cm]{eprl3.eps} \end{array}&=& \begin{array}{c}\psfrag{w}{$$} \includegraphics[width=6cm]{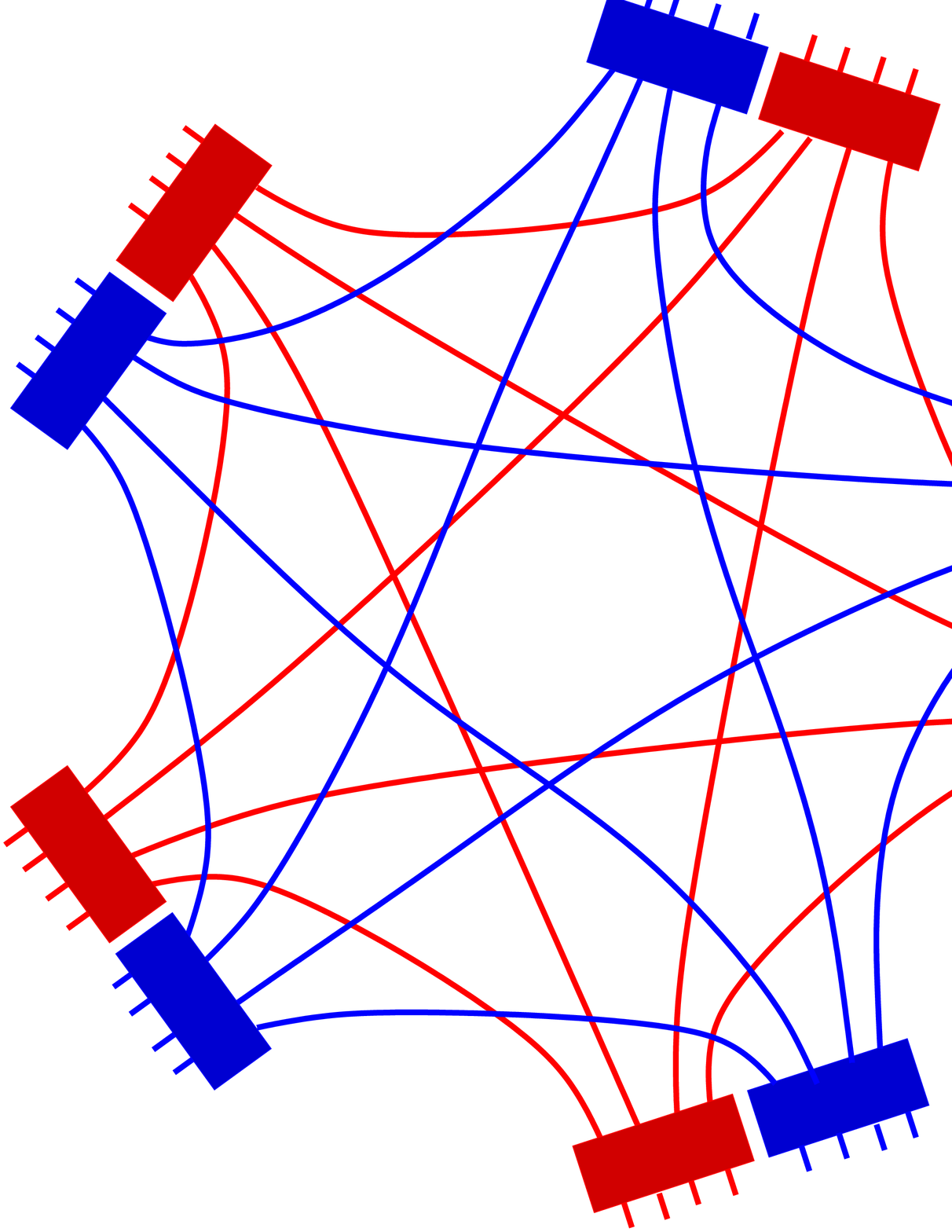}\label{dosf}
\end{array} =\n \\ 
&=& \sum_{\iota }\begin{array}{c}
\psfrag{a}{$\van \iota$}
\psfrag{b}{$\! \van \bar \iota$}
\psfrag{w}{}
\includegraphics[width=6cm]{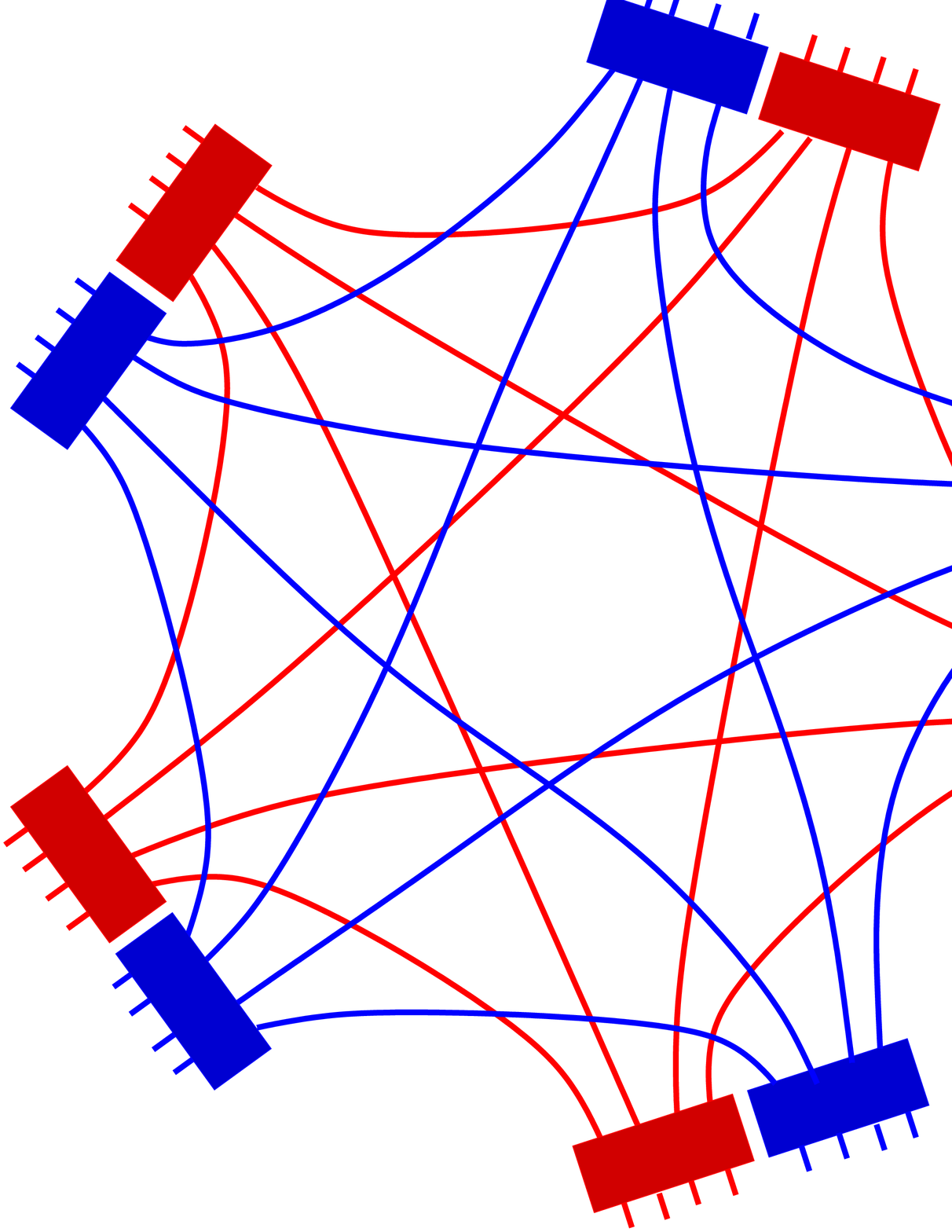}
\end{array}
\label{dosf}
\ea
which follows basically from the invariance of the Haar measure (\ref{invariance}) (in the last line we have used (\ref{cab-3d})). More presicely, the integration of the subgroup $SU(2)\in Spin(4)$, represented by the green box on the right,
can be absorbed by suitable redefinition of the integration on the right and left copies of $SU(2)$, represented by the red and blue boxes respectively.
With this we can already write the spin foam representation of the EPRL model, namely  
\be\label{eprl-sf}
Z^{E}_{eprl}(\Delta)=\sum \limits_{ j_f} \sum_{\iota_e}  \\ \prod\limits_{f \in \Delta^{\star}} {\rm d}_{|1-\gamma|\frac{j}{2}}{\rm d}_{(1+\gamma)\frac{j}{2}} 
\prod_{v\in \Delta^{\star}}\begin{array}{c}\psfrag{a}{$\van \iota_1$}
\psfrag{b}{$\van  \iota_2$}
\psfrag{c}{$\van  \iota_3$}
\psfrag{d}{$\van  \iota_4$}
\psfrag{e}{$\van  \iota_5$}
\includegraphics[width=5cm]{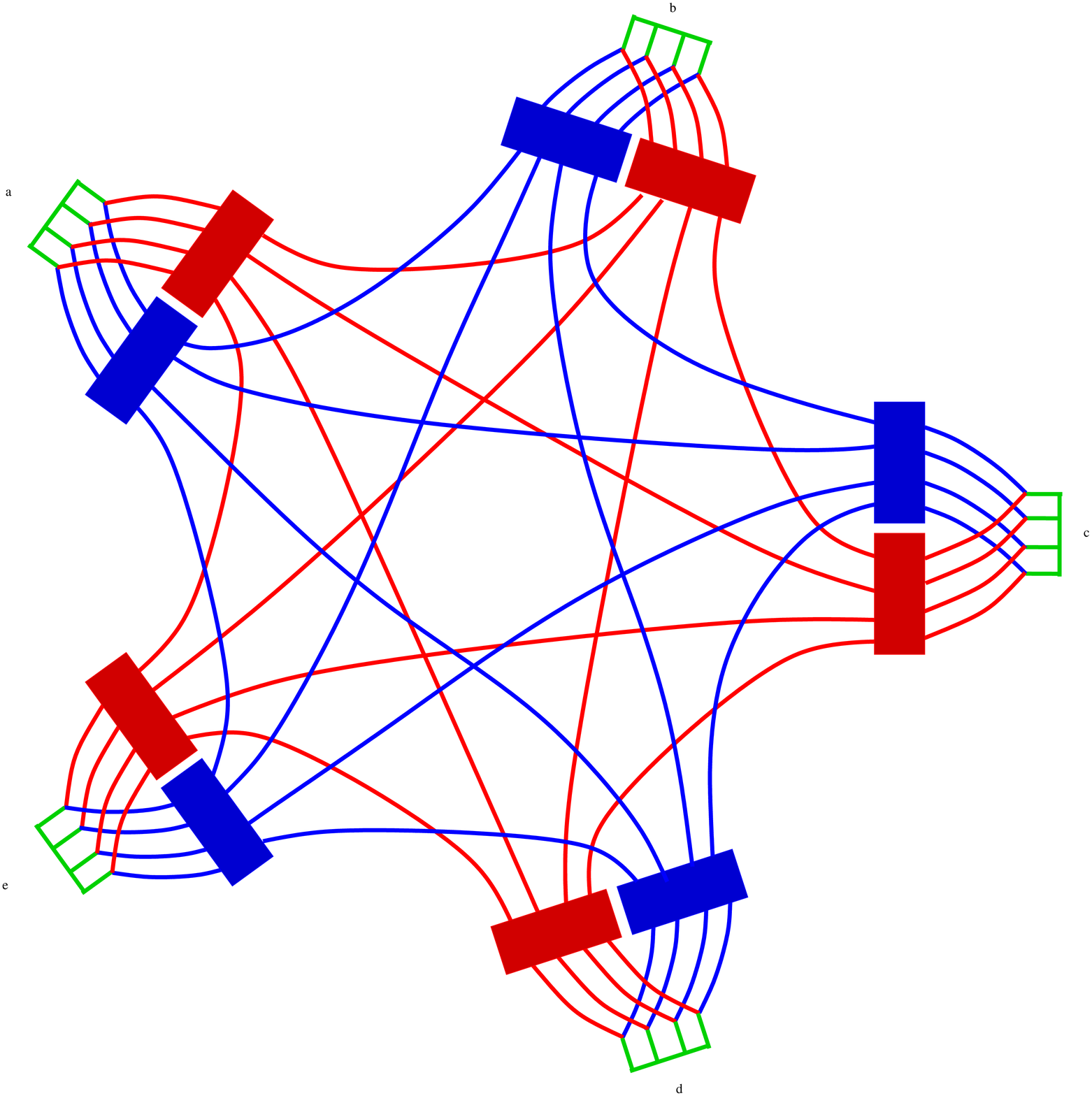}
\end{array},
\ee 
where the vertex amplitude (graphically represented) depends on the 10 spins $j$ associated to the face-wires and the 5 intertwiners associated to the five edges (tetrahedra).
As in previous equations we have left the spin labels of wires implicit for notational simplicity.
We can write the previous spin foam amplitude in another form by  integrating out all the projectors (boxes) explicitly.  Using, (\ref{cab-3d})  we get \be
\begin{array}{c}
\psfrag{w}{$= \sum\limits_{\iota_{\va +}\iota_{\va -}\iota} $}
\psfrag{x}{$\!\!\!\!\!\!\!\!
 \vani \iota_{\va +}\, \ \ \ \bar\iota_{\va +}$}
\psfrag{z}{$\!\!\!\!\!\!\!\! \vani \iota_{\va -}\,\ \ \ \bar\iota_{\va-}$}
\psfrag{y}{$\!\!\, \vani \iota\, \bar\iota$}
\includegraphics[width=6cm]{eprl.eps}\ \ \ \ \  \includegraphics[width=6cm]{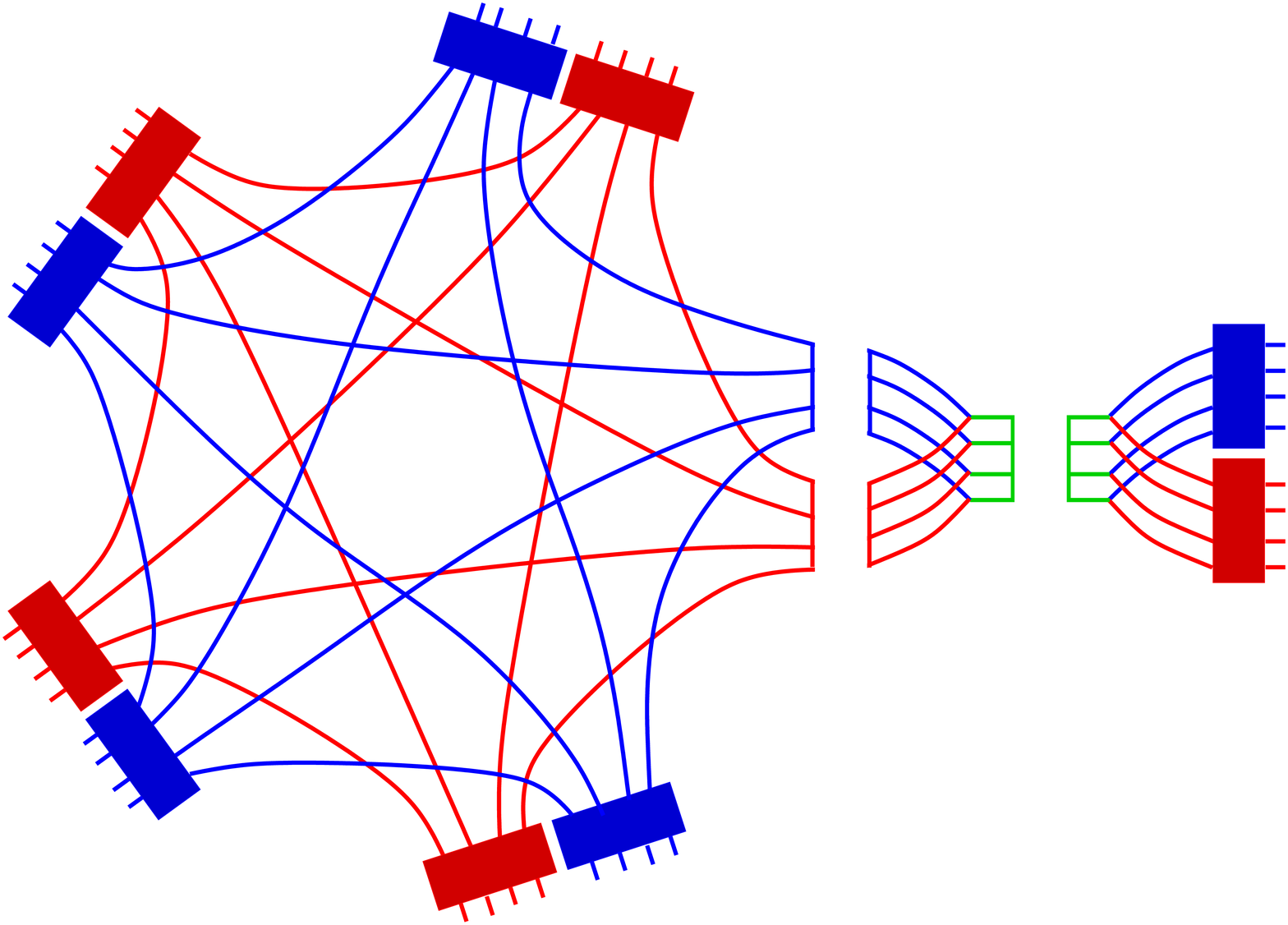}
\end{array}
\label{unof}
\ee
thus replacing this in (\ref{eprl-so4}) we get 
\ba 
&&
Z^{E}_{eprl}(\Delta)=\sum \limits_{ j_f }  \ \prod\limits_{f \in \Delta^{\star}}   {\rm d}_{|\gamma-1|\frac{j}{2}}{\rm d}_{(\gamma+1)\frac{j}{2}} \sum \limits_{\iota_e } \ \prod\limits_{v\in \Delta^{\star}}  \\ &&   \sum_{\iota^{-}_1\cdots \iota^-_5}\sum_{\iota^{+}_1\cdots \iota^+_5} \prod\limits_{a=1}^{5} f^{\iota_a}_{\iota^{-}_a,\iota^{+}_{a}}
\begin{array}{c}
\psfrag{a}{$\van \iota^{-}_1$}
\psfrag{b}{$\van \iota^{-}_2$}
\psfrag{c}{$\van \iota^{-}_3$}
\psfrag{d}{$\van \iota^{-}_4$}
\psfrag{e}{$\van \iota^{-}_5$}
\psfrag{A}{$\!\!\!\!\!\!\!\!\!\!\van |1-\gamma|\frac{j_1}{2}$}
\psfrag{B}{$\!\!\!\!\!\!\!\!\!\! \van |1-\gamma|\frac{j_2}{2}$}
\psfrag{C}{$\!\!\!\!\!\!\!\!\!\!\!\! \van |1-\gamma|\frac{j_3}{2}$}
\psfrag{D}{$\!\!\!\!\!\!\!\!\!\!\!\! \van |1-\gamma|\frac{j_4}{2}$}
\psfrag{E}{$\!\!\!\!\!\!\!\!\!\!\!\! \van |1-\gamma|\frac{j_5}{2}$}
\psfrag{F}{$\!\!\!\!\!\!\!\! \van |1-\gamma|\frac{j_6}{2}$}
\psfrag{G}{$\!\!\!\!\!\!\!\! \van |1-\gamma|\frac{j_7}{2}$}
\psfrag{H}{$\!\!\!\!\!\!\!\! \van |1-\gamma|\frac{j_8}{2}$}
\psfrag{I}{$\!\!\!\!\!\!\!\!\!\!\!\! \van |1-\gamma|\frac{j_9}{2}$}
\psfrag{J}{$\!\!\!\!\!\!\!\!\!\!\!\! \van |1-\gamma|\frac{j_{10}}{2}$}
\includegraphics[height=4.7cm]{BF4V-g.eps}
\end{array}\ \ \ \ \ 
\begin{array}{c}
\psfrag{ap}{$\van \iota^{-}_1$}
\psfrag{bp}{$\van \iota^{-}_2$}
\psfrag{cp}{$\van \iota^{-}_3$}
\psfrag{dp}{$\van \iota^{-}_4$}
\psfrag{ep}{$\van \iota^{-}_5$}
\psfrag{Ap}{$\!\!\!\!\!\!\!\! \van |1+\gamma|\frac{j_1}{2}$}
\psfrag{Bp}{$\!\!\!\!\!\!\!\!\!\! \van |1+\gamma|\frac{j_2}{2}$}
\psfrag{Cp}{$\!\!\!\!\!\!\!\!\!\!\!\! \van |1+\gamma|\frac{j_3}{2}$}
\psfrag{Dp}{$\!\!\!\!\!\!\!\!\!\!\!\! \van |1+\gamma|\frac{j_4}{2}$}
\psfrag{Ep}{$\!\!\!\!\!\!\!\!\!\!\!\! \van |1+\gamma|\frac{j_5}{2}$}
\psfrag{Fp}{$\!\!\!\!\!\!\!\! \van |1+\gamma|\frac{j_6}{2}$}
\psfrag{Gp}{$\!\!\!\!\!\!\!\! \van |1+\gamma|\frac{j_7}{2}$}
\psfrag{Hp}{$\!\!\!\!\!\!\!\! \van |1+\gamma|\frac{j_8}{2}$}
\psfrag{Ip}{$\!\!\!\!\!\!\!\!\!\!\!\! \van |1+\gamma|\frac{j_9}{2}$}
\psfrag{Jp}{$\!\!\!\!\!\!\!\!\!\!\!\! \van |1+\gamma|\frac{j_{10}}{2}$}
\includegraphics[height=4.7cm]{BF4V-b.eps}
\end{array}\n 
\label{SF-eprl}
\ea
where the coefficients $f^{\iota}_{\iota^+\iota^{-}}$ are the so-called fusion coefficients
which appear in their graphical form already in (\ref{unof}), more explicitly
\be f^{\iota}_{\iota^+\iota^{-}}(j_1,\cdots,j_4)=
\begin{array}{c}
\psfrag{x}{$ \van \iota_{ +}$}
\psfrag{z}{$\van  \iota_{\va -}$}
\psfrag{y}{$\van \!\!\, \iota$}
\psfrag{a}{$\van |1-\gamma| \frac{j_1}{2}$}
\psfrag{b}{$\van |1-\gamma| \frac{j_2}{2}$}
\psfrag{c}{$\van |1-\gamma| \frac{j_3}{2}$}
\psfrag{d}{$\van |1-\gamma| \frac{j_4}{2}$}
\psfrag{ap}{$\van |1+\gamma| \frac{j_1}{2}$}
\psfrag{bp}{$\van |1+\gamma| \frac{j_2}{2}$}
\psfrag{cp}{$\van |1+\gamma| \frac{j_3}{2}$}
\psfrag{dp}{$\van |1+\gamma| \frac{j_4}{2}$}
\psfrag{A}{$\van {j_1}$}
\psfrag{B}{$\van {j_2}$}
\psfrag{C}{$\van {j_3}$}
\psfrag{D}{$\van {j_4}$}
\includegraphics[height=5cm]{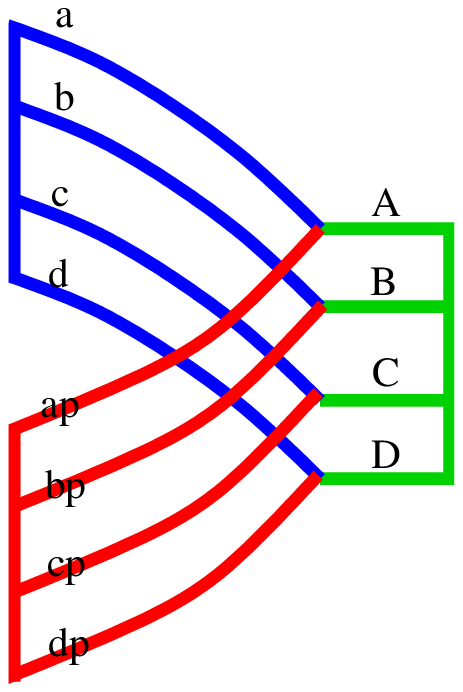}
\end{array}
\ee
The previous is the form of the EPRL model as derived in \cite{Engle:2007wy}.

\subsection{Proof of validity of the Plebanski constraints}\label{cuadratiqui}
In this section we prove that the quadratic constraints are satisfied in the sense 
that their path integral expectation value and fluctuation vanish in the appropriate 
semiclassical limit.

\subsubsection{The quadratic Plebanski constraints}

The quadratic Plebanski constraints are
\be \epsilon_{IJKL} B^{IJ}_{\mu\nu}B^{KL}_{\rho\sigma}-e \
 \epsilon_{\mu\nu\rho\sigma}\approx 0. \label{dual}\ee
The constraints in this form are more suitable for the translation into the 
discrete formulation. More precisely, according to (\ref{Bdisc}), the smooth fields $B_{\mu\nu}^{IJ}$ is now associated with
the discrete quantities $B_{\va {\rm triangles}}^{IJ}$, or equivalently $B^{IJ}_{f}$ as, we recall, faces $f\in \Delta^{\star}$ are in one-to-one correspondence to 
triangles in four dimensions.  The constraints (\ref{dual}) are local constraints valid at every spacetime point. 
In the discrete setting, spacetime points are represented by four-simplexes or (more addapted to our discussion) vertices $v\in \Delta^{\star}$.
With all this the constraints (\ref{dual}) are discretized as follows:
\be\label{2s}
\mbox{\bf Triangle (or diagonal) constraints: \ \ \ \ \ \ \ $ \epsilon_{IJKL} B^{IJ}_{f}B^{KL}_{f}=0$,}
\ee
for all  $f\in v$, i.e., for each and every face of the 10 possible faces touching the vertex $v$.
\be\label{3s}
\mbox{\bf Tetrahedron constraints: \ \ \ \ \ \ \  $\epsilon_{IJKL} B^{IJ}_{f}B^{KL}_{f'}=0$,}
\ee
for all $f,f'\in v$ such that they are dual to triangls sharing a one-simplex, i.e., belonging to the same tetrahedron out of the five possible ones. 
\be\label{4s}
\mbox{\bf 4-simplex constraints: \ \ \ \ \ \ \  $\epsilon_{IJKL} B^{IJ}_{f}B^{KL}_{\bar f}=e_v$,}
\ee
for any pair of faces $f,\bar f\in v$ that are dual to triangles sharing a single point. The last constraint will require a more detailed discussion. At this point let us point out that the 
constraint (\ref{4s}) is interpreted as a definition of the four volume $e_v$ of the four-simplex. The constraint requires that such definition be consistent, i.e., the true condition is
\be\label{4strue}
\epsilon_{IJKL} B^{IJ}_{f}B^{KL}_{\bar f}=\epsilon_{IJKL} B^{IJ}_{f'}B^{KL}_{\bar f'}=\epsilon_{IJKL} B^{IJ}_{f''}B^{KL}_{\bar f''}=\cdots=e_v
\ee
for all five different possible pairs of $f$ and $\bar f$ in a four simplex, and where we assume the pairs $f$-$\bar f$ are ordered 
in agreement with the orientation of the complex $\Delta^{\star}$.

\subsubsection{The path integral expectation value of the Plebanski constraints}

Here we prove that the Plebanski constraint are satisfied by the EPRL amplitudes in the 
path integral expectation value sense.

\subsubsection*{The triangle constraints:} 

We start from the simplest case:  the triangle (or diagonal) constraints (\ref{2s}).
We choose a face $f\in v$ (dual to a triangle) in the cable-wire-diagram of Equation (\ref{eprl-so4}).
This amounts to choosing a pair of wires (right and left representations) connecting two nodes in the vertex cable wire diagram. 
The two nodes are dual to the two tetrahedra---in the four simplex dual to the vertex---sharing the chosen triangle.
From equation (\ref{defidefi}) can show that
\be
 \epsilon_{IJKL} B^{IJ}_{f}B^{KL}_{f}\propto (1+\gamma)^2 J_{f}^{-}\cdot J_{f}^{-} -(1-\gamma)^2 J_{f}^{+}\cdot J_{f}^{+},
\ee
where $J_{f}^{\pm}$ denotes the self-dual and anti-self-dual parts of $\Pi^{IJ}_{f}$.  
The path integral expectation value of the triangle constraint is then
\ba \label{pipo}
 && \langle (1+\gamma)^2 J_{f}^{-}\cdot J_{f}^{-} -(1-\gamma)^2 J_{f}^{+}\cdot J_{f}^{+}\rangle \propto \\ 
&& \n (1+\gamma)^2 \begin{array}{c}
\includegraphics[width=4.5cm]{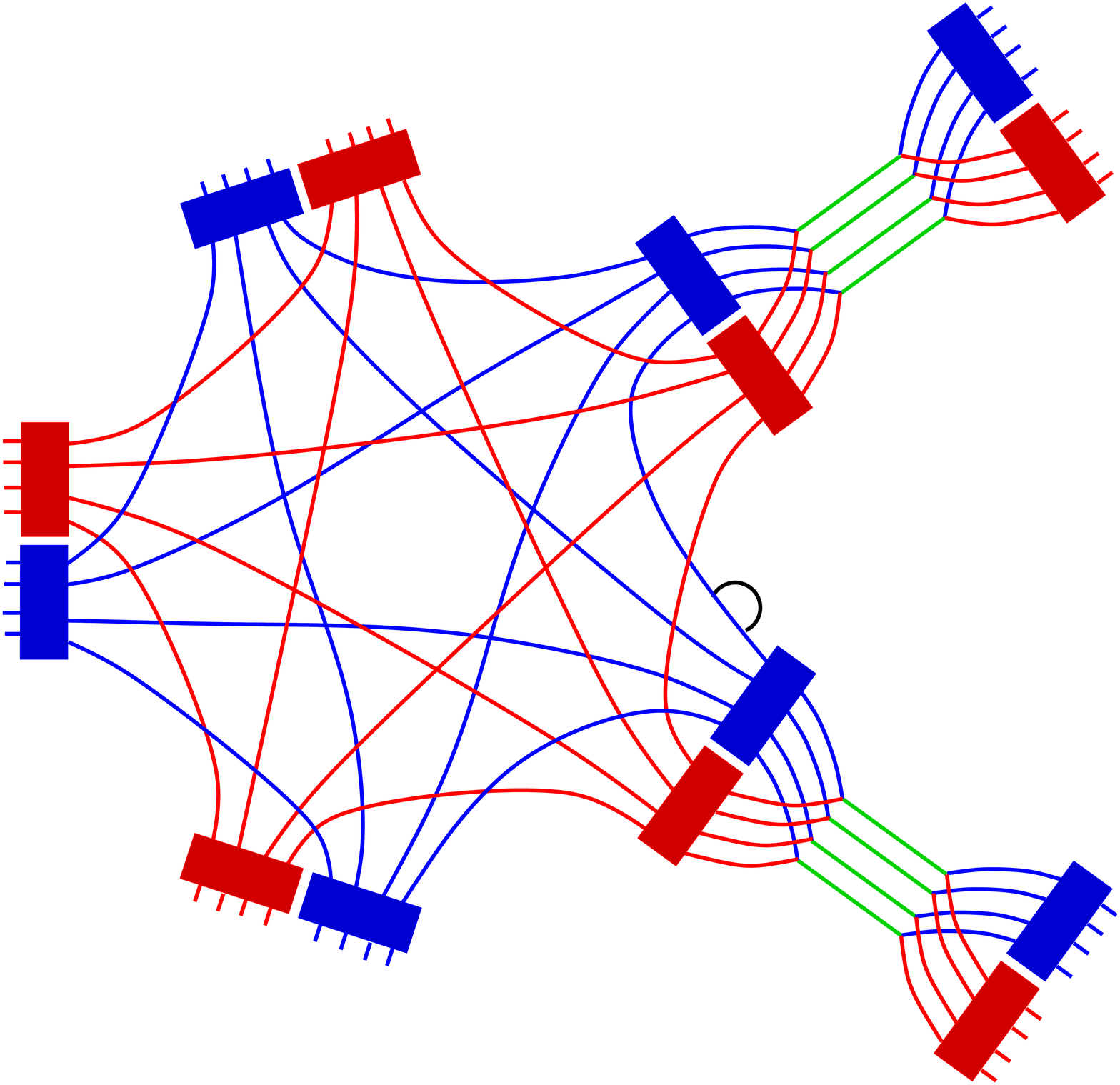}
\end{array}
-(1-\gamma)^2
\begin{array}{c}
\includegraphics[width=4.5cm]{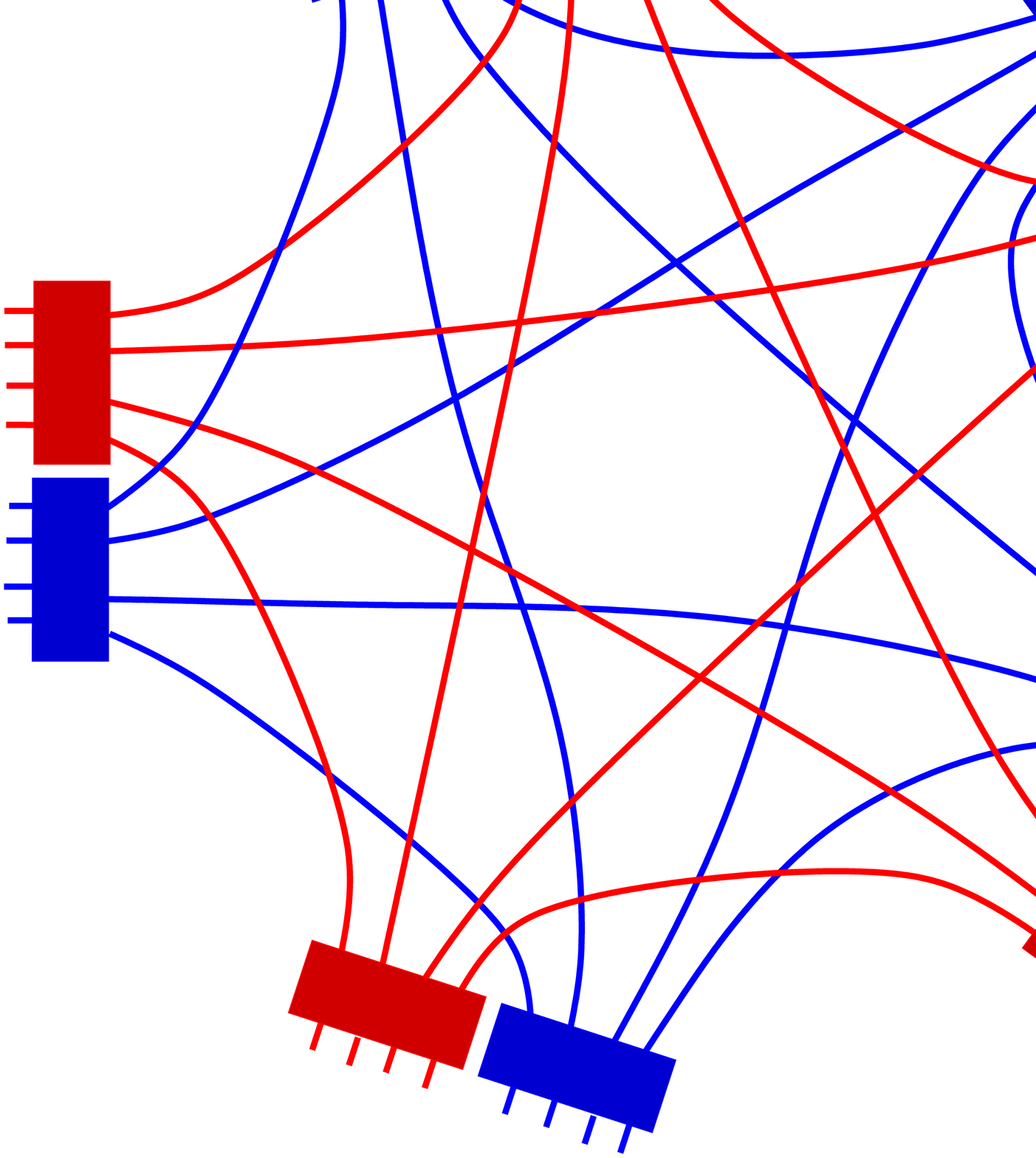}
\end{array}=\sO_{sc},
\ea
where the double graspings on the anti-self-dual (blue) wire and the self-dual (red) wire represent the action of the Casimirs
$J_{f}^{-}\cdot J_{f}^{-}$ and $J_{f}^{+}\cdot J_{f}^{+}$ on the cable-wire diagram of the corresponding vertex. Direct evaluation
shows that the previous diagram is proportional to $\hbar^2 j_f$ which vanishes in the semiclassical limit $\hbar\to 0$, $j\to \infty$
with $\hbar j=$constant. We use the notation already adopted in (\ref{needed}) and call such quantity $\sO_{sc}$. This concludes the proof that the triangle Plebanski constraints 
are satisfied in the semiclassical sense.

 \subsubsection*{The tetrahedra constraints:} 
 
The proof of the validity of the tetrahedra constraints (\ref{3s}). In this case we also have
\be
(1+\gamma)^2 \begin{array}{c}
\includegraphics[width=4.5cm]{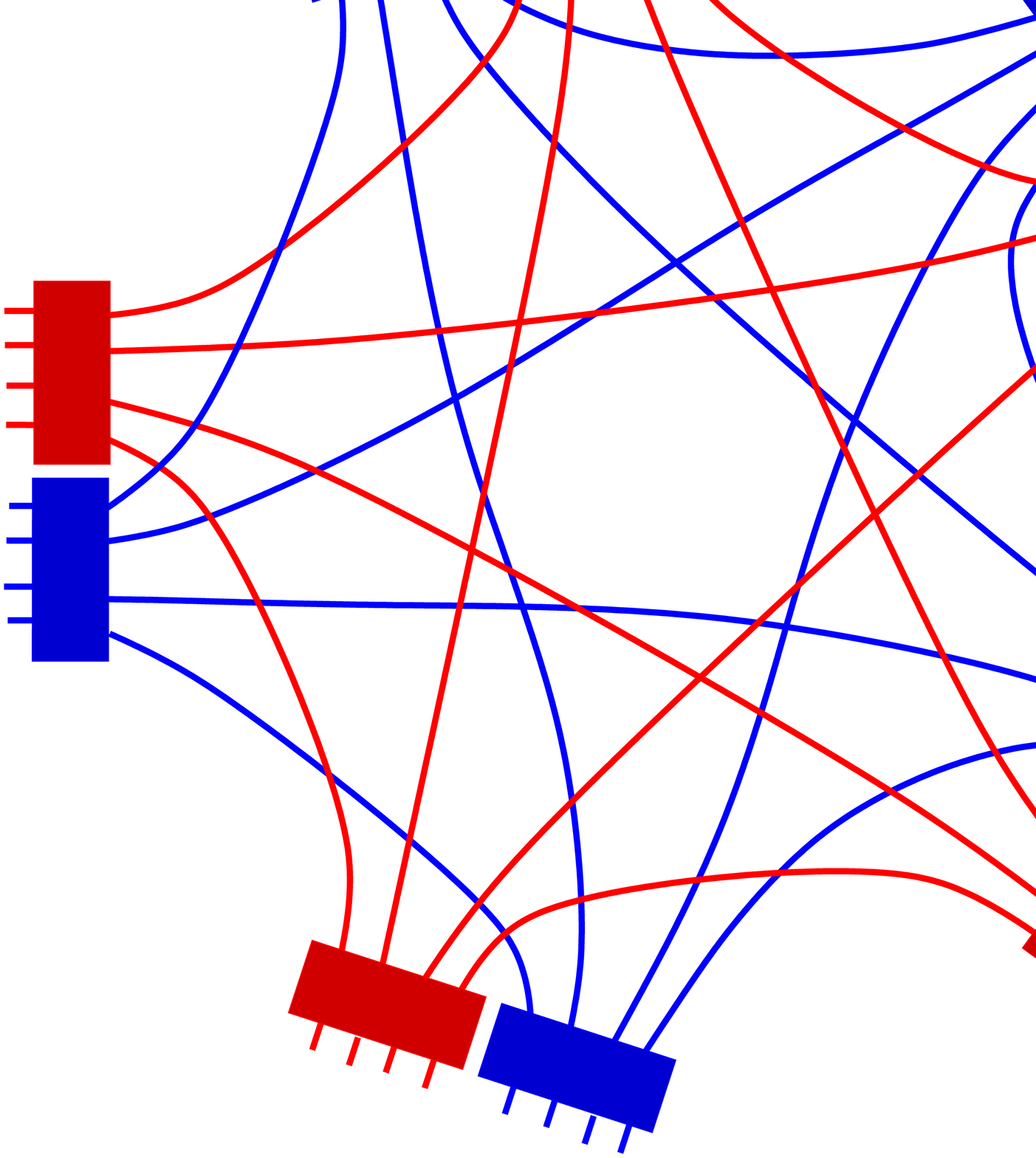}
\end{array}
-(1-\gamma)^2
\begin{array}{c}
\includegraphics[width=4.5cm]{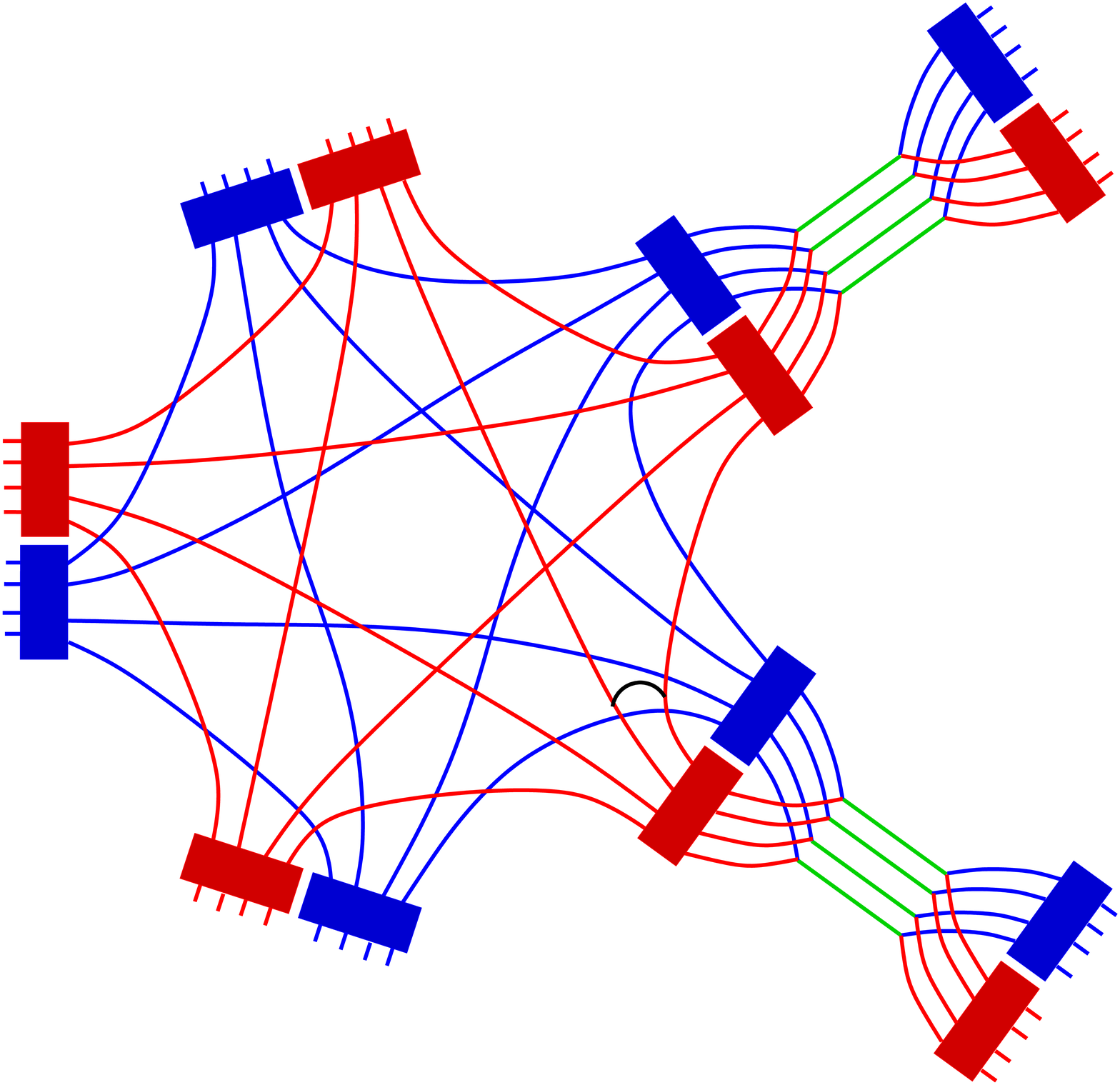}
\end{array}=\sO_{sc}.\label{tetra}
\ee
where we have chosen an arbitrary pair of faces. 
In order to prove this let us develop the term on the right. The result follows from 
\ba &&
\begin{array}{c}
\includegraphics[width=2cm]{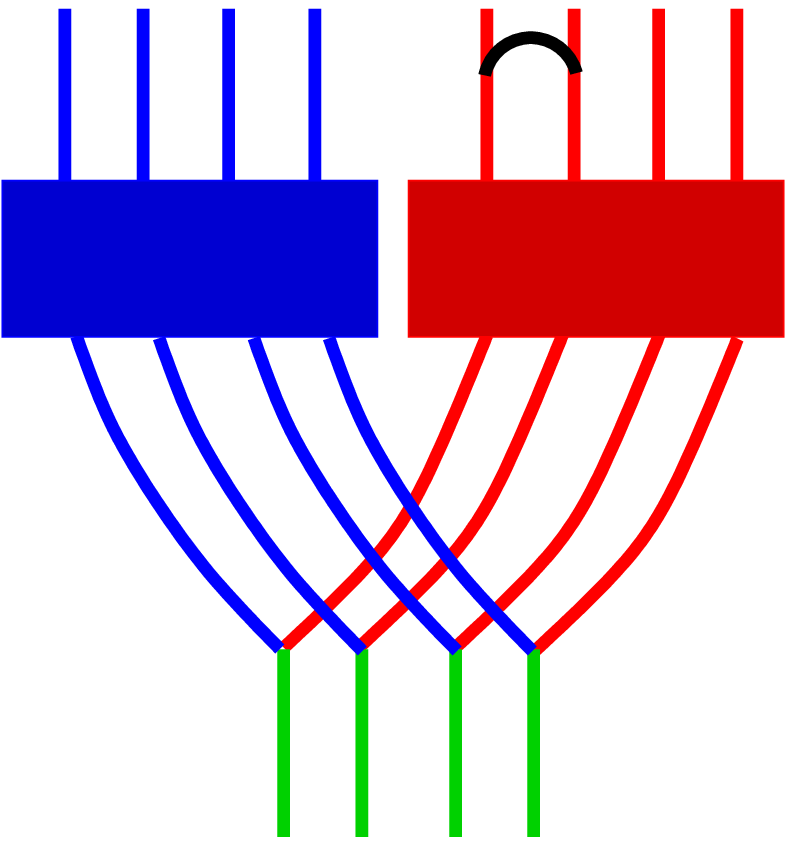}
\end{array}
=
\begin{array}{c}
\includegraphics[width=2cm]{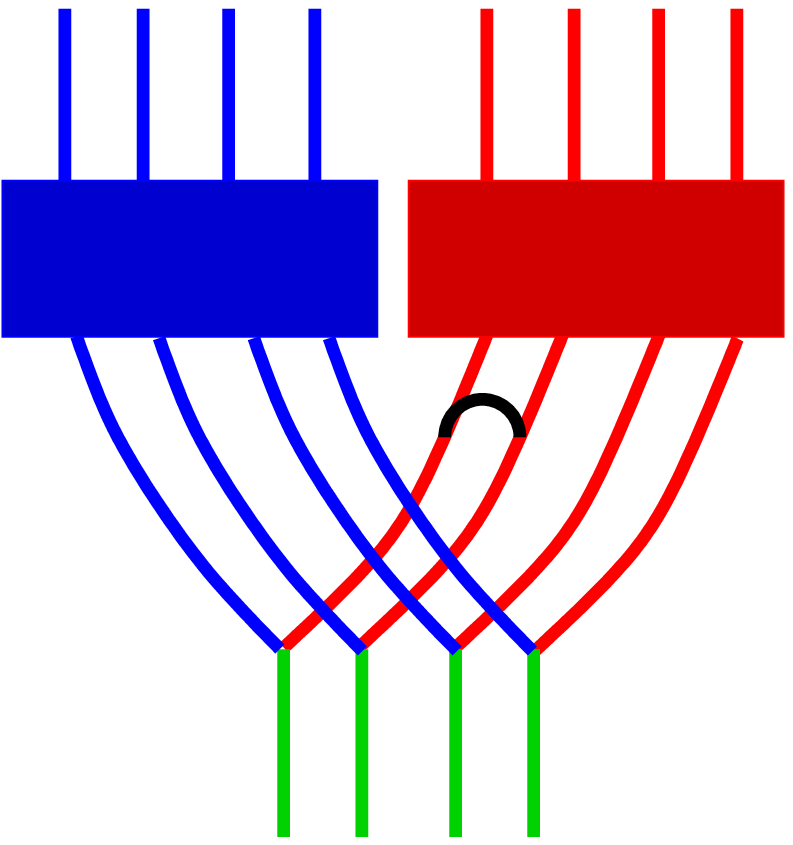}
\end{array}=\n \\
&& =\frac{(1+\gamma)}{|1-\gamma|}
\begin{array}{c}
\includegraphics[width=2cm]{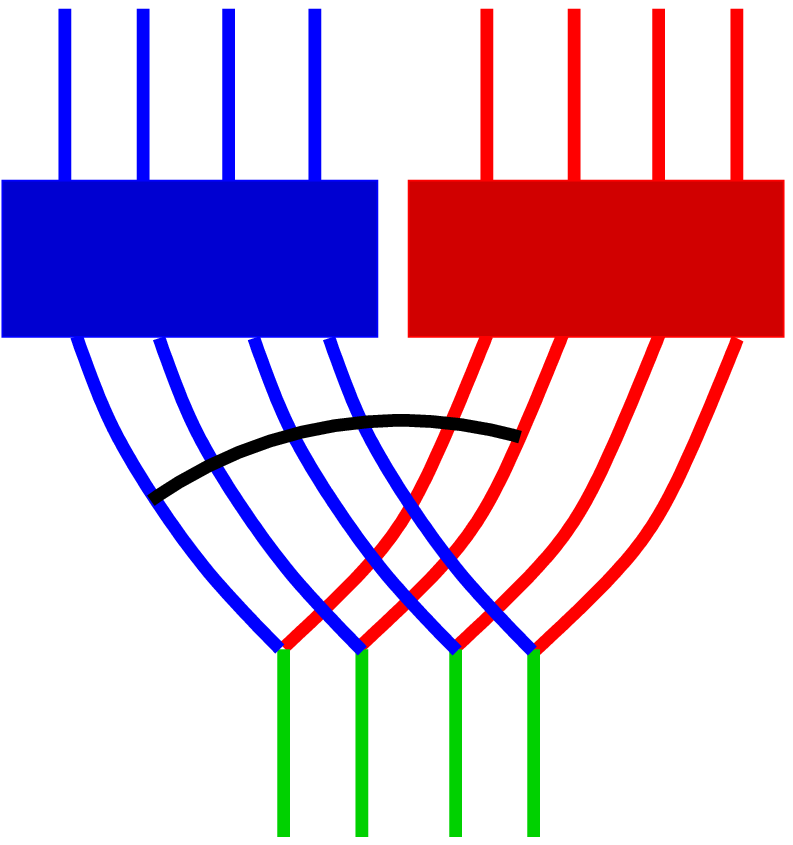}
\end{array}+\sO_{sc}
=\frac{(1+\gamma)^2}{(1-\gamma)^2}
\begin{array}{c}
\includegraphics[width=2cm]{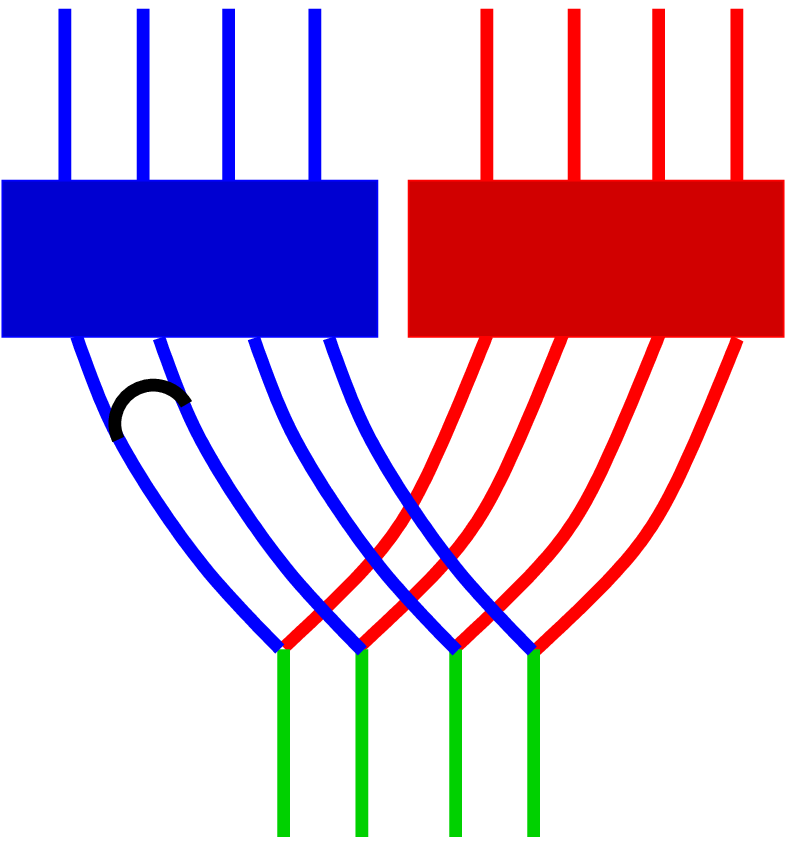}
\end{array}+\sO_{sc}=\n \\ &&=\frac{(1+\gamma)^2}{(1-\gamma)^2} \begin{array}{c}
\includegraphics[width=2cm]{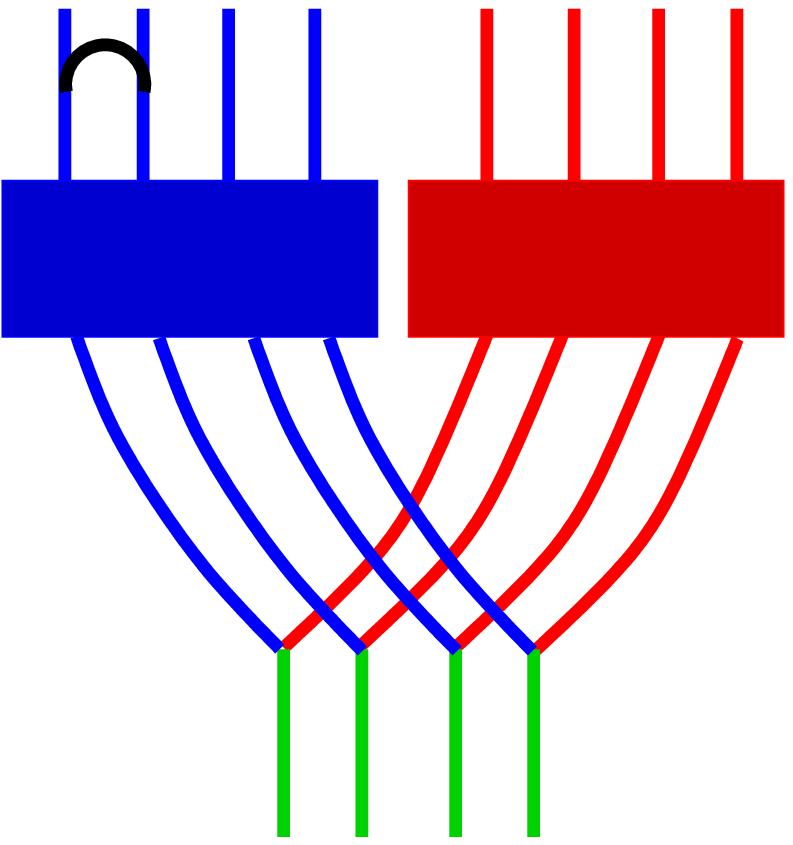}
\end{array}+\sO_{sc}, \label{pipisi}
\ea
where in the first line we have used the fact that the double grasping can be shifted through the group integration (due to gauge invariance (\ref{invariance})), and
in the first and second terms on the second line we have used Equation (\ref{trival}) to move the graspings on self-dual wires to the corresponding anti-self-dual wires. 
Equation (\ref{tetra}) follows immediately from the pervious one; the argument works in the same way for any other pair of faces.
Notice that the first equality in Equation (\ref{pipisi}) implies that we can view the Plebanski constraint as applied in the frame of the tetrahedron as well as in a Lorentz invariant 
framework (the double grasping defines an intertwiner operator commuting with the projection $P^e_{inv}$ represented by the box). An analogous statement also holds 
for the triangle constraints (\ref{pipo}).

\subsubsection*{The 4-simplex constraints}

Now we show the validity of the four simplex constraints in their form (\ref{4strue}). As we show below, this last set of constraints 
follow from the $Spin(4)$ gauge invariance of the EPRL node (i.e., the validity of the Gauss law) plus the validity of the tetrahedra constraints (\ref{3s}).
Gauge invariance of the node  takes the following form in graphical notation
\be
\begin{array}{c}
\includegraphics[width=2cm]{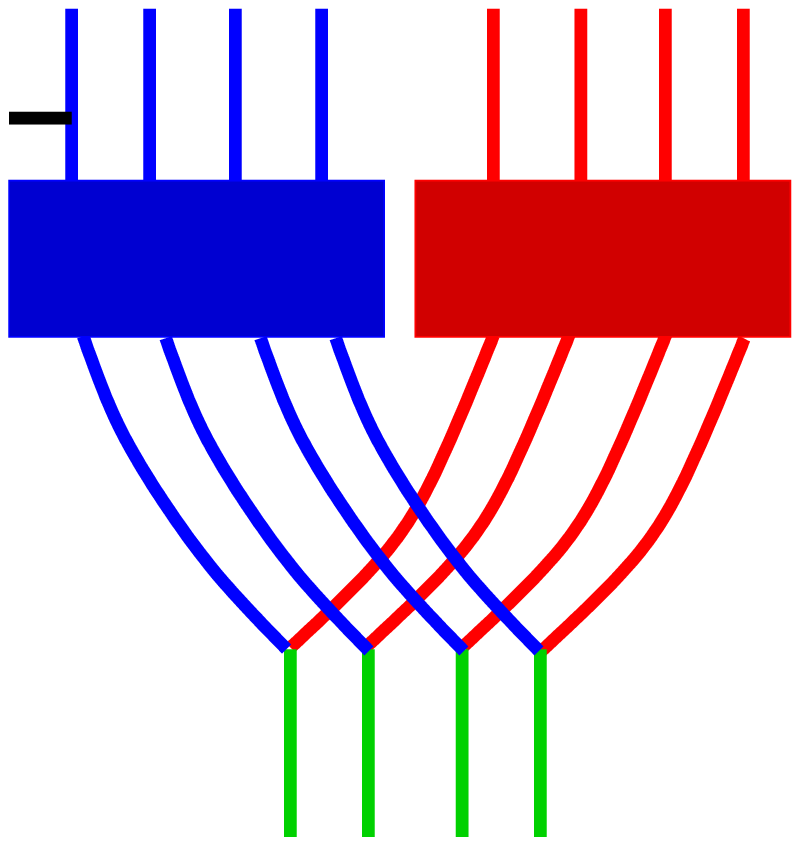}
\end{array}
+
\begin{array}{c}
\includegraphics[width=2cm]{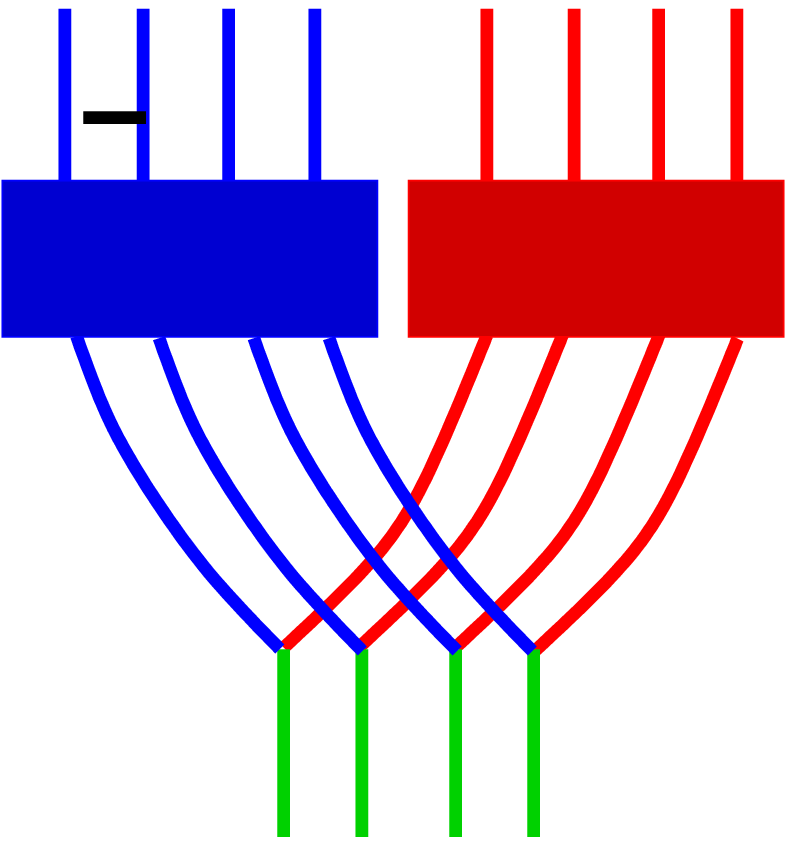}
\end{array}+\begin{array}{c}
\includegraphics[width=2cm]{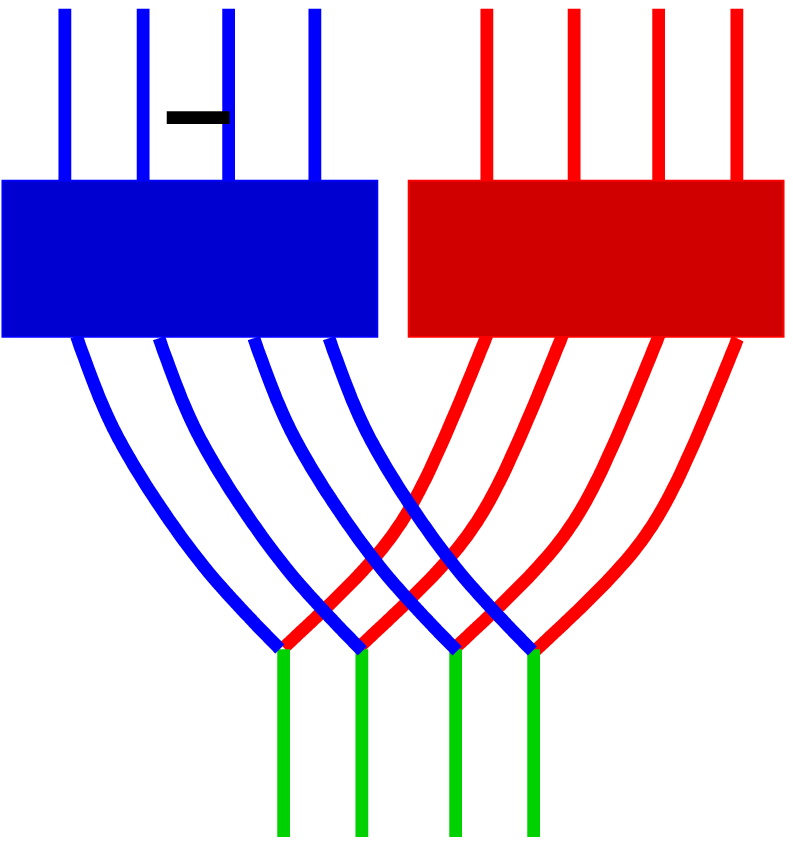}
\end{array}+\begin{array}{c}
\includegraphics[width=2cm]{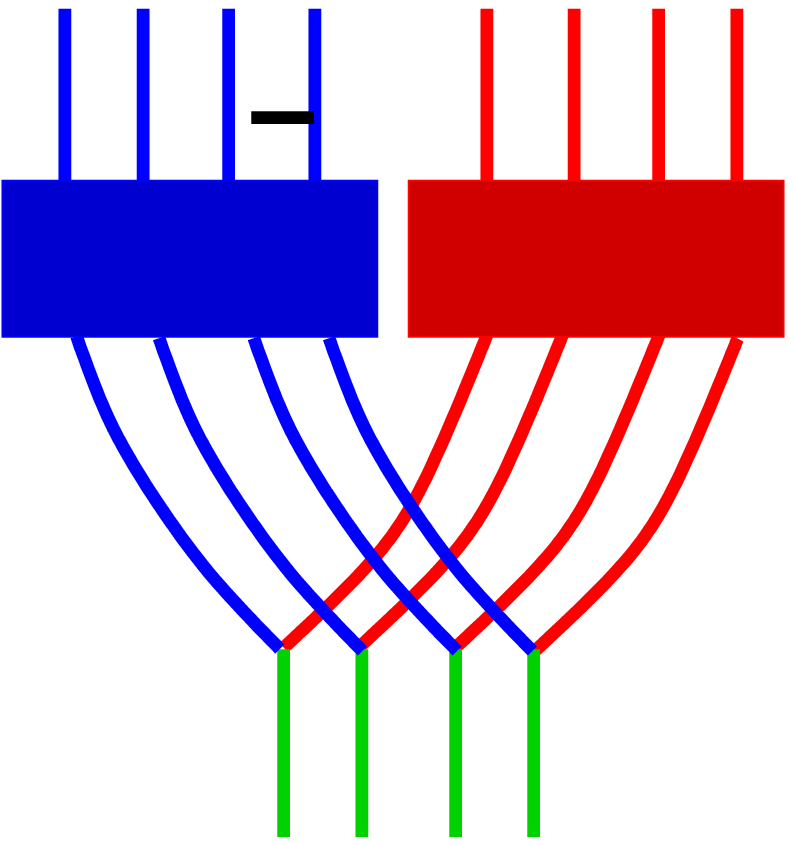}
\end{array}=0,\label{eprl-gauss}
\ee
where the above equation represents the gauge invariance under infinitesimal left $SU(2)$ rotations.
An analogous equation with insertions on the right is also valid. The validity of the previous equation can 
again be related to the invariance of the Haar measure used in the integration on the gauge group that 
defines the boxes (\ref{invariance}).

Now we chose an arbitrary pair $f$ and $\bar f$ (where, recall, $\bar f$ is one of the three possible faces whose dual triangle only shares a point with the 
corresponding to $f$) and will show how the four volumen $e_v$ defined by it equals the one defined by any other admissible pair. The first  step is to show that
we get the same result using the pair $f$-$\bar f$ and $f$-$\bar{\bar f}$, where $\bar{\bar f}$ is another of the three admissible faces opposite to $f$. 
The full result follows from applying the same procedure iteratively to reach any admissible pair. It will be obvious from the treatment given below that this is possible.
Thus, for a given pair of admissible faces we have
\ba && \n
\!\!\!\!\!\!\!\!\!\!\!\!e_v=(1+\gamma)^2 \begin{array}{c}
\includegraphics[width=4cm]{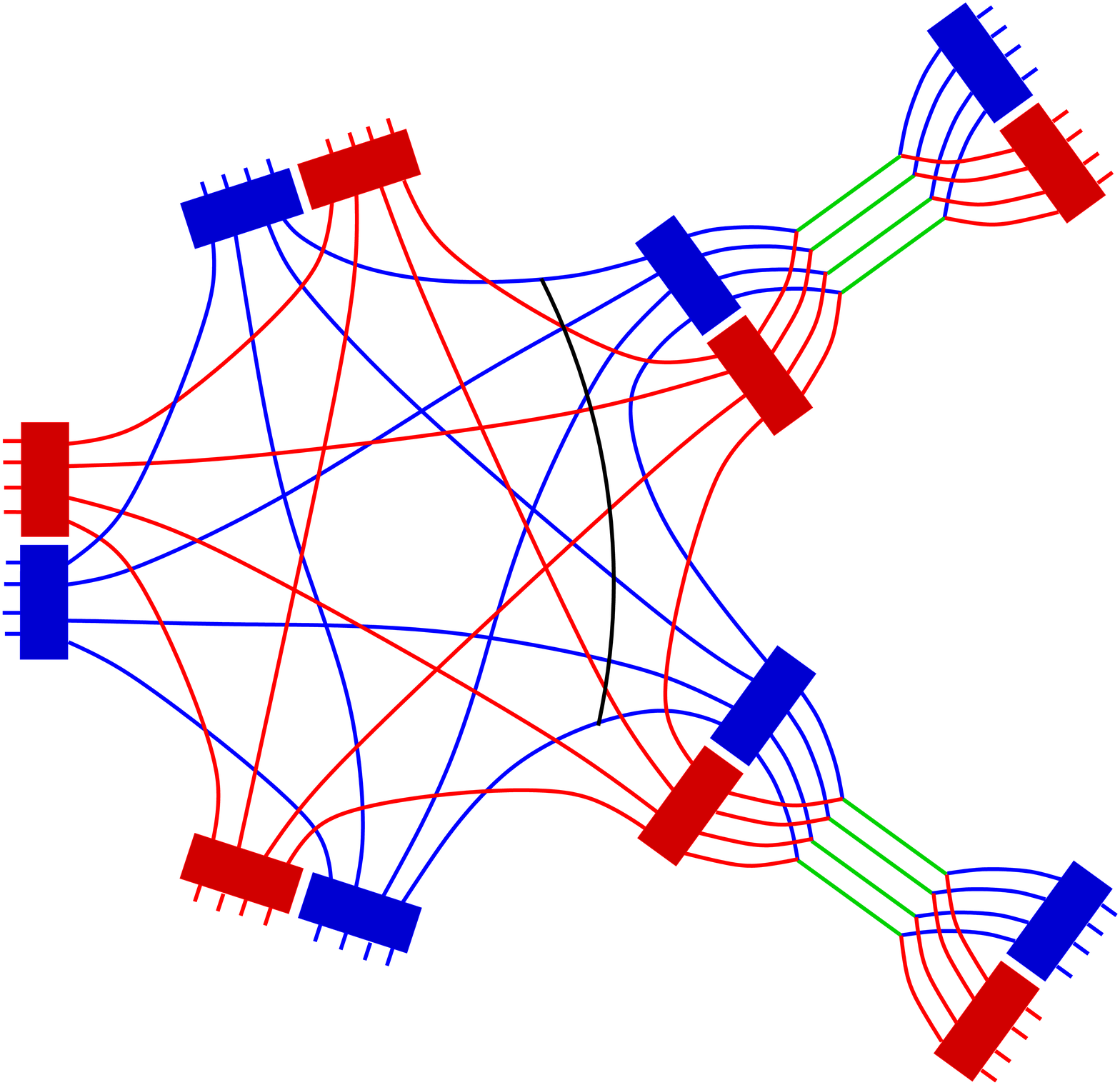}
\end{array}
-(1-\gamma)^2
\begin{array}{c}
\includegraphics[width=4cm]{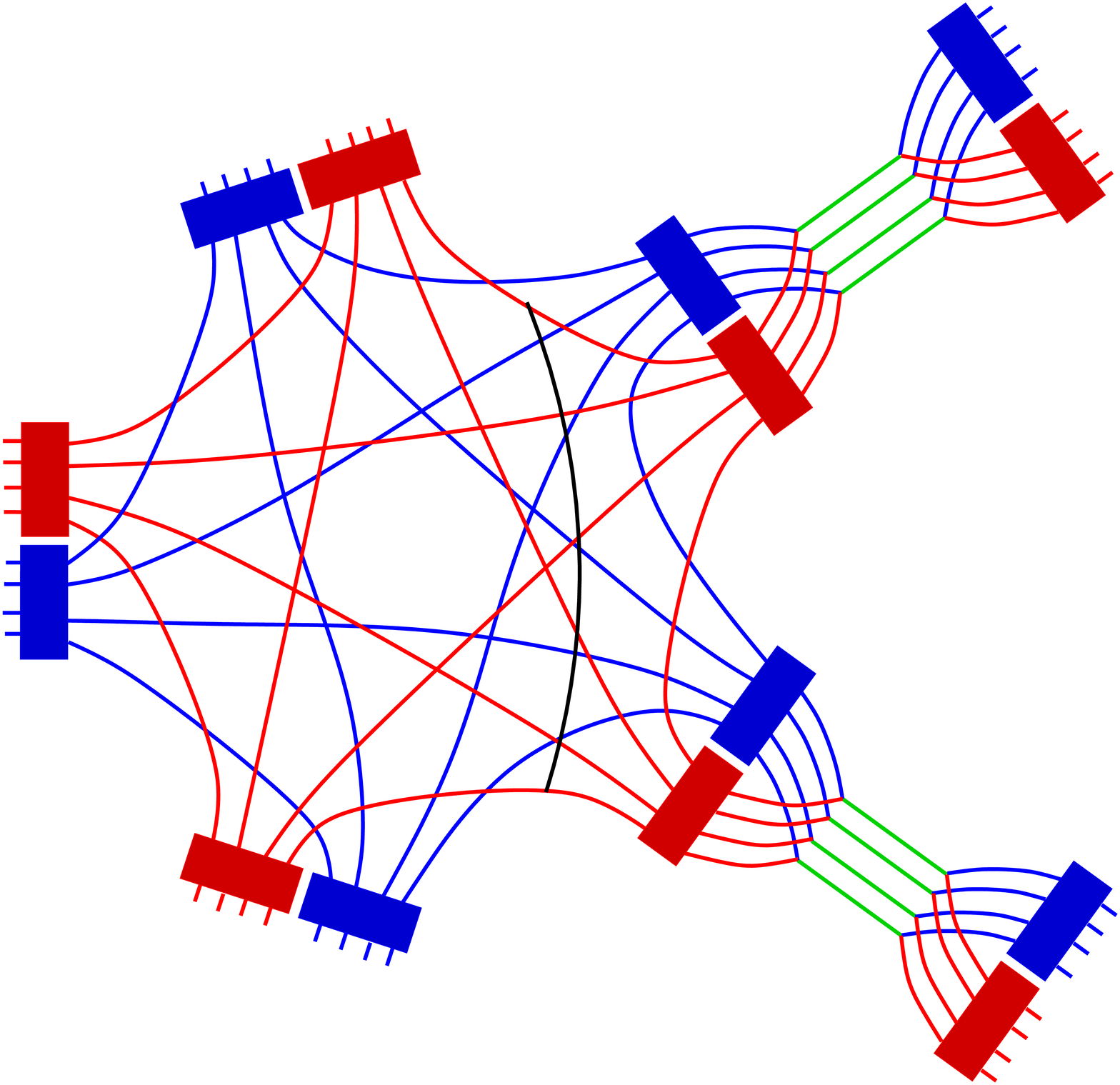}
\end{array} =\\ 
&& \!\!\!\!\!\!\!\!\!\!\!\! -(1+\gamma)^2\left[ \begin{array}{c}
\includegraphics[width=4cm]{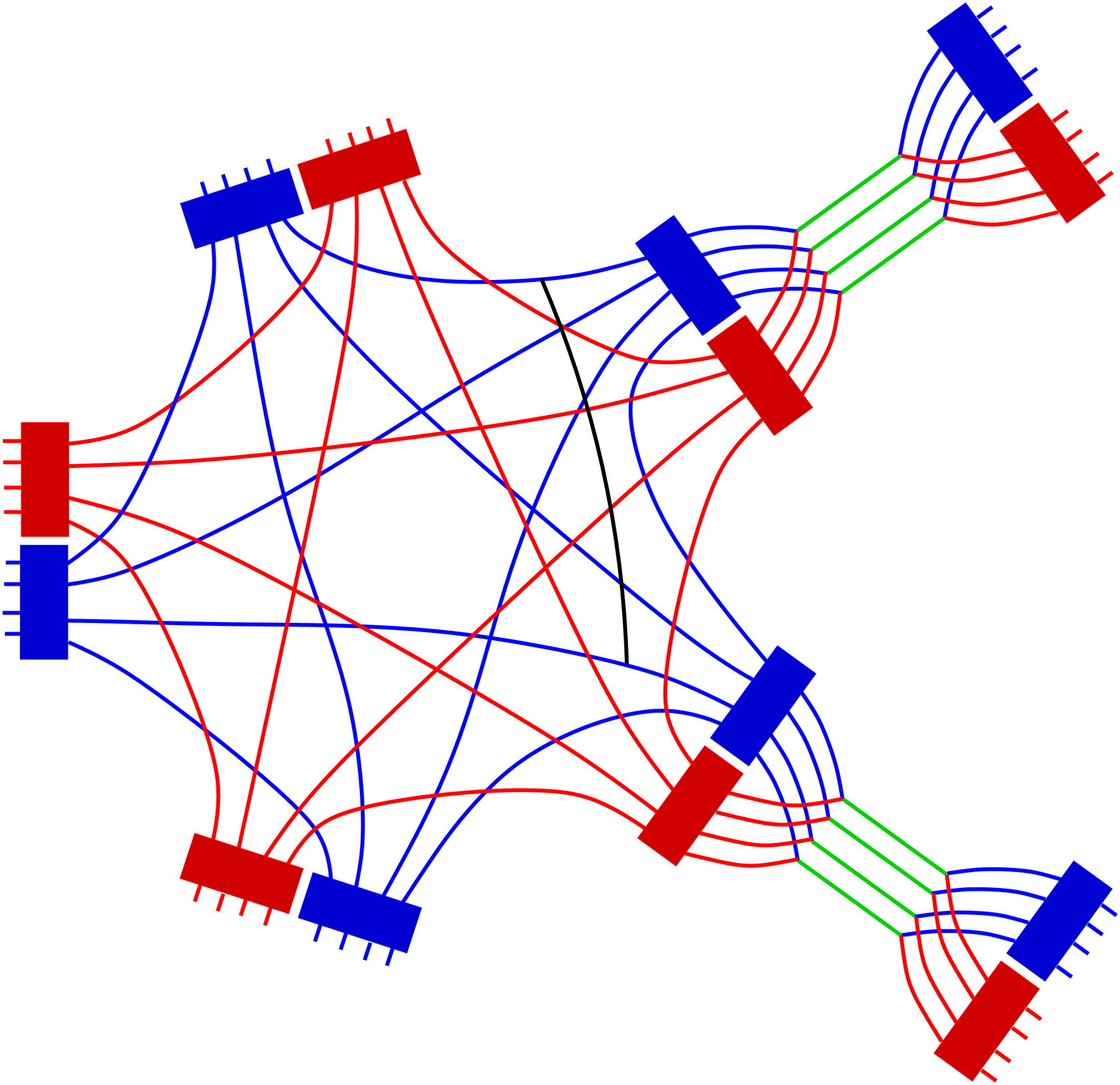}
\end{array}+\begin{array}{c}
\includegraphics[width=4cm]{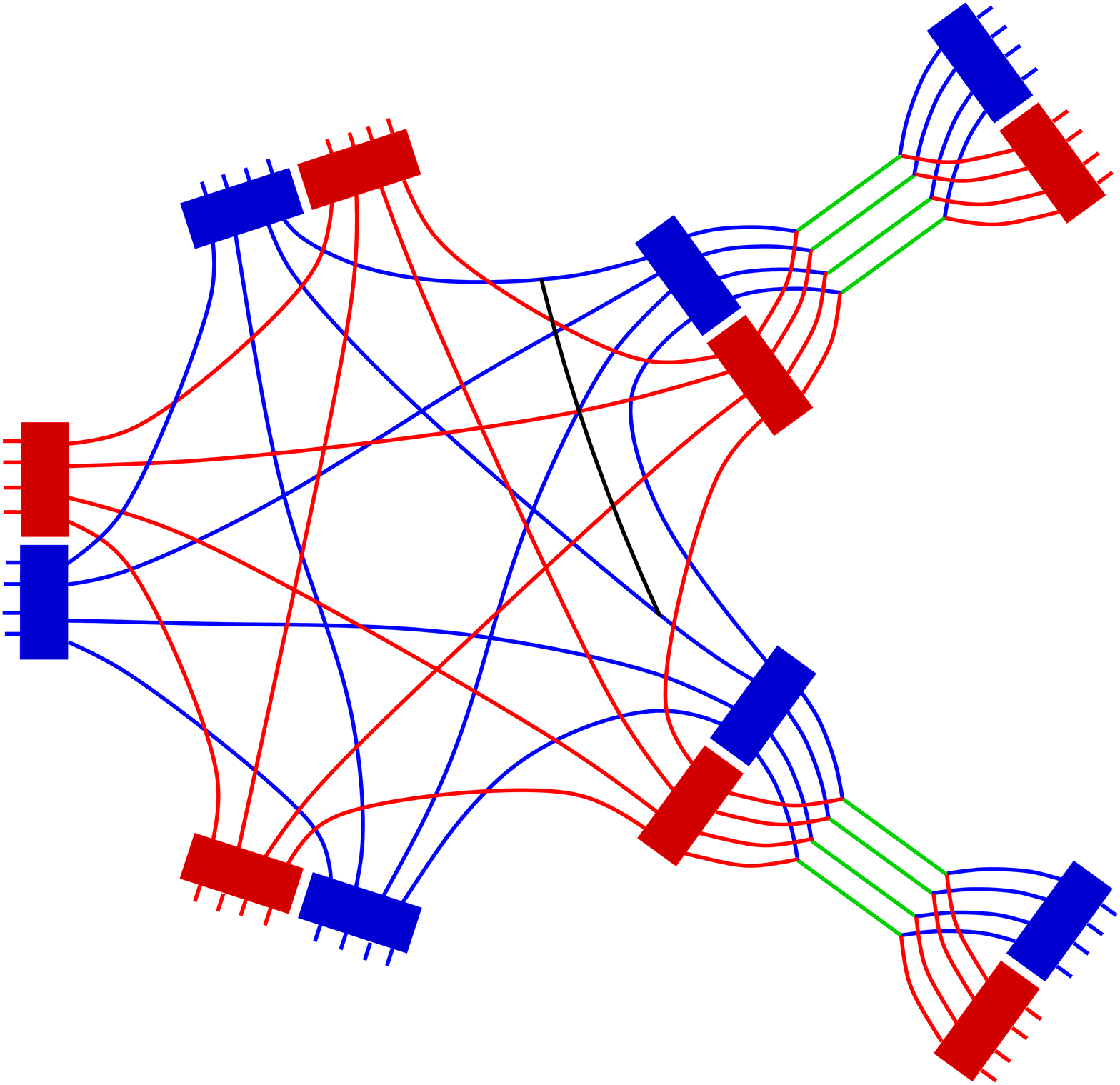}
\end{array}+\begin{array}{c}
\includegraphics[width=4cm]{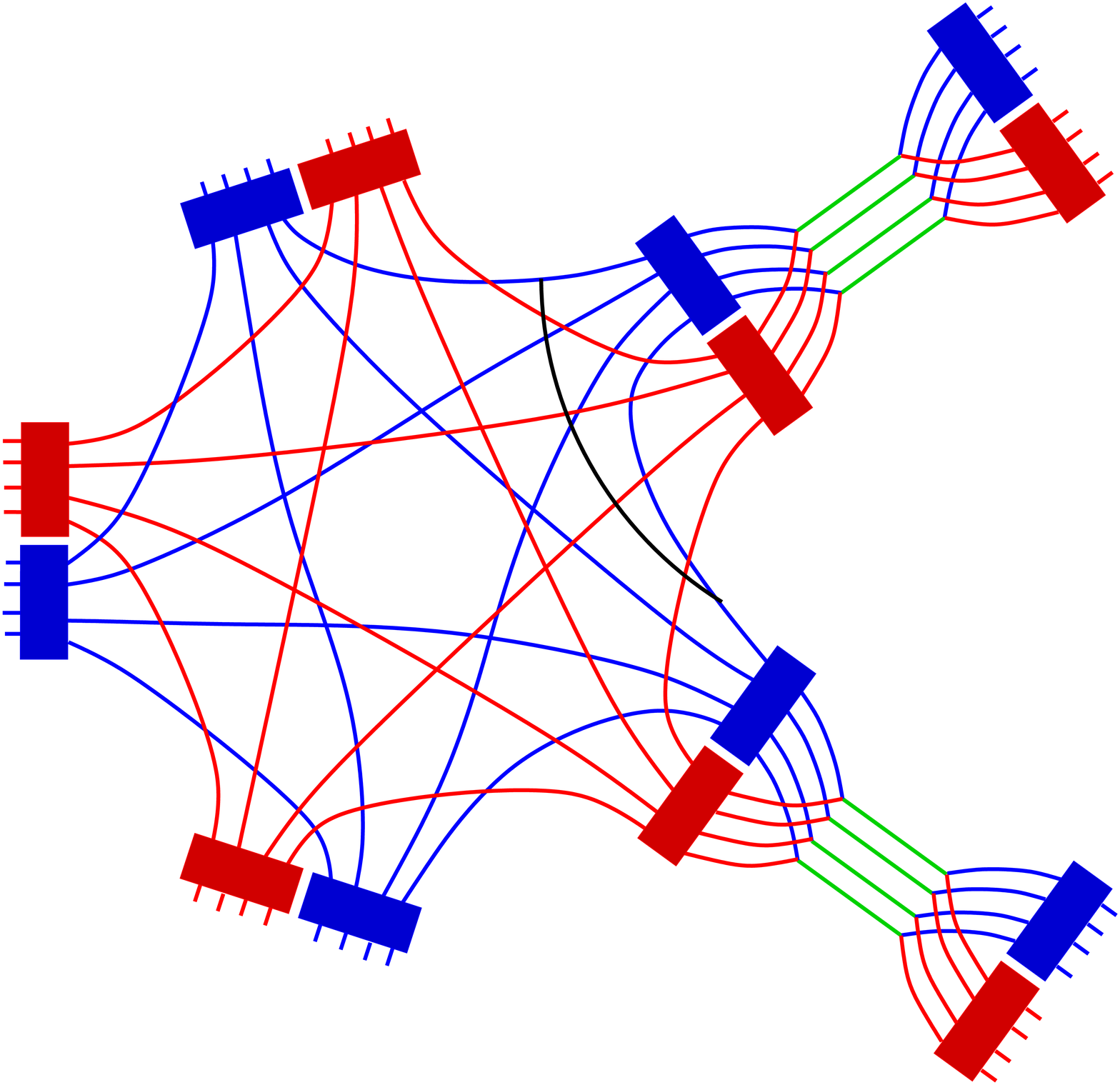}
\end{array}\right]+\n \\
&& \!\!\!\!\!\!\!\!\!\!\!\! +
(1-\gamma)^2\left[ \begin{array}{c}
\includegraphics[width=4cm]{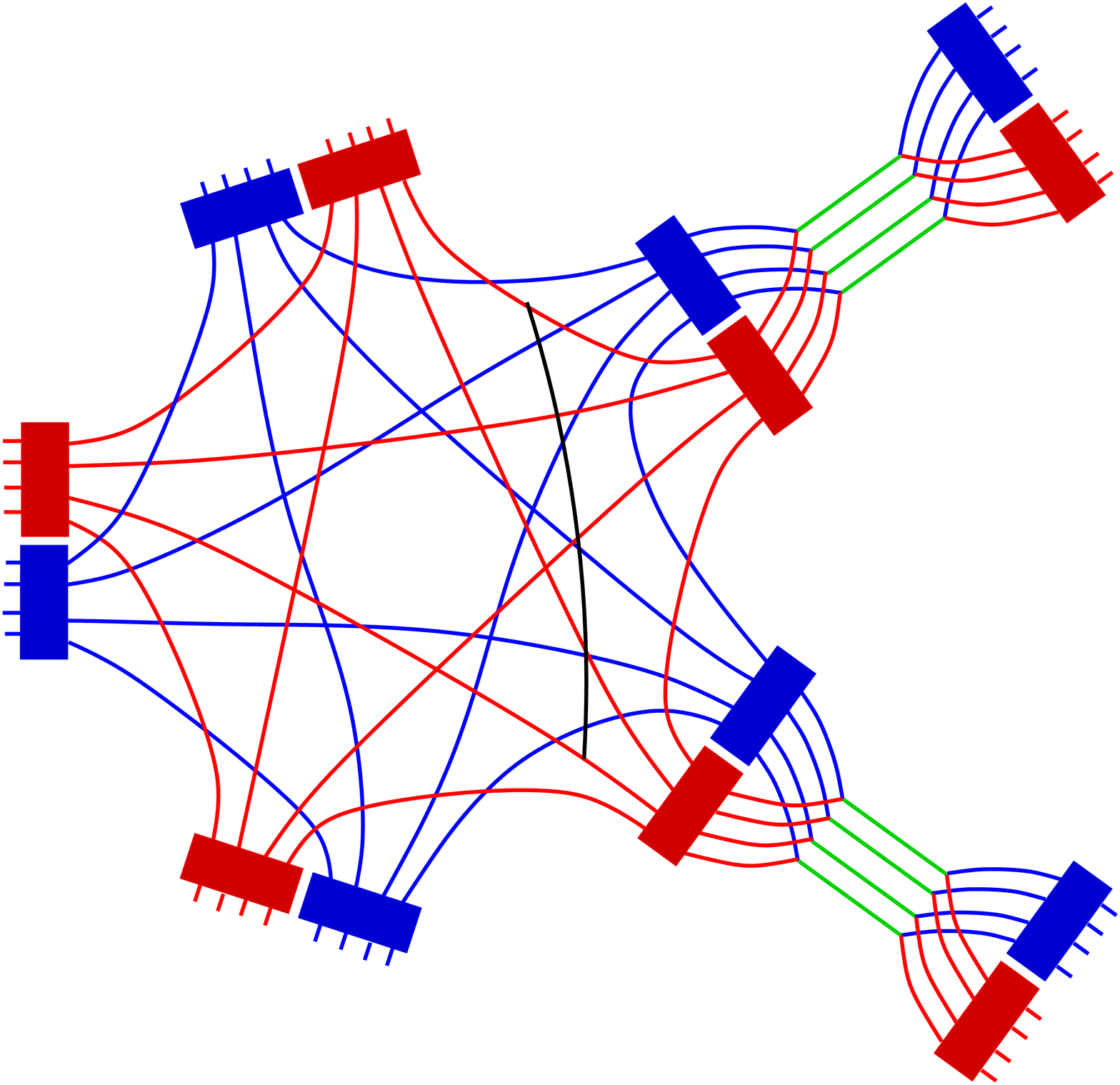}
\end{array}+\begin{array}{c}
\includegraphics[width=4cm]{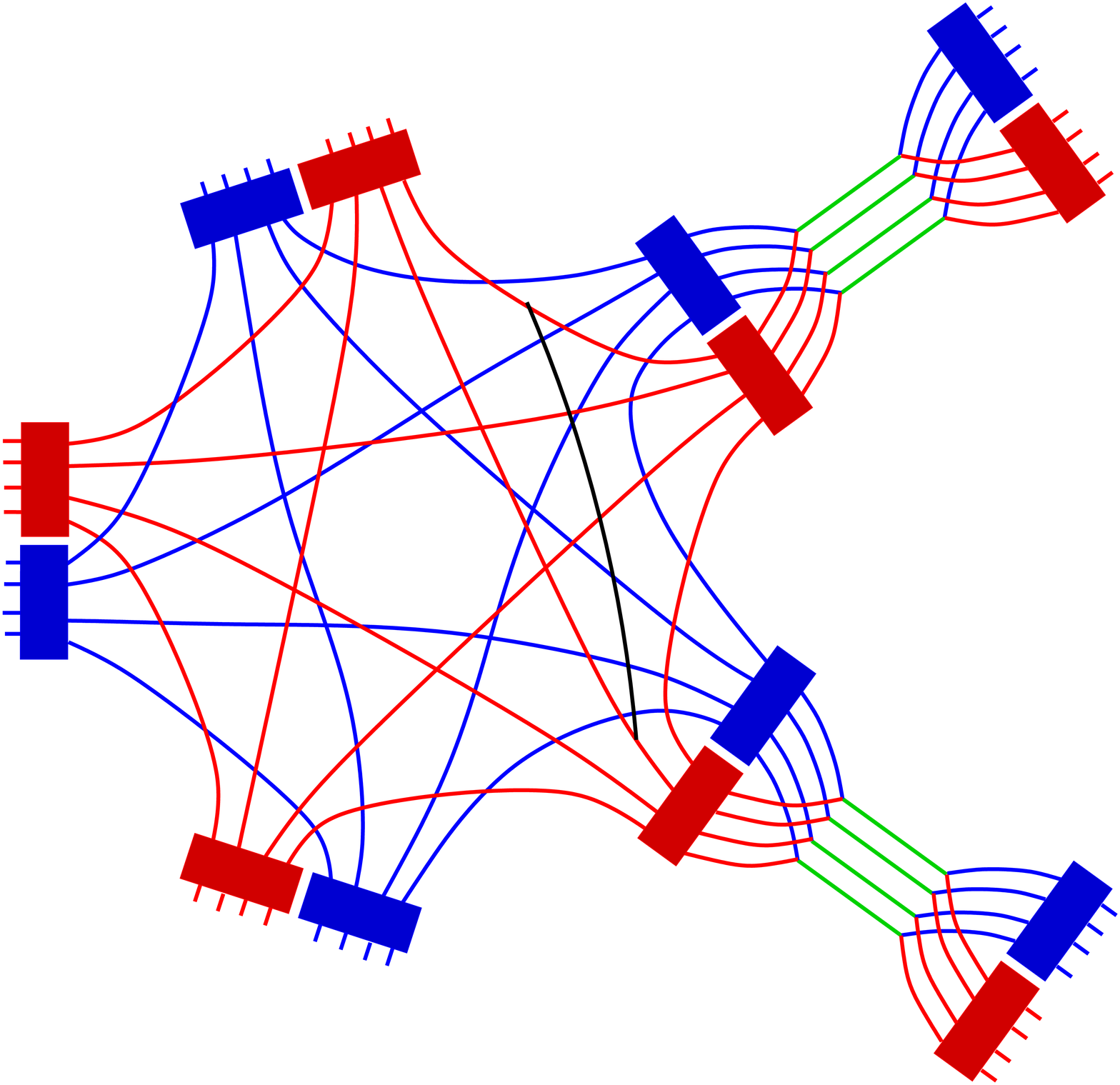}
\end{array}+\begin{array}{c}
\includegraphics[width=4cm]{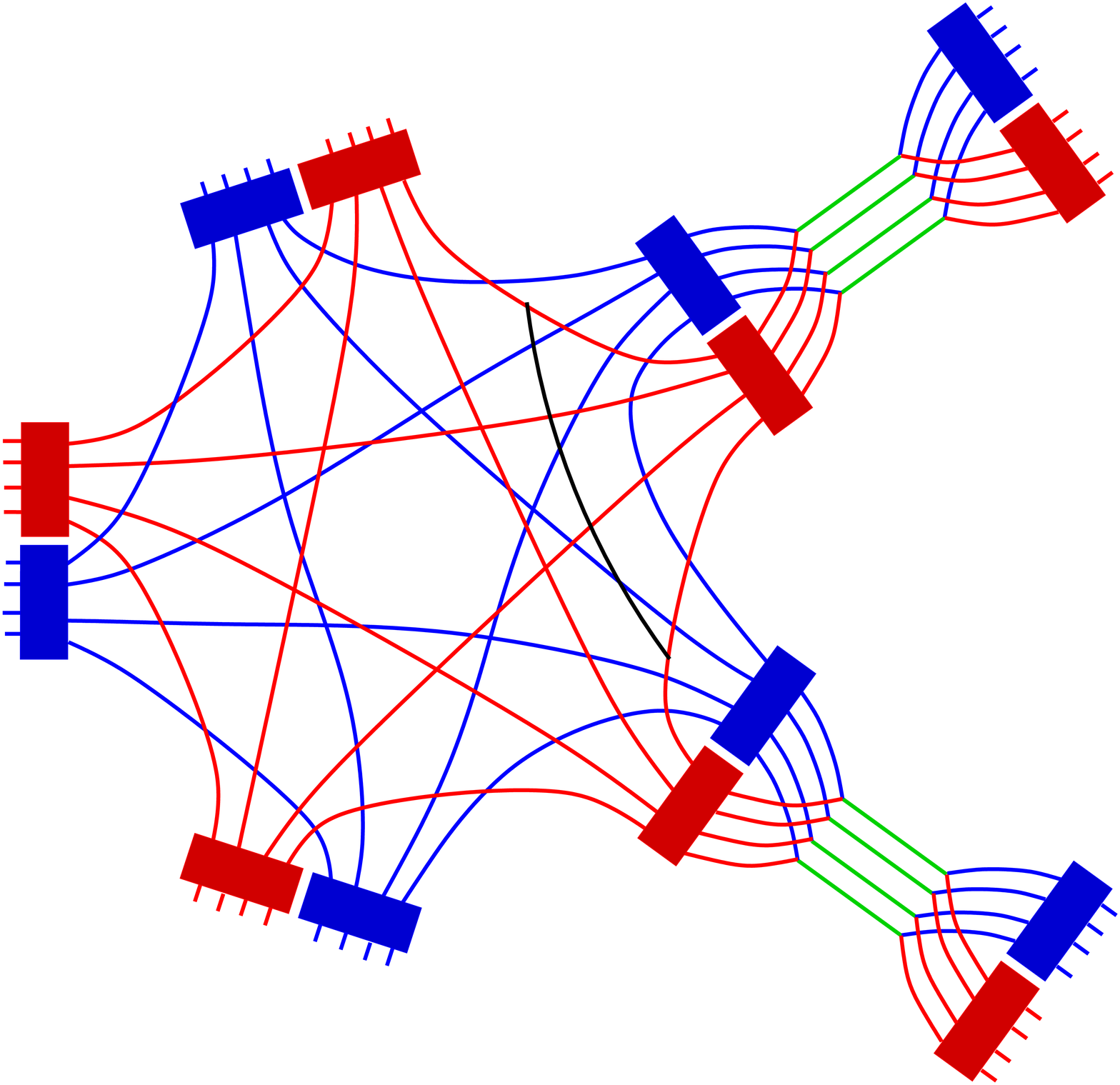}
\end{array}\right]=\n \\ 
&& \!\!\!\!\!\!\!\!\!\!\!\!
-(1+\gamma)^2 \begin{array}{c}
\includegraphics[width=4cm]{5-1.eps}
\end{array}
+(1-\gamma)^2
\begin{array}{c}
\includegraphics[width=4cm]{6-1.eps}
\end{array} +\sO_{sc},
\ea
where going from the first line to the second and third lines we have simply used (\ref{eprl-gauss}) on the bottom graspings on the right and left wires.
The last line results from the validity of (\ref{3s}): notice that the second terms in the second and third lines add up to $\sO_{sc}$ as well as the third terms in the second and third line. There is an overall minus sign which amounts for an orientation factor. It should be clear that we can apply the same procedure to arrive at any admissible pair. 


\subsubsection{$P_{eprl}$ is not a projector}

Let us study in a bit more detail the object $P_{eprl}^e(j_1,\cdots, j_4)$. We see that it is made of two ingredients. The first one is the 
projection to the maximum  weight subspace $\sH_{j}$ for $\gamma>1$  in the decomposition of $\sH_{j^+,j^-}$  for $j^{\pm}=(1\pm\gamma)j/2$ ($j^{\pm}=(\gamma\pm 1)j/2$ for $\gamma>1$) 
in terms of irreducible representations of an arbitrarily chosen $SU(2)$ subgroup of $Spin(4)$. The second ingredient is to eliminate the dependence on the choice of subgroup by 
group averaging with respect to the full gauge group $Spin(4)$. This is diagramaticaly represented in (\ref{eprl-projection}). However $P_{eprl}^e(j_1,\cdots, j_4)$ is not a projector, namely
\be
P_{eprl}^e(j_1,\cdots, j_4)^2\not=P_{eprl}^e(j_1,\cdots, j_4).
\ee
Technically this follows from (\ref{regalo}) and the fact that  
 \be [P^e_{inv} (\rho_{1}\cdots \rho_4),( \sY_{j_1}\otimes\cdots\otimes\sY_{j_4} )]\not=0\ee 
i.e., the projection imposing the linear constraints (defined on the frame of a tetrahedrom or edge) and the $Spin(4)$ (or Lorentz) group 
averaging---rendering the result gauge invariant---do not commute. The fact the  $P_{eprl}^e(j_1,\cdots, j_4)$  is not a projection operator has important 
consequences in the mathematical structure of the model:
\begin{enumerate}
\item 
From (\ref{eprl-so4}) one can  immediately obtain the following expression for the EPRL amplitude
\be\label{eprl-p}
Z_{eprl}(\Delta)=\sum \limits_{ \rho_f \in \sK}  \ \prod\limits_{f \in \Delta^{\star}} {\rm d}_{|1-\gamma|\frac{j}{2}}{\rm d}_{(1+\gamma)\frac{j}{2}}
\prod \limits_{e} P_{eprl}^e(j_1,\cdots, j_4).
\ee 
This expression has the formal structure of expression (\ref{bf4}) for BF theory. The formal similarity however is broken by the fact that  $P_{eprl}^e(j_1,\cdots, j_4)$ is not a projection operator.
From the formal perspective is the possibility that the amplitudes  be defined in term of a network of projectors (as in BF theory) might provide an interesting structure that might be of relevance in the definition of a discretization independent model. On the contrary, the failure of $P_{eprl}^e(j_1,\cdots, j_4)$ to be a projector may lead, in my opinion, to difficulties in the limit where the
complex $\Delta$ is refined: the increasing of the number of edges might produce either trivial or divergent amplitudes \footnote{This is obviously not clear from the form of
(\ref{eprl-p}). We are extrapolating the properties of $(P_{eprl}^e)^{N}$ for large $N$ to those of the amplitude (\ref{eprl-p}) in the large number of edges limit implied by the continuum limit.}. 

\item Another difficulty associated with $P_{eprl}^e(j_1,\cdots, j_4)^2\not=P_{eprl}^e(j_1,\cdots, j_4)$ is the failure of the amplitudes of the EPRL model, as defined here, to be consistent with
the abstract notion of spin foams as defined in \cite{baez7}. This is a point of crucial importance under current discussion in the community. The point is that the cellular decomposition $\Delta$
has no physical meaning and is to be interpreted as a subsidiary regulating structure to be removed when computing physical quantities. Spin foams configurationa can fit in different ways on a given 
$\Delta$, yet any of these different embeddings represent the same physical process (like the same gravitational field in different coordinates). Consistency requires the spin foam amplitudes to be independent of the embedding, i.e., well defined on the equivalence classes of spin foams as defined by Baez in \cite{baez7} (the importance of these consistency requirements was emphasized in \cite{myo}). The amplitude (\ref{eprl-p}) fails this requirement due to  $P_{eprl}^e(j_1,\cdots, j_4)^2\not=P_{eprl}^e(j_1,\cdots, j_4)$. 

\end{enumerate}
\subsubsection{The Warsaw proposal}

If one sees the above  as difficulties then there is  a simple solution, at least in the Riemannian case. As proposed in \cite{Bahr:2010bs, Kaminski:2009cc} one can obtain a consistent modification of the EPRL model 
by replacing $P^e_{eprl}$ in (\ref{eprl-p}) by a genuine projector $P^{e}_{w}$, graphically
\ba 
 P_{w}^{e}(j_1\cdots j_4)=\sum_{\alpha\beta}  {\rm Inv} \left( \begin{array}{c}\psfrag{a}{$\!\!\!\! \van \alpha$}\psfrag{b}{$\van \beta$}
\includegraphics[height=1.5cm]{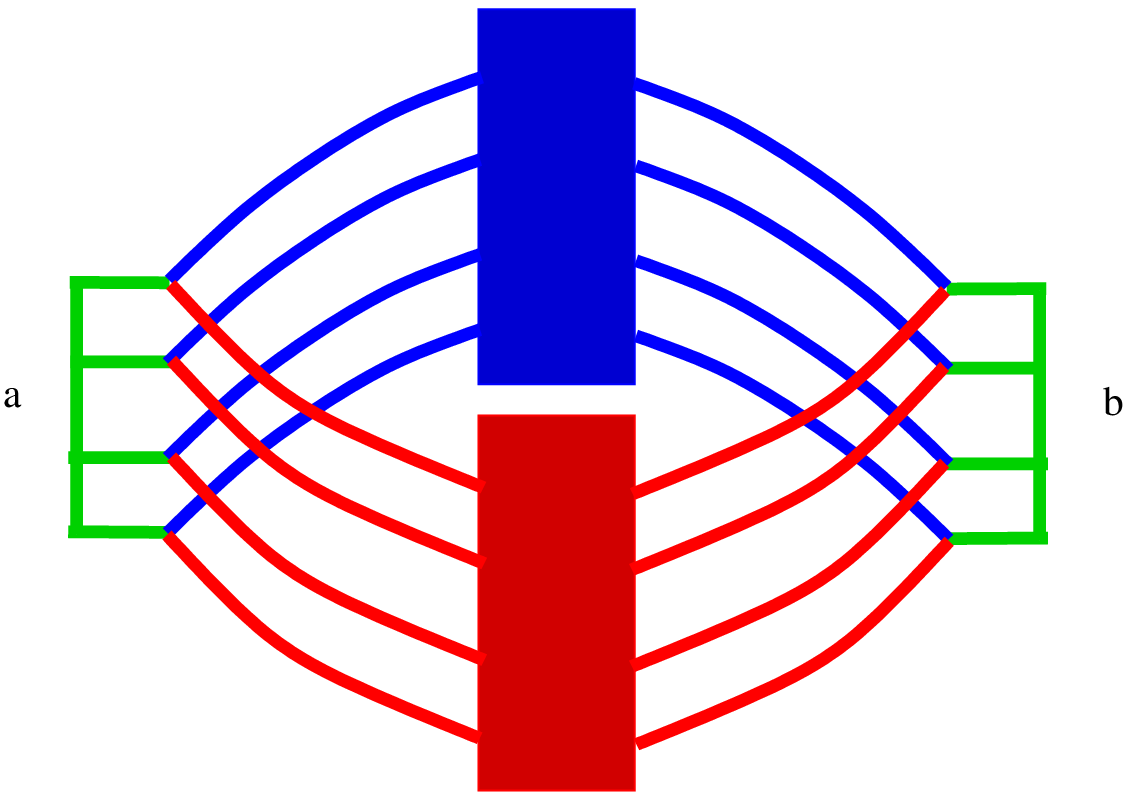}
\end{array}\right)
\!\!\!\!\! \begin{array}{c}\psfrag{a}{$\! \van \alpha$}\psfrag{b}{$\! \van \beta$}
\includegraphics[height=3cm]{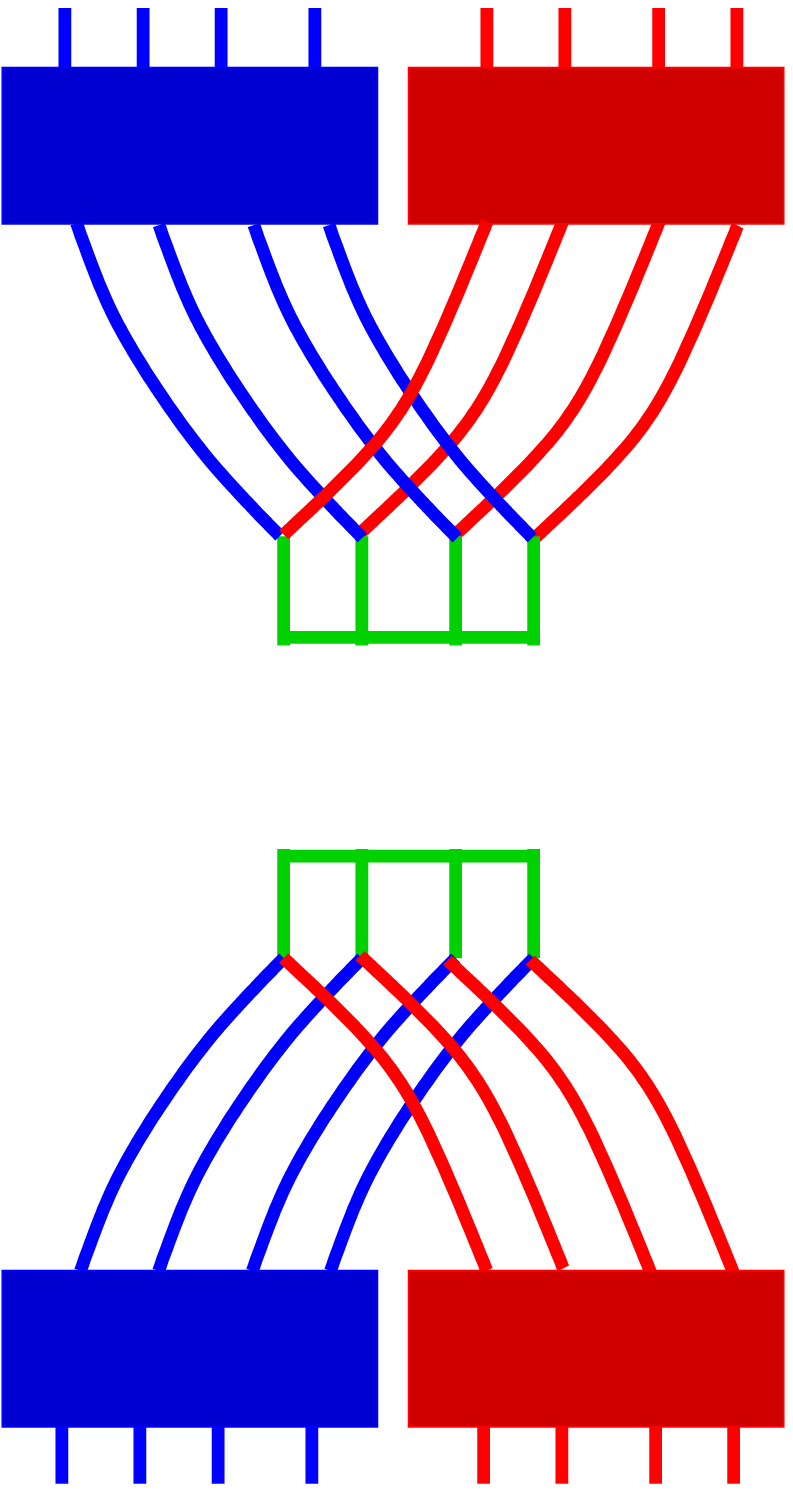}
\end{array},
\ea
It is easy to check that by construction \be
(P_{w}^{e}(j_1\cdots j_4))^2=P_{w}^{e}(j_1\cdots j_4).
\ee
The variant of the EPRL model proposed in \cite{Bahr:2010bs, Kaminski:2009cc} takes then the form
\ba
\label{eprl-warsaw}  Z_{eprl}(\Delta)&=&\sum \limits_{ j_f}  \ \prod\limits_{f \in \Delta^{\star}} {\rm d}_{|1-\gamma|\frac{j}{2}}{\rm d}_{(1+\gamma)\frac{j}{2}}
\prod \limits_{e} P_{w}^e(j_1,\cdots, j_4)\\
&=&\sum \limits_{ j_f }\sum \limits_{ \iota_{ev} }  \ \prod\limits_{f \in \Delta^{\star}} {\rm d}_{|1-\gamma|\frac{j}{2}}{\rm d}_{(1+\gamma)\frac{j}{2}} \prod_{e\in \Delta^{\star}} g^e_{\iota^e_{v_s}\iota^e_{v_t}} \prod_{v\in \Delta^{\star}}\begin{array}{c}\psfrag{a}{$\van \iota^1_{v}$}
\psfrag{b}{$\van  \iota^2_{v}$}
\psfrag{c}{$\van  \iota^3_{v}$}
\psfrag{d}{$\van  \iota^4_{v}$}
\psfrag{e}{$\van  \iota^5_{v}$}
\includegraphics[width=5cm]{eprl-vertex.eps}
\end{array}. \n \ea
Thus in the modified EPRL model edges $e\in \Delta^{\star}$ are assigned pairs of intertwiner quantum numbers $\iota^e_{v_s}$ and $\iota^e_{v_t}$  
and an edge amplitude given by the matrix elements $g^e_{\iota^e_{v_s},\iota^{e}_{v_t}}$  (where $v_s$ and $v_t$ stand for the source and target vertices of the given oriented edge).  
The fact that edges are not assigned a single quantum number is not really significative; one could go to a basis of normalized eigenstates of $P^e_{w}$ and rewrite
the modified model above as a spin foam model where edges are assigned a single (basis element) quantum number.  As the nature of such basis and the quantum 
geometric interpretation of its elements is not clear at this stage, it seems simpler to represent the amplitudes of the  modified model in the above form.

The advantages of the modified model are important,; however, a generalization of the above modification of the EPRL model in the Lorentzian case is still lacking.
Notice that this modification does not interfere with the results on the semiclassical limit (to leading order) as reviewed in Section \ref{semiclas}. The reason is that the 
matrix elements $g^e_{\alpha\beta}\to \delta_{\alpha\beta}$ in that limit \cite{Alesci:2008un}.

 \subsection{The coherent states representation}\label{7-6}

We have written the amplitude defining the EPRL model by constraining the 
state sum of BF theory.  For semiclassical studies that we will review in 
Section \ref{semiclas} it is convenient to express the EPRL amplitude in terms of the 
coherent states basis. The importance of coherent states in spin foam models
was put forward in \cite{Livine:2007vk} and explicitly used to re-derive the EPRL model in \cite{Livine:2007ya}. The coherent state technology was used by Freidel and Krasnov
in \cite{Freidel:2007py} to introduce a new kind of spin foam models for gravity: the FK models.
In some cases the FK model is equivalent to the EPRL model; we will review this in detail in 
Section \ref{fk}. 

The coherent state representation of the EPRL model is obtained by replacing (\ref{patacul}) in each of the intermediate 
$SU(2)$ (green) wires in the expression (\ref{eprl-so4}) of the EPRL amplitudes, namely 
\ba && \begin{array}{c}
\psfrag{w}{$$}
\psfrag{a}{$n_{\va 1}$}
\psfrag{A}{$\va n_1$}
\psfrag{b}{$\va n_2$}
\psfrag{B}{$\va n_2$}
\psfrag{c}{$\va n_3$}
\psfrag{C}{$\va n_3$}
\psfrag{d}{$\va n_4$}
\psfrag{D}{$\va n_4$}
\includegraphics[width=6cm]{eprl3.eps}\end{array}=\n \\ && \ \ \ \ \ \ \ \ \ \ \ \ \ \ \ \ \ \ \ \ \ \ \ \ \ \ \ =\int\limits_{[S^2]^4} \prod\limits_{\va I=1}^{\va 4}{\rm d}_{j_I}  dn_{\va I}  \begin{array}{c}\psfrag{w}{$$}
\psfrag{a}{$\va n_1$}
\psfrag{A}{$\va n_1$}
\psfrag{b}{$\va n_2$}
\psfrag{B}{$\va n_2$}
\psfrag{c}{$\va n_3$}
\psfrag{C}{$\va n_3$}
\psfrag{d}{$\va n_4$}
\psfrag{D}{$\va n_4$}
\includegraphics[width=6cm]{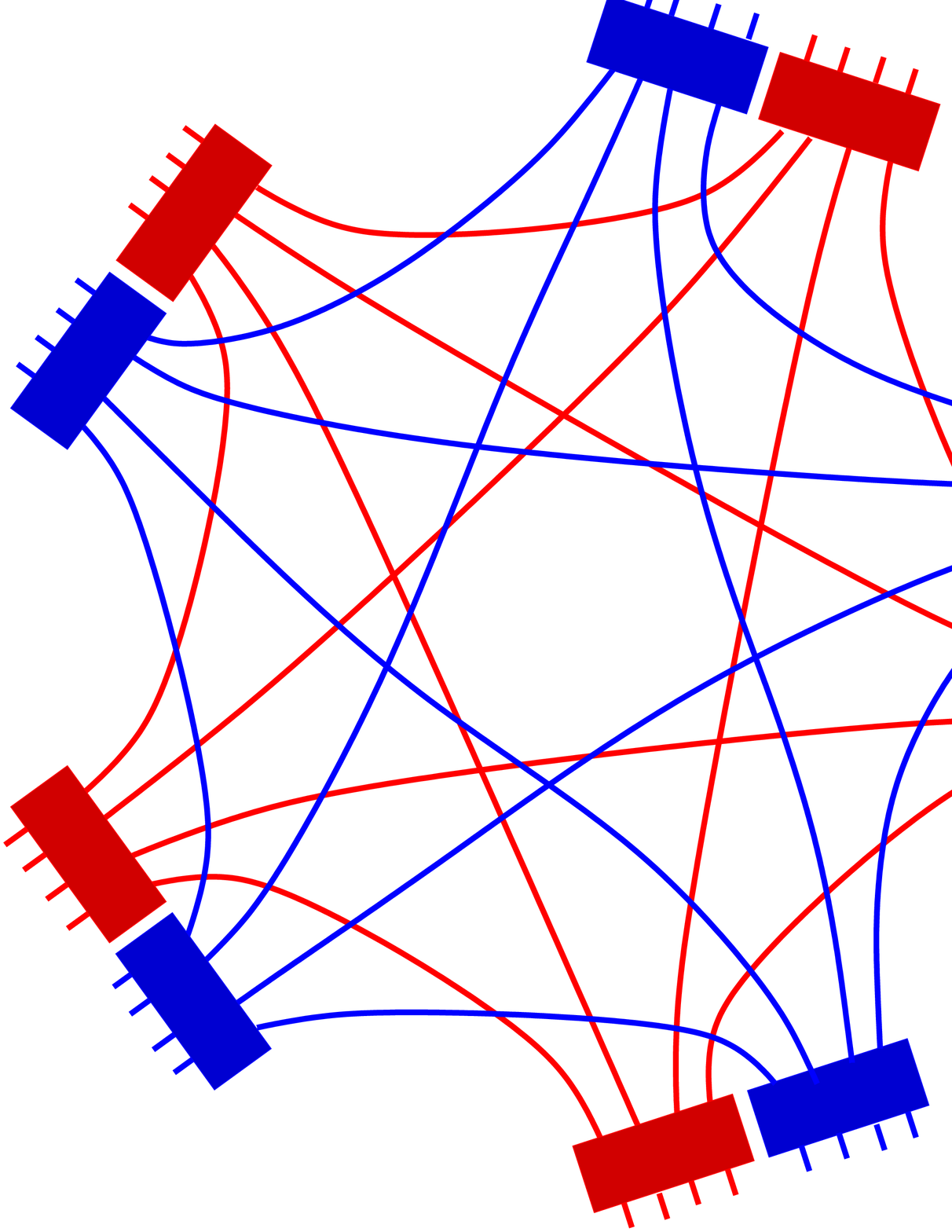}\end{array}\label{tresf}
\ea
\subsubsection*{The case $\gamma<1$}
In this case the coherent state 
property (\ref{exp}) implies 
\be\begin{array}{c}
\psfrag{w}{$=\int\limits_{[S^3]^4} \prod\limits_{\va I=1}^{\va 4} dn_{\va I}$}
\psfrag{a}{$\va n_1$}
\psfrag{A}{$\va n_1$}
\psfrag{b}{$\va n_2$}
\psfrag{B}{$\va n_2$}
\psfrag{c}{$\va n_3$}
\psfrag{C}{$\va n_3$}
\psfrag{d}{$\va n_4$}
\psfrag{D}{$\va n_4$}
\psfrag{x}{$$}
\includegraphics[width=6.5cm]{eprl4.eps}\end{array}= \begin{array}{c}\psfrag{w}{$=\int\limits_{[S^3]^4} \prod\limits_{\va I=1}^{\va 4} dn_{\va I}$}
\psfrag{a}{$\va n_1$}
\psfrag{A}{$\va n_1$}
\psfrag{b}{$\va n_2$}
\psfrag{B}{$\va n_2$}
\psfrag{c}{$\va n_3$}
\psfrag{C}{$\va n_3$}
\psfrag{d}{$\va n_4$}
\psfrag{D}{$\va n_4$}
\psfrag{x}{$$}
\includegraphics[width=6.5cm]{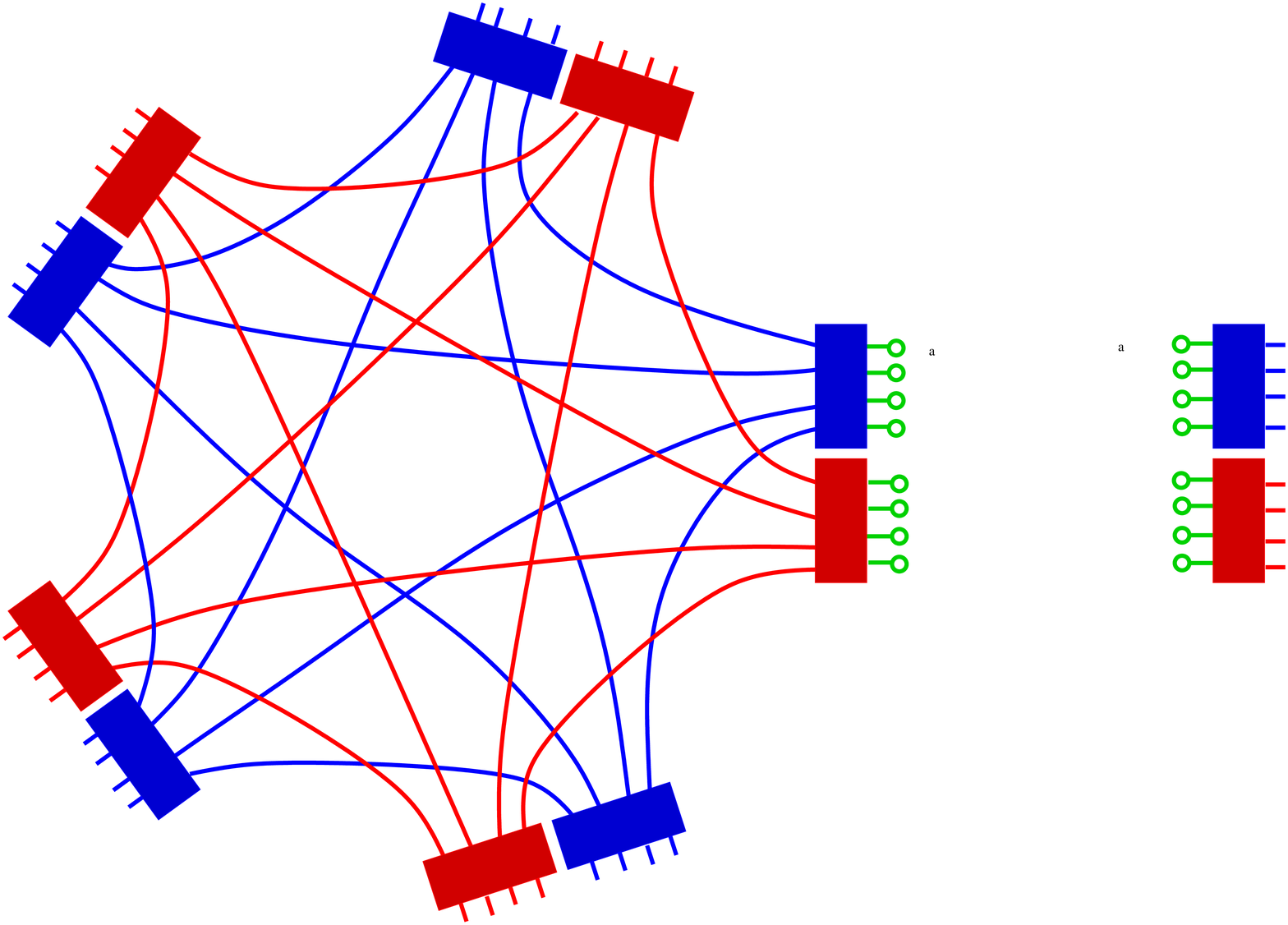}\end{array}
\label{cuatrofy},\ee
where we used in the last line the fact that for $\gamma<1$ the representations $j$ of the subgroup $SU(2)\in Spin(4)$  are maximum weight, i.e., $j=j^++j^-$.
Doing this at each edge we get
\ba\label{eprl-cohe}
&& Z^{E}_{eprl}(\Delta)=\sum \limits_{j_f}  \ \prod\limits_{f \in \Delta^{\star}} {\rm d}_{j_f^{-}}{\rm d}_{j_f^{+}} \n \\ && \int \prod_{e\in  \in \Delta^{\star}} {\rm d}_{j_{ef}} dn_{ef}
\begin{array}{c}\psfrag{w}{$$}
\psfrag{a}{$\va n_{ 1}$}
\psfrag{A}{$\va n_1$}
\psfrag{b}{$\va n_2$}
\psfrag{B}{$\va n_2$}
\psfrag{c}{$\va n_3$}
\psfrag{C}{$\va n_3$}
\psfrag{d}{$\va n_4$}
\psfrag{D}{$\va n_4$}
\includegraphics[width=7cm]{coherent-v.eps}
\end{array},
\ea
where we have explicitly written the $n\in S^2$ integration variables on a single cable. The expression above is very similar to the coherent states representation of $Spin(4)$ BF theory given in Equation (\ref{bf-cohe}). In fact one would get the above
expression if one would start form the expression (\ref{bf-cohe}) and would set $n^{+}_{ef}=n^{-}_{ef}=n_{ef}$ while dropping for example all the sphere integrations corresponding to the $n^{+}_{ef}$ (or equivalently $n^{-}_{ef}$).  
Moreover, by construction the coherent states participating in the previous amplitude satisfy the linear constraints (\ref{constrainty-e}) in expectation values, namely
\ba\n 
\langle j, n_{ef} | D^i_f| j, n_{ef}\rangle &=&\\ &=& 
\langle j, n_{ef} | (1-\gamma) J^{+i}_ f+ (1+\gamma) J^{-i}_f  |j, n_{ef}\rangle=0.
\ea
Thus the coherent states participating in the above representation of the EPRL amplitudes solve the linear simplicity constraints in the usual semiclassical sense.
The same manipulations leading to (\ref{discrete-action}) in Section \ref{BF} lead to a discrete effective action for the EPRL model, namely
\ba
\label{eprl-cohe}Z^{\va \gamma<1}_{eprl}=\sum \limits_{ j_f }  \ \prod\limits_{f \in \Delta^{\star}} {\rm d}_{(1-\gamma)\frac{j_f}{2}}{\rm d}_{(1+\gamma)\frac{j_f}{2}} \int \prod_{e\in  \Delta^{\star}} {\rm d}_{j_{ef}} dn_{ef} dg^{-}_{ef}dg^{+}_{ef}
\ \exp{(S^{\va \gamma<1}_{j^{\pm},\bn}[g^{\pm}])}, \ea
where the discrete action \be\label{discrete-action}
S^{\va \gamma<1}_{j^{\pm},\bn}[g^{\pm}]=\sum_{v\in\Delta^{\star}} (S^v_{(1-\gamma)\frac{j_f}{2},\bn}[g^{-}]+S^v_{(1+\gamma)\frac{j_f}{2},\bn}[g^{+}])\ee with 
\be
\label{v-action}
S^v_{j,\bn}[g] = \sum\limits^{5}_{a < b=1} 2j_{ab} \ln \, \la n_{ab}| g^{-1}_a g_b| \, n_{ba} \ra,
\ee
and the indices $a,b$ label the five edges of a given vertex. The previous expression is exactly equal to the form (\ref{coloring4}) of the BF amplitude. In the case of the gravity models presented here,
the coherent state path integral representation (analogous to (\ref{cspig})) will be the basic tool for the study of the semiclassical limit of the models and the relationship with Regge discrete formulation of general relativity. 

\subsubsection*{The case $\gamma>1$}

The case $\gamma>1$ is more complicated \cite{Barrett:2009gg}. The reason is that the step (\ref{cuatrofy}) directly leading  to the discrete action in the previous case is no longer valid as the
representations of the subgroup $SU(2)\in Spin(4)$ are now minimum instead of maximum weight. However, the representations $j^{+}=j^{-}+j$ are maximum weight. We can therefore insert 
coherent states resolution of the identity on the right representations and get:
\ba && \n \begin{array}{c}
\psfrag{w}{$=\int\limits_{[S^3]^4} \prod\limits_{\va I=1}^{\va 4} dn_{\va I}$}
\psfrag{a}{$\va n_1$}
\psfrag{A}{$\va n_1$}
\psfrag{b}{$\va n_2$}
\psfrag{B}{$\va n_2$}
\psfrag{c}{$\va n_3$}
\psfrag{C}{$\va n_3$}
\psfrag{d}{$\va n_4$}
\psfrag{D}{$\va n_4$}
\psfrag{x}{$$}
\includegraphics[width=2.7cm]{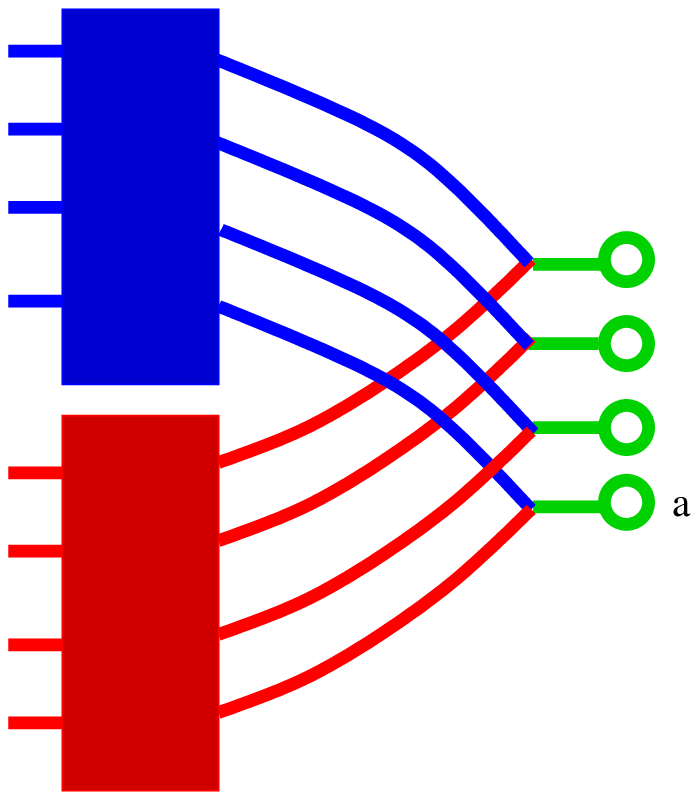}\end{array}= \int\limits_{[S^3]^4} \prod \limits_{\va I=1}^{\va 4} {\rm d}_{(1+\gamma)\frac{j_{I}}{2}}\ dm_{\va I} \begin{array}{c}
\psfrag{w}{$=\int\limits_{[S^3]^4} \prod\limits_{\va I=1}^{\va 4} dn_{\va I}$}
\psfrag{m1}{$\va m_1$}
\psfrag{m2}{$\va m_2$}
\psfrag{m3}{$\va m_3$}
\psfrag{m4}{$\va m_4$}
\psfrag{a}{$\va n_1$}
\psfrag{A}{$\va n_1$}
\psfrag{b}{$\va n_2$}
\psfrag{B}{$\va n_2$}
\psfrag{c}{$\va n_3$}
\psfrag{C}{$\va n_3$}
\psfrag{d}{$\va n_4$}
\psfrag{D}{$\va n_4$}
\psfrag{x}{$$}
\includegraphics[width=2.7cm]{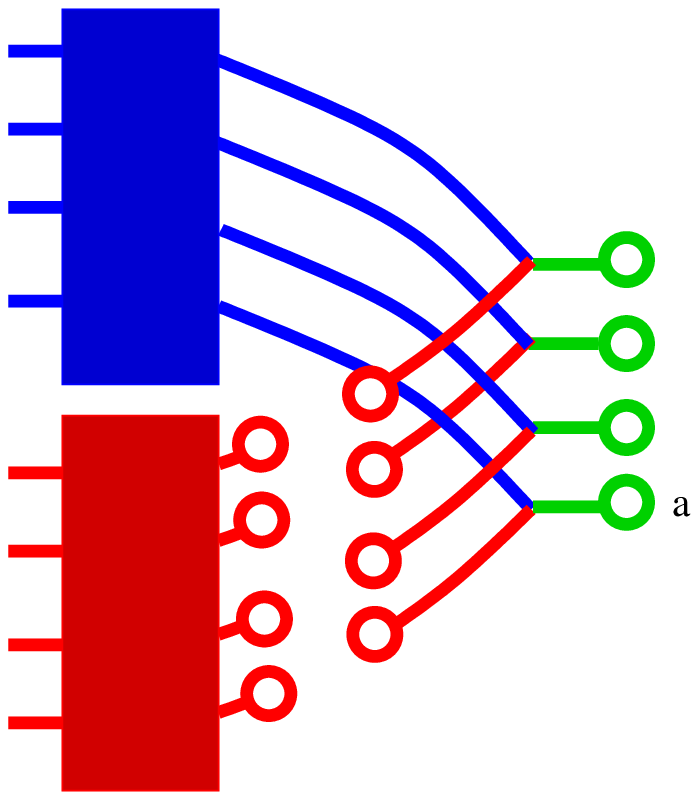}\end{array}=\\ && =\int\limits_{[S^3]^4} \prod \limits_{\va I=1}^{\va 4} {\rm d}_{(1+\gamma)\frac{j_{I}}{2}}\ dm_{\va I}  \begin{array}{c}
\psfrag{w}{$=\int\limits_{[S^3]^4} \prod\limits_{\va I=1}^{\va 4} dn_{\va I}$}
\psfrag{m1}{$\va m_1$}
\psfrag{m2}{$\va m_2$}
\psfrag{m3}{$\va m_3$}
\psfrag{m4}{$\va m_4$}
\psfrag{a}{$\va n_1$}
\psfrag{A}{$\va n_1$}
\psfrag{b}{$\va n_2$}
\psfrag{B}{$\va n_2$}
\psfrag{c}{$\va n_3$}
\psfrag{C}{$\va n_3$}
\psfrag{d}{$\va n_4$}
\psfrag{D}{$\va n_4$}
\psfrag{x}{$$}
\includegraphics[width=2.7cm]{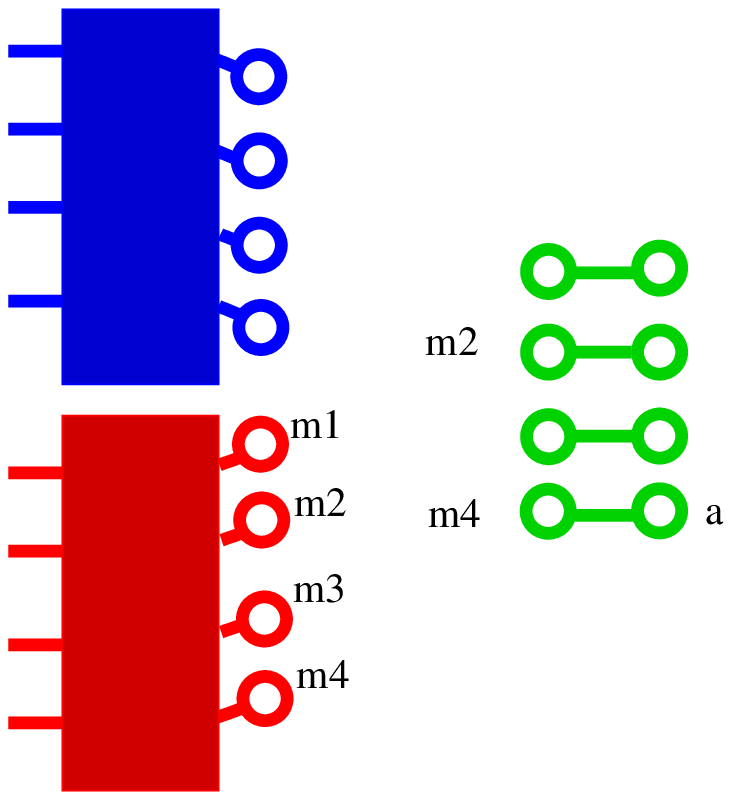}\end{array}
\label{cuat},\ea
where we are representing the relevant part of the diagram appearing in equation (\ref{tresf}). In the last line we have used that $j^+=j+j^-$ (i.e. maximum weight), and the graphical notation
$\begin{array}{c}\psfrag{a}{$\! \!\! \, m$}
\psfrag{b}{$n$}
\includegraphics[width=1cm]{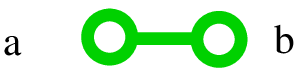}
\end{array}\equiv \langle m |n\rangle$ as it follows from our previous conventions.
With all this, one gets
\ba
\label{eprl-cohe-g}&& Z^{\va \gamma> 1}_{eprl}= \\ \n &&\sum \limits_{ j_f }  \ \prod\limits_{f \in \Delta^{\star}} {\rm d}_{(1-\gamma)\frac{j_f}{2}}{\rm d}_{(1+\gamma)\frac{j_f}{2}} \int \prod_{e\in  \Delta^{\star}} {\rm d}_{j_{ef}} {\rm d}_{(1+\gamma)\frac{j_{ef}}{2}} dn_{ef} dm_{ef} dg^{-}_{ef}dg^{+}_{ef}
\ \exp{(S^{\va \gamma>1}_{j^{\pm},\bn,\bm} [g^{\pm}])}, \ea
where the discrete action \be\label{discrete-action-g<}
S^{\va \gamma>1}_{j^{\pm},\bn,\bm}[g^{\pm}]=\sum_{v\in\Delta^{\star}} S^v_{j^{\pm},\bn,\bm}[g^{\pm}]\ee with
\ba\label{gamma-g}
S^v_{j^{\pm},\bn,\bm}[g^{\pm}]=\sum_{1\le a<b\le 5} j_{ab} (1+\gamma)\log( \langle m_{ab} |g^{+}_{ab}|m_{ba} \rangle)+j_{ab} (\gamma-1) \log(\langle m_{ab} |g^{-}_{ab}|m_{ba}\rangle)+\n \\ 
\ \ \ \ \ \ \ \ \ \ \ \ \ \ \ \ \ \ \ \ \ +2j_{ab} \left(\log(\langle n_{ab} |m_{ab} \rangle)+\log(\langle m_{ba} |n_{ba} \rangle)\right).
\ea

\subsubsection{Some additional remarks}

It is important to point out that the commutation relations of basic fields---reflecting the simple algebraic structure of $spin(4)$---used here is the one induced by the canonical analysis 
of BF theory presented previously. The presence of constraints generally modifies canonical commutation relations in particular in the presence of second class constraints. 
For some investigation of the issue in the context of the EPRL and FK  models see \cite{Alexandrov:2010pg}. 
In \cite{Alexandrov:2008da} it is pointed out that the presence of secondary constraints in the canonical analysis of Plebanski action should translated in additional constraints
in the holonomies of the spin foam models here considered (see also \cite{Alexandrov:2007pq}). A possible view is  that the simplicity constraints are here imposed  {\em for all times} and thus
secondary constraints should be imposed automatically. 

There are alternative derivations of the models presented in the previous sections. In particular 
one can derive them from a strict Lagrangean approach of Plebanski's action. Such viewpoint is taken in 
\cite{Bonzom:2009hw, Bonzom:2009wm, Bonzom:2008ru}.  
The path integral formulation of Plebansky theory using commuting $B $-fields was studied in \cite{Han:2010rb}, where it is shown that only in the appropriate semiclassical limit
the amplitudes coincide with the ones presented in the previous sections (this is just another indication that the construction of the models have a certain semiclassical input; see below). 
The spin foam quantization of the  Holst formulation of gravity via cubulations was investigated in 
\cite{Baratin:2008du}.
The simplicity constraints can also be studied from the perspective of the $U(N)$ formulation of quantum geometry 
 \cite{Dupuis:2010iq}. Such $U(N)$ treatment is related to previous work \cite{Freidel:2010tt, Freidel:2009ck} which has been extended to a completely new perspective on quantum geometry 
 with possible advantageous features \cite{Borja:2010rc, Livine:2011gp}. For additional discussion on the simplicity constraints see \cite{Dittrich:2010ey}.


\subsection{Presentation of the EPRL Lorentzian model}\label{alala}

As briefly discussed in Section \ref{eprl-r}, unitary irreducible representations of $SL(2,\C)$ are infinite dimensional and labelled by a  positive real number $p\in \R^{+}$ and a half-integer $k\in \N/2$. These representation are the ones that intervene in the harmonic analysis of square integrable functions of $SL(2,\C)$ \cite{gelfand}. Consequently, 
one has an explicit expression of the delta function distribution (defined on such test function), namely
\be
\delta(g)=\sum_{k}\int_{\R^+} dp\   (p^2+k^2) \ \sum_{j,m} D^{p,k}_{jmjm}(g)
\ee
 where $D^{p,k}_{jmj'm'}(g)$  with $j\ge k$ and $j\ge m \ge-j$ (similarly for the primed indices) are the matrix elements of the unitary representations $p-k$ in the so-called canonical basis \cite{ruhl}. One can use the previous expression the Lorentzian version of Equation (\ref{coloring4}) in order to introduce a formal definition of the BF amplitudes, which now would involve integration of the continuous labels $p_f$ in addition of sums over discrete quantum numbers such as $k$, $j$ and $m$.
 The Lorentzian version of the EPRL model can be obtained from the imposition of the linear simplicity constraints to this formal expression.
As the continuum labels $p_f$ are restricted to $p_f=\gamma j_f$ the Lorentzian EPRL model becomes a state-sum model as its Riemannian relative.
Using the following graphical notation 
\be
D^{p,k}_{jmj'm'}(g)=\ \ \ \begin{array}{c}\psfrag{a}{$\van p$}\psfrag{b}{$\van k$}\psfrag{c}{$\van \!\!\! j',m'$}\psfrag{d}{$\van \!\!\!\!\!\!\!j,m$}
\includegraphics[width=2.5cm]{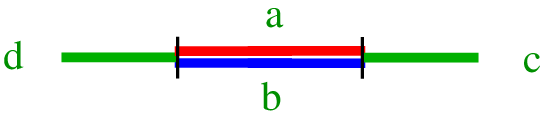}
\end{array}
\ee
the amplitude is
\ba\label{eprl-sl2c}\n
Z^{L}_{eprl}(\Delta)= \sum \limits_{ j_f}  \ \prod\limits_{f \in \Delta^{\star}} (1+\gamma^2) j_f^2 \begin{array}{c}
\includegraphics[width=5cm]{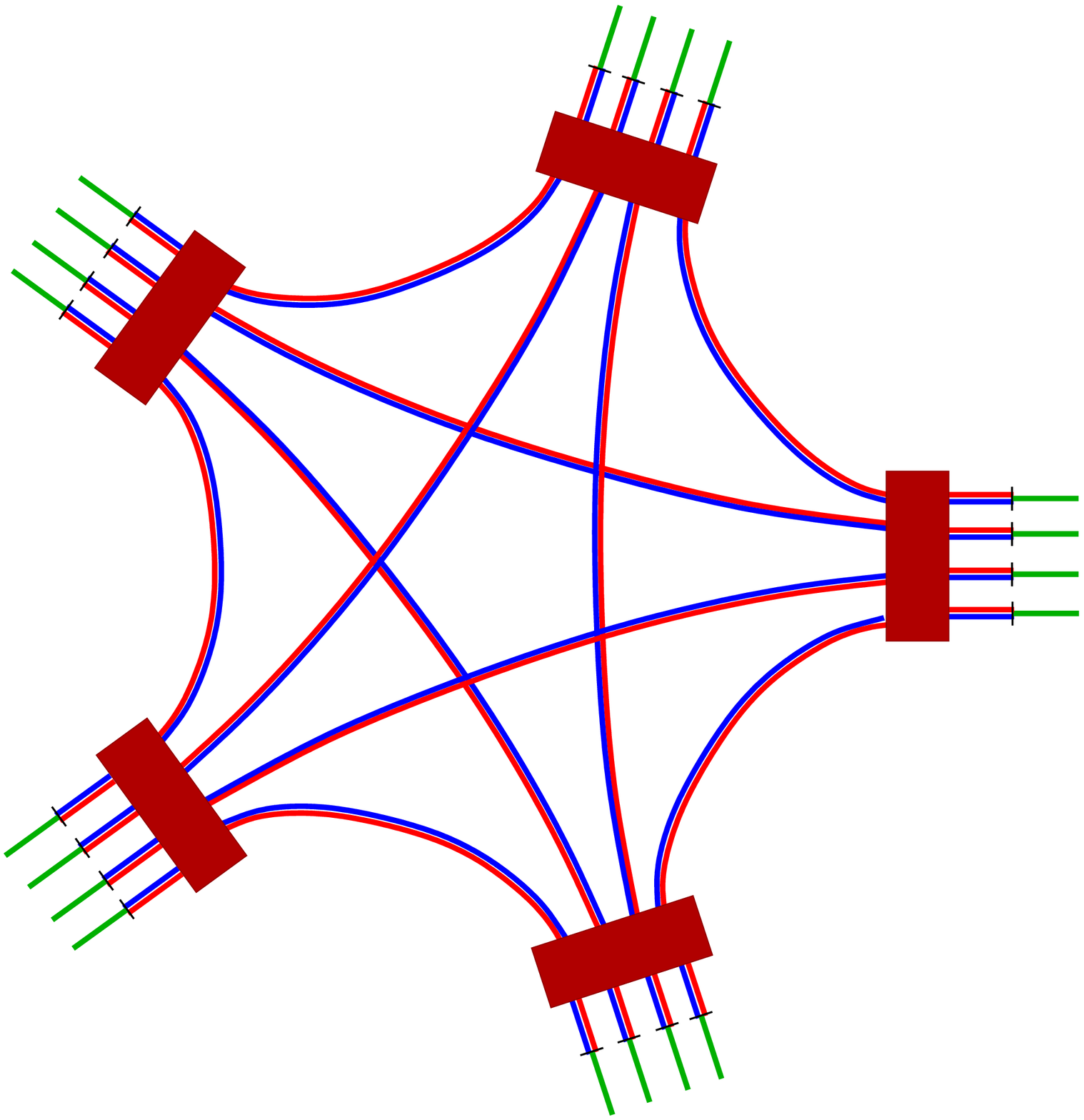}
\end{array},
\ea
where the boxes now represent $SL(2,\C)$ integrations with the invariant measure. The previous amplitude is equivalent to the its spin foam representation
\ba\label{eprl-sf-sl2c}\n
Z^{L}_{eprl}(\Delta)= \sum \limits_{ j_f}\sum \limits_{ \iota_e}  \ \prod\limits_{f \in \Delta^{\star}} (1+\gamma^2) j_f^2 \prod_{v\in \Delta^{\star}} \begin{array}{c}\psfrag{a}{$\iota_1$}\psfrag{b}{$\iota_2$}\psfrag{c}{$\iota_3$}\psfrag{d}{$\iota_4$}\psfrag{e}{$\iota_5$}
\includegraphics[width=5cm]{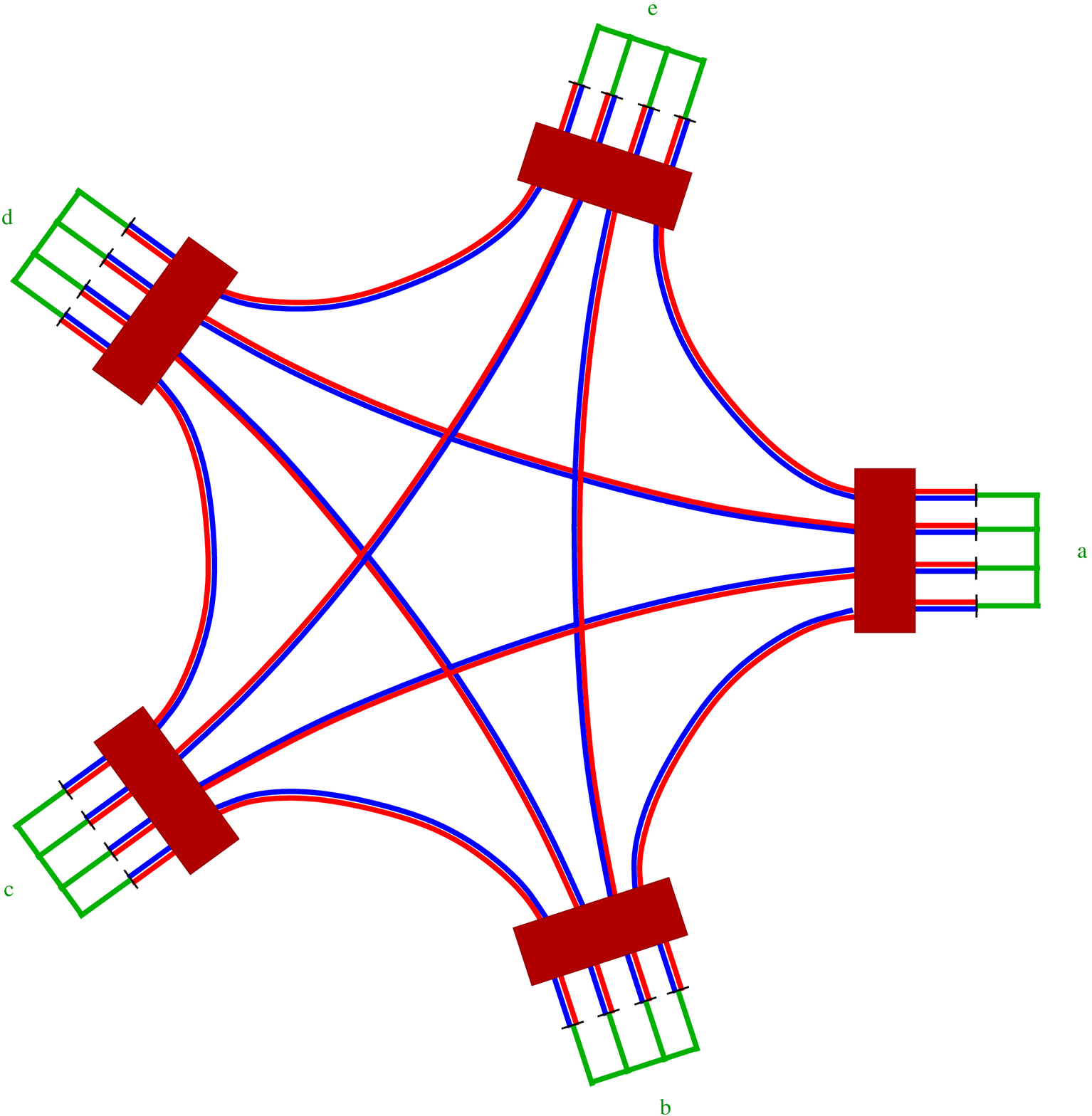}
\end{array},
\ea
The vertex amplitude is divergent due to the presence of a redundant integration over $SL(2,\C)$, it becomes finite by dropping an arbitrary integration, i.e. removing any of the 5 boxes in the vertex expression \cite{Engle:2008ev}.

\subsubsection{The coherent state representation}

It is immediate to obtain the coherent states representation of the Lorentzian models. As in the Riemannian case, one simply inserts resolution of the identities 
(\ref{ident-coherent}) on the intermediate $SU(2)$  (green) wires in (\ref{eprl-sl2c}) from where it results 
\ba\label{eprl-cohe-l}
&& Z^{L}_{eprl}(\Delta)=\sum \limits_{j_f}  \ \prod\limits_{f \in \Delta^{\star}} (1+\gamma^2) j^2 \n \\ && \int \prod_{e\in  \in \Delta^{\star}} {\rm d}_{j_{ef}} dn_{ef}
\begin{array}{c}\psfrag{w}{$$}
\psfrag{a}{$\va n_{ 1}$}
\psfrag{A}{$\va n_1$}
\psfrag{b}{$\va n_2$}
\psfrag{B}{$\va n_2$}
\psfrag{c}{$\va n_3$}
\psfrag{C}{$\va n_3$}
\psfrag{d}{$\va n_4$}
\psfrag{D}{$\va n_4$}
\includegraphics[width=7cm]{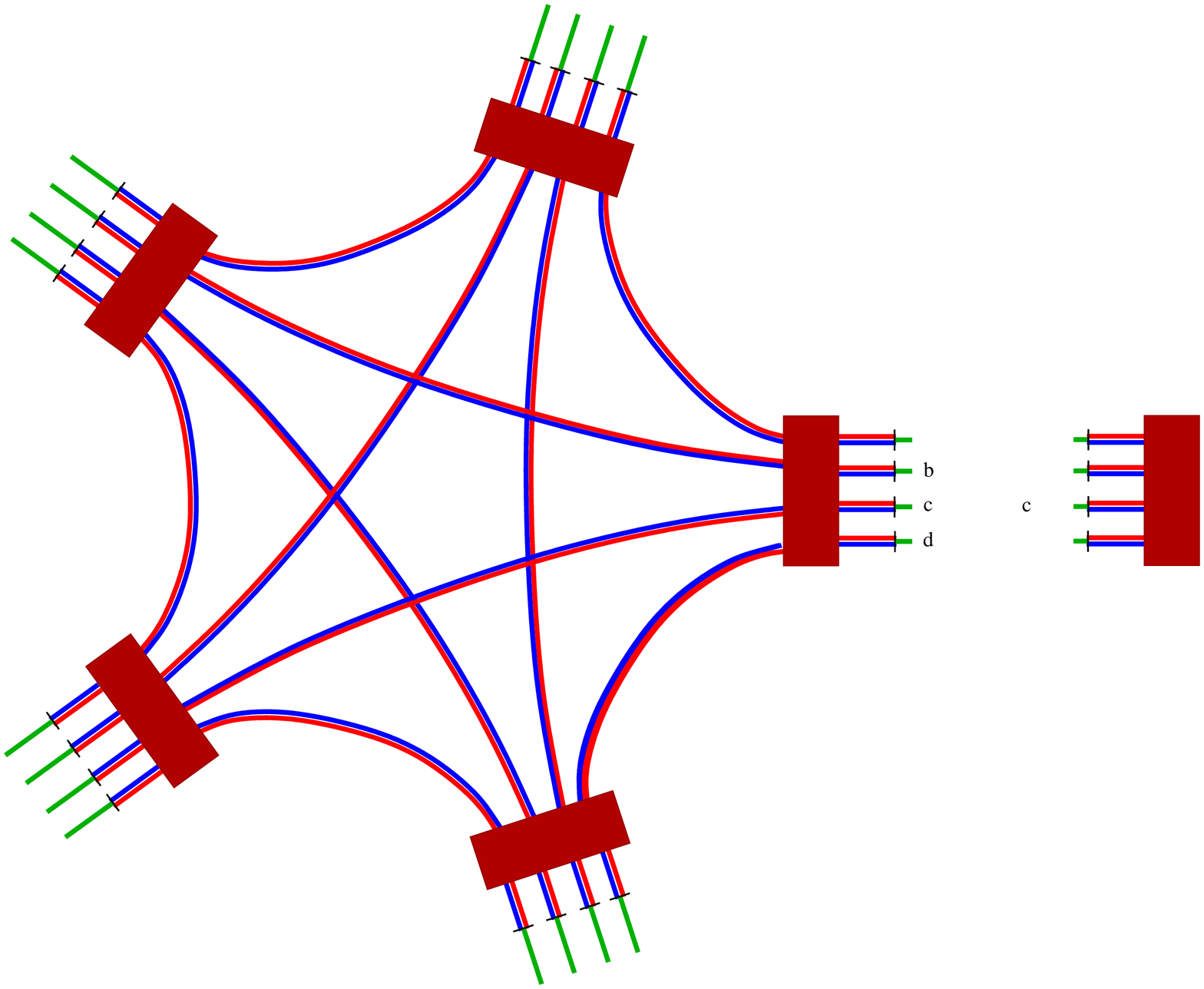}
\end{array},
\ea

%

\section{The Freidel-Krasnov (FK) model}\label{fk}

Shortly after the appearance of the paper \cite{Engle:2007uq}, Freidel and Krasnov \cite{Freidel:2007py} introduced a set of new spin foam models 
for four dimensional gravity using the coherent state basis of the quantum tetrahedron of Livine and Speziale \cite{Livine:2007vk}.
The idea is to impose the linearized simplicity constraints (\ref{constrainty-e}) directly as a semiclassical condition on the coherent state basis.
As we have seen above, coherent states are quantum states of the right and left tetrahedra in BF theory which have a clear-cut semiclassical interpretation 
through their property  (\ref{geo}).  We have also seen that the imposition of the linear constraints (\ref{constrainty-e}) {\em a la} EPRL is in essence semiclassical as they are strictly 
valid only in the large spin limit. In the FK approach one simply accept from the starting point that, due to their property of not defining set that is closed 
under commutation relations, the Plebansky  are to be imposed semiclassically. One defines new models by restricting 
the set of coherent states entering in the coherent state representation of $Spin(4)$ BF theory (\ref{bf-cohe}) to those that satisfy condition 
(\ref{constrainty-e}) in expectation values.  They also emphasize how the model \cite{Engle:2007uq} corresponds indeed to the sector $\gamma=\infty$ which 
has been shown to be topological \cite{Liu:2009em}.  

\subsubsection*{The case $\gamma<1$} For $\gamma<1$ the vertex amplitude is identical to the EPRL model. This is apparent in the coherent state expression of the EPRL model (\ref{eprl-cohe}).
Thus we have
\ba\label{FK<}&& 
Z_{fk}^{\va \gamma<1}(\Delta)=\sum \limits_{ j_f}   \ \prod\limits_{f \in \Delta^{\star}} {\rm d}_{|1-\gamma|\frac{j}{2}}{\rm d}_{(1+\gamma)\frac{j}{2}} 
\n \\ && \ \ \ \ \ \ \ \ \ \ \ \ \  \prod_{e\in \Delta^\star} \int {\rm d}_{(1+\gamma)\frac{j}{2}}{\rm d}_{(\gamma-1)\frac{j_{ef}}{2}} dn_{ef} 
\begin{array}{c}\psfrag{w}{$=\int\limits_{[S^3]^4} \prod\limits_{\va I=1}^{\va 4} dn_{\va I}$}
\psfrag{a}{$\va n_1$}
\psfrag{A}{$\va n_1$}
\psfrag{b}{$\va n_2$}
\psfrag{B}{$\va n_2$}
\psfrag{c}{$\va n_3$}
\psfrag{C}{$\va n_3$}
\psfrag{d}{$\va n_4$}
\psfrag{D}{$\va n_4$}
\psfrag{x}{$$}
\includegraphics[width=6.5cm]{eprl5.eps}\end{array}.
\ea 
From the previous expression we conclude that the vertex amplitudes of the FK and EPRL model coincide for $\gamma<1$
\be
A_{v\ fk}^{\va \gamma<1}=A_{v \ eprl}^{\va \gamma<1}.
\ee
Notice however that different weights are assigned to edges in the FK model.  This is due to the fact that one is restricting  the $Spin(4)$ resolution of identity
in the coherent basis in the previous expression, while in the EPRL model the coherent state resolution of the identity is used for $SU(2)$ representations.
This difference is important and has to do with the still un-settled discussion concerning the measure in the path integral representation.

\subsubsection*{The case $\gamma>1$} 
For the case $\gamma>1$ the FK amplitude is given by
\ba\label{FK>}&& 
Z_{fk}^{\va \gamma>1}(\Delta)=\sum \limits_{ j_f}   \ \prod\limits_{f \in \Delta^{\star}} {\rm d}_{|1-\gamma|\frac{j}{2}}{\rm d}_{(1+\gamma)\frac{j}{2}} 
\n \\ && \ \ \ \ \ \ \ \ \ \ \ \ \  \prod_{e\in \Delta^\star} \int {\rm d}_{(1+\gamma)\frac{j}{2}}{\rm d}_{(\gamma-1)\frac{j_{ef}}{2}} dn_{ef} 
\begin{array}{c}\psfrag{w}{$=\int\limits_{[S^3]^4} \prod\limits_{\va I=1}^{\va 4} dn_{\va I}$}
\psfrag{a}{$\va n_1$}
\psfrag{A}{$\va -n_1$}
\psfrag{b}{$\va n_2$}
\psfrag{B}{$\va -n_2$}
\psfrag{c}{$\va n_3$}
\psfrag{C}{$\va -n_3$}
\psfrag{d}{$\va n_4$}
\psfrag{D}{$\va -n_4$}
\psfrag{x}{$$}
\includegraphics[width=6.5cm]{eprl5.eps}\end{array}.
\ea 
The study of the coherent state representation of the FK model for $\gamma>1$,  and comparison with  equation (\ref{cuat}) for the EPRL model, clearly shows the difference between the two models in this regime. 
\ba
\label{fk-cohe}Z^{\va \gamma}_{fk}=\sum \limits_{ j_f }  \ \prod\limits_{f \in \Delta^{\star}} {\rm d}_{(1-\gamma)\frac{j_f}{2}}{\rm d}_{(1+\gamma)\frac{j_f}{2}} \int \prod_{e\in  \Delta^{\star}} {\rm d}_{|1-\gamma|\frac{j_{ef}}{2}} {\rm d}_{(1+\gamma)\frac{j_{ef}}{2}} dn_{ef} dg^{-}_{ef}dg^{+}_{ef}
\ \exp{(S^{\va fk\ \gamma}_{j^{\pm},\bn}[g^{\pm}])}, \ea
where the discrete action \be\label{fk-discrete-action}
S^{\va fk\ \gamma}_{j^{\pm},\bn}[g^{\pm}]=\sum_{v\in\Delta^{\star}} (S^v_{(1-\gamma)\frac{j_f}{2},\bn}[g^{-}]+S^v_{(1+\gamma)\frac{j_f}{2},s(\gamma) \bn}[g^{+}]), \ee where $s(\gamma)={\rm sign}(1-\gamma)$ and  
\be
\label{fk-v-action}
S^v_{j,\bn}[g] = \sum\limits^{5}_{a < b=1} 2j_{ab} \ln \, \la n_{ab}| g^{-1}_a g_b| \, n_{ba} \ra,
\ee
with the indices $a,b$ labelling the five edges of a given vertex. 

\section{Boundary data for the new models and relationship with the canonical theory}\label{boundarya}

So far we have considered cellular complexes with no boundary. Transition amplitudes
are expected to be related to the definition of the physical scalar product.
In order to define them one needs to consider complexes with boundaries. Boundary states are defined on the boundary of the dual two-complex $\Delta^{\star}$ that we denote
$\partial \Delta^{\star}$.  The object $\partial \Delta^{\star}$ is a one-complex (a graph). According to the construction of the model (Section \ref{eprl-r}) boundary states 
are in one-to-one correspondence with $SU(2)$ spin networks.
This comes simply from the fact that links (one-cells) $\ell\in \partial\Delta^{\star}$ inherit the
 spins labels (unitary irreducible representations of the subgroup $SU(2)$) of the boundary faces while 
 nodes (zero-cells) $n\in \partial \Delta^{\star}$ inherit the intertwiner levels of boundary edges.
 
 At this stage one can associate the boundary data with elements of a Hilbert space. Being in one-to-one correspondence with $SU(2)$ spin networks, a 
 natural possibility is to associate to them an element of the kinematical Hilbert space of LQG. More precisely, with a given coloured boundary graph $\gamma$ with  links labelled by spins $j_{\ell}$ and nodes labelled by interwiners $\iota_n$  
 we associate a cylindrical function $\Psi_{\gamma,\{j_\ell\},\{\iota_n\}}\in {\sL}^2(SU(2)^{N_{\ell}})$, where here $N_{\ell}$ denotes number of links in the graph $\gamma$. 
 In this way, the boundary Hilbert space associated with $\partial \Delta^{\star}$ is isomorphic (if one used the natural AL measure)
 with the Hilbert space of LQG truncated to that fixed graph. Moreover, geometric operators such as volume and area
 defined in the covariant context are shown to coincide with the corresponding operators defined in the canonical formulation \cite{Ding:2010ye, Ding:2009jq}.
 Now, if cellular complexes are  dual to triangulations then  the boundary spin networks can have at most four valent nodes. 
 This limitation can be easily overcome: as in BF theory the EPRL amplitudes can be generalized to arbitrary complexes with boundaries given by graphs with
 nodes of arbitrary valence. The extension of the model to arbitrary complexes has been first studied in \cite{Kaminski:2009qb, Kaminski:2009fm}, it has also been revisited in 
 \cite{Ding:2010fw}.
 
 Alternatively, one can associate the boundary states with elements of $\sL^2(Spin(4)^{N_{\ell}})$ (in the Riemannian models)---or 
 carefully define the analog of spin network states as distributions in the Lorentzian case (see \cite{Freidel:2002xb} for some insights on the problem of defining a gauge invariant Hilbert space of graphs for non compact gauge groups).
 In this case one gets special kind of spin network states that are a subclass of the so-called projected spin networks introduced in
 \cite{Alexandrov:2002br, Livine:2002ak} in order to define an heuristic quantization of the (non-commutative and very complicated) Dirac algebra of a Lorentz connection formulation of the phase space of gravity 
 \cite{Alexandrov:2006wt, Alexandrov:2005ng, Alexandrov:2002br, Alexandrov:2002xc,Alexandrov:2001wt, Alexandrov:2000jw, Alexandrov:1997yk}.
 The fact that these special subclass of projected spin networks  appear naturally as boundary states of the new spin foams is shown in \cite{Dupuis:2010jn}. 
  
Due to their similarity for $\gamma<1$ the same relationship between boundary data and elements of the kinematical Hilbert space hold for
the FK model. However, the such simple relationship does not hold for the model in the case $\gamma>1$.

It is important to mention that the knotting properties of boundary spin network do not seem to play a role in present definitions of transition amplitudes \cite{Bahr:2010my}.

\section{Further developments and related models} \label{further}


The spin foam amplitudes discussed in the previous sections have been introduced by 
constraining the BF histories through the simplicity constraints. However, in the 
path integral formulation, the presence of constraints has the additional effect of 
modifying the weights with which  those histories are to be summed: second class constraints
modify the path integral measure (in the spin foam context this issue was raised in \cite{myo}).
As pointed out before, this question has not been completely settled yet in the spin foam community.
The explicit modification of the formal measure in terms of continuous variables 
for the Plebansky action was presented in \cite{karim3}.  
A systematic investigation of the measure in the spin foam context was attempted in  \cite{Engle:2009ba} and \cite{Han:2009bb}.
As pointed out in \cite{myo}, there are restrictions in the manifold of possibilities coming from the requirement of
background independence. The simple BF measure chosen in the presentation of the amplitudes in the previous sections
satisfy these requirements.   There are other consistent possibilities; see for instance \cite{Bianchi:2010fj}  for
a modified measure which remains extremely simple and is suggested from the structure of LQG.


An important question is the relationship between the spin foam amplitudes and the canonical operator formulation. The question of wether one can reconstruct the Hamiltonian constraints out of spin foam amplitudes 
has been analysed in detail in three dimensions. For the study of quantum three dimensional gravity from the BF perspective see \cite{a20}, we will in fact present this perspective in detail in the three dimensional part of this article.
For the relationship with the canonical theory using variables that are natural from the Regge gravity perspective see \cite{Bonzom:2011hm, Bonzom:2011tf}
There are generalizations of Regge variables moro adapted to the interpretation of spin foams \cite{Dittrich:2008va}.
In four dimensions the question has been investigated in \cite{Alesci:2008yf} in the context of the new spin foam models.
In the context of group field theories  this issue is explored in \cite{Livine:2011yb}. 
Finally, spin foams can in principle be obtained directly from the implementation of the Dirac program 
using path integral methods this avenue has been explored in \cite{Han:2009aw, Han:2009ay} 
from which a discrete path integral formulation followed \cite{Han:2009az}.
The question of the relationship between covariant and canonical formulations in the discrete setting has been analyzed
also in \cite{Dittrich:2009fb}.


By construction all tetrahedra in the FK and EPRL models are embedded in a spacelike
hypersurface and hence have only spacelike triangles. It seem natural to ask the question of whether a more 
general construction allowing for timelike faces is possible. 
The models described in previous sections have been generalized in order to include timelike faces in the work of F. Conrady 
\cite{Conrady:2010sx,
Conrady:2010vx,
Conrady:2010kc}. An earlier attempt to define such models  in the context of the Barrett-Crane model can be found in \cite{a8}.


The issue of the coupling of the new spin foam models to matter  remains to a large extend un-explored territory.
Nevertheless some results can be found in the literature. The coupling of the Barrett-Crane model (the $\gamma\to \infty$ limit of the EPRL model) to Yang-Mills fields 
was studied in \cite{ori4}. More recently the coupling of the EPRL model to fermions has been investigated in \cite{Han:2011as, Bianchi:2010bn}. 
A novel possibility of unification of the gravitational and gauge fields was recently proposed in \cite{Alexander:2011jf}.


The introduction of a cosmological constant in the construction of four dimensional  spin foam models has a long history. 
Barrett and Crane introduced a vertex amplitude \cite{BC1} in terms of the Crane and Yetter model \cite{crane0} 
for BF theory with cosmological constant. The Lorentzian quantum deformed version of the previous model was 
studied in \cite{Noui:2002ag}.   For the new models the coupling  with a cosmological constant is explored in terms of the quantum deformation 
of the internal gauge symmetry in \cite{Ding:2011hp,Han:2010pz} as well as (independently) in \cite{Fairbairn:2010cp}.  The asymptotics of the vertex amplitude are shown to be consistent with a cosmological constant term in the semiclassical limit in \cite{Han:2011aa}.


The spin foam approach applied to quantum cosmology has been explored in
\cite{Bianchi:2011ym, Vidotto:2010kw, Henderson:2010qd, Bianchi:2010zs,
Rovelli:2009tp, :2008dx}.
The spin foam formulation can also be obtained from the canonical picture provided by loop quantum cosmology (see \cite{Bojowald:2006da} and references therein).
This has been explored systematically in  \cite{Ashtekar:2010gz, Ashtekar:2010ve, Ashtekar:2009dn, Campiglia:2010jw}.


As we have discussed in the introduction of the new models, Heisenberg uncertainty principle precludes the strong imposition of the Plebanski constraints that reduce BF theory 
to general relativity. The results on the semiclassical limit of these models seem to indicate that metric gravity should be recovered in the low energy limit. However, its seems likely that the semiclassical limit could be related to
certain  modifications of Plebanski's formulation of gravity \cite{Krasnov:2006du, Krasnov:2007cq, Krasnov:2008fm, Krasnov:2009iy, Krasnov:2009ip}. A simple interpretation of the new models in the
context of the bi-gravity  paradigm proposed in \cite{Speziale:2010cf} could be of interest.


As already pointed out in \cite{baez7} spin foams can be
interpreted in close analogy to Feynman diagrams. Standard Feynman
graphs are generalized to $2$-complexes and the labeling of
propagators by momenta to the assignment of spins to faces.
Finally, momentum conservation at vertices in standard
feynmanology is now represented by spin-conservation at edges,
ensured by the assignment of the corresponding intertwiners. In
spin foam models the non-trivial content of amplitudes is
contained in the vertex amplitude  which in the language of Feynman
diagrams can be interpreted as an interaction. This analogy is indeed realized in the formulation of spin foam
models in terms of a group field theory (GFT) \cite{reis1,reis2}.

The GFT formulation resolves by definition the two fundamental conceptual
problems of the spin foam approach: diffeomorphism gauge symmetry and
discretization dependence. The difficulties are shifted to the question of the
physical role of $\lambda$ and the convergence of the corresponding perturbative series.

In three dimensions this idea has been studied in more detail. 
In \cite{Magnen:2009at} scaling properties of the modification of the Boulatov group field theory 
introduced in \cite{fre10} was studied in detail. In a further modification of the previous model (known as coloured tensor models \cite{Gurau:2009tw}) 
new techniques based on a suitable $1/N$ expansion imply that amplitudes are dominated by spherical topology  
\cite{Gurau:2010ba}; moreover, it seem possible that the continuum limit  might be critical as in certain matrix models
\cite{Gurau:2011tj, Bonzom:2011zz, Gurau:2011xq, Gurau:2011aq, Ryan:2011qm}. However, it is not yet clear if there is a sense 
in which these models correspond to a physical theory. The naive interpretation of the models is that they correspond to a
formulation of 3d quantum gravity including a dynamical topology.

\section{Results on the semiclassical limit of  EPRL-FK models}
\label{semiclas}

Having introduced the relevant spin foam models in the previous sections we now 
present the results on the large spin asymptotics of the spin foam amplitudes suggesting
that on a fixed discretization  the semiclassical limit of the EPRL-FK models is 
given by Regge's discrete formulation of general relativity \cite{Barrett:2009gg, Barrett:2009mw}.

The semiclassical limit of spin foams is based on the  study of the the large spin limit asymptotic behaviour of coherent state spin foam amplitudes. The notion of large spin can be 
defined by the rescaling of quantum numbers and Planck constant according to $j\to \lambda j$ and  $\hbar \to \hbar/\lambda$ and taking $\lambda >>1$.
In this limit the quantum geometry approximates the classical one when tested with suitable states (e.g. coherent states).
However, the geometry remains discrete during this limiting process as the limit is taken on a fixed regulating cellular structure. That is why one usually makes a clear distinction between  
semiclassical limit and the continuum limit. In the semiclassical analysis presented here one can only hope to make contact with 
discrete formulations of classical gravity; hence the importance of Regge calculus in the discussion of this section. 

The key technical ingredient in this analysis is the representation of spin foam amplitudes 
in terms of the coherent state basis introduced in Section \ref{cohecohe}.  Here we follow \cite{Barrett:2009gg, Barrett:2009mw, Barrett:2009cj, Barrett:2009as, Barrett:2010ex}.
The idea of using coherent states and discrete effective actions for the study of the large spin asymptotics of spin foam amplitudes was
put forward in \cite{Conrady:2008mk, Conrady:2008ea}.
The study of the large spin asymptotics has a long tradition in the context of quantum gravity dating back  
to the studied of Ponzano-Regge \cite{ponza}.  More directly related to our discussion here are the early works \cite{Barrett:2002ur, Barrett:1998gs}.
The key idea is to use asymptotic stationary phase methods for the amplitudes written in terms of the discrete actions presented in the previous section.


In this section we review the results of the analysis of the large spin asymptotics of
the EPRL vertex amplitude for both the Riemannian and Lorentztian models. We follow the notation and terminology 
of \cite{Barrett:2009gg} and related papers. 

\subsubsection{$\SU(2)$ $\ftj$-symbol asymptotics}

 As $SU(2)$ BF theory is quite relevant for the construction of the EPRL-FK models, the study of the
 large spin asymptotics of the $SU(2)$ vertex amplitude is a key ingredient in the analysis of \cite{Barrett:2009gg}. 
 The coherent state vertex amplitude is
 \ba
 \ftj(j,\nb)=\int \prod_{a=1}^5 dg_{a} \prod_{1\le a\le b\le 5} \langle n_{ab} |g^{-1}_{a} g_{b}| n_{ba}\rangle^{2j_{ab}},\ea
 which depends on $10$ spins $j_{ab}$ and $20$ normals $n_{ab}\not=n_{ba}$. The previous amplitude can be expressed 
 as
  \ba\label{qq}
 \ftj(j,\nb)=\int \prod_{a=1}^5 dg_{a} \prod_{1\le a\le b\le 5} \exp S_{j,\bn}[g],\ea
\be \label{v-a}
S_{j,\bn}[g] = \sum\limits^{5}_{a < b=1} 2j_{ab} \ln \, \la n_{ab}| g^{-1}_a g_b| \, n_{ba} \ra,
\ee
and the indices $a,b$ label the five edges of a given vertex. The previous expression is exactly equal to the form (\ref{coloring4}) of the BF amplitude. In the case of the EPRL model studied in Sections \ref{eprl-r} 
the coherent state representation---see equations \ref{eprl-cohe}, \ref{eprl-cohe-g}, and \ref{eprl-cohe-l}---is  the basic tool for the study of the semiclassical limit of the models and the relationship with Regge discrete formulation of general relativity. 

In order to study the asymptotics of  (\ref{qq}) one needs to use extended stationary phase methods due to the fact the the action (\ref{v-a}) is complex (see \cite{Conrady:2008mk, Conrady:2008ea}).
The basic idea is that in addition to stationarity one requires real part of the action to be maximal. Points satisfying these two conditions are called {\em critical points}. 
As the real part of the action is negative definite, the action at critical points is purely imaginary.  

Notice that the action (\ref{v-a}) depends parametrically on 10 spins $j$ and $20$ normals $\bn$. These parameters define the so-called boundary data for the
four simplex $v\in \Delta^{\star}$. Thus, there is an action principle for every given boundary data. The number of critical points and their properties depend on these 
boundary data, hence the asymptotics of the vertex amplitude is a function of the boundary data.  Different cases are studied in detail in \cite{Barrett:2009gg}, here we present their results in the
special case where the boundary data describe a non-degenerate Regge geometry for the boundary of a four simplex, these data are referred to as Regge-like, and satisfy the 
gluing constraints. For such boundary data the action (\ref{v-a}) has exactly two critical points leading to the asymptotic formula
\be 
15j(\lambda j,\bn) \sim  \frac{1}{\lambda^{12}} \left[ N_+ \exp ( i \sum\limits_{a < b}\lambda  j_{ab} \Theta_{ab}^E )+ N_- \exp ( -i  \sum\limits_{a < b}\lambda  j_{ab} \Theta_{ab}^E )\right],
\ee
where 
 $\Theta_{ab}$ the appropriate diahedral angles
defined by the four simplex geometry; finally the 
$N_{\pm}$ are constants that do not scale with $\lambda$.

\subsubsection{The Riemannian EPRL vertex asymptotics} 

The previous result together with the fact that the  EPRL amplitude for $\gamma<1$ is a product of
$SU(2)$ amplitudes with the same $\bn$ in the coherent state representation (\ref{eprl-cohe}) implies the asymptotic formula for the vertex amplitude to be given by the unbalanced square of the above formula \cite{Barrett:2009cj}, namely
\ba \n && 
A^{\va eprl}_v\sim \frac{1}{\lambda^{12}} \left[ N_+  e^{ i \frac{(1-\gamma)}{2} \sum\limits_{a < b}\lambda  j_{ab} \Theta_{ab}^E  } + N_- \ e^{- i  \frac{(1-\gamma)}{2} \sum\limits_{a < b}\lambda  j_{ab} \Theta_{ab}^E } \right]\times \\ && \n \ \ \ \ \ \ \ \ \ \ \ \ \ \ \ \  \ \ \ \ \ \ \ \ \ \ \ \ \ \ \ \  \left[ N_+ \ e^{ i \frac{(1+\gamma)}{2}\sum\limits_{a < b}\lambda  j_{ab} \Theta_{ab}^E  } + N_- \ e^{- i \frac{(1+\gamma)}{2} \sum\limits_{a < b}\lambda  j_{ab} \Theta_{ab}^E  } \right].
\ea
One can write the previous expression as
\ba
A^{\va eprl}_v\sim \frac{1}{\lambda^{12}} \left[2 N_{+}N_{-} \cos \left(   S^{\va E}_{\va Regge} \right) + N_{+}^2  \ e^{i \frac{1}{\gamma}S^{\van E}_{\va Regge} } + N_{-}^2\  e^{ -i \frac{1}{\gamma} S^{\va E}_{\va Regge} } \right].
\ea
where
\be
S^{\van E}_{\va Regge}=\sum_{a < b}\lambda \gamma  j_{ab} \Theta_{ab}^E 
\ee
is the Regge like action for  $\lambda\gamma j_{ab}=A_{ab}$ the ten triangle areas (according to the LQG area spectrum \cite{bookt, book}).
Remarkably, the above asymptotic formula is also valid for the case $\gamma>1$ \cite{Barrett:2009gg}. The first term in the vertex asymptotics is 
in essence the expected one: it is the analog of the $6j$ symbol asymptotics in three dimensional spin foams. Due to their explicit dependence on the Immirzi parameter, the last two 
terms are somewhat strange from the point of view of the continuum  field theoretical view point. However, this seems to be a 
peculiarity of the Riemannian theory alone as the results of \cite{Barrett:2009mw} for the Lorentzian models show.
{ Non geometric configurations are exponentially surpressed}

\subsubsection{Lorentzian EPRL model}

To each solution one can associate a second solution corresponding to a parity related $4$-simplex and, consequently, the asymptotic formula has two terms. It is given, up to a global sign, by the expression
\be 
A^{\va eprl}_{v} \sim  \frac{1}{\lambda^{12}} \left[ N_+ \exp \left( i \lambda \gamma \sum\limits_{a < b} j_{ab} \Theta_{ab}^L \right) + N_- \exp \left(- i \lambda \gamma \sum\limits_{a < b} j_{ab} \Theta_{ab}^L \right) \right],
\ee
where $N_{\pm}$ are constants that do not scale. 
{Non geometric configurations are exponentially surpressed}


In \cite{Conrady:2008ea} Freidel and Conrady gave a detailed description of the coherent state representation of the various spin foam models described so far.
In particular they provided the definition of the effective discrete actions associated to each case which we presented in (\ref{fk-cohe}). This provides the basic elements for setting up the asymptotic 
analysis presented in \cite{Conrady:2008mk} (the first results on the semiclassical limit of the new spin foam models) which is similar to the studies of the asymptotic of the vertex amplitude reviewed above but more general
in the sense that the semiclassical limit of a full spin foam configuration (involving many vertices) is studied.
The result is technically more complex as one studies now critical points of the action associated to a coloured complex which in addition of depending on group variables  
$g$ it depends on the coherent state parameters $\bn$. The authors of \cite{Conrady:2008mk} write Equation (\ref{fk-cohe}) in the following way
\ba
\label{fk-semi}Z^{\va \gamma}_{fk}=\sum \limits_{ j_f }  \ \prod\limits_{f \in \Delta^{\star}} {\rm d}_{(1-\gamma)\frac{j_f}{2}}{\rm d}_{(1+\gamma)\frac{j_f}{2}} W_{\Delta^{\star}}^{\gamma} (j_f), \ea
where
\be
W_{\Delta^{\star}}^{\gamma}(j_f)=\int \prod_{e\in  \Delta^{\star}} {\rm d}_{|1-\gamma|\frac{j_{ef}}{2}} {\rm d}_{(1+\gamma)\frac{j_{ef}}{2}}  dn_{ef} dg^{-}_{ef}dg^{+}_{ef}
\ \exp{(S^{\va fk\ \gamma}_{j^{\pm},\bn}[g^{\pm}])}.
\ee
They show that those solutions of the equations of motion of the effective discrete action that are non geometric (i.e. the contrary of Regge like) are not critical and hence exponentially suppressed in the scaling $j_{f}\to \lambda j_f$ with $\lambda >>1$. If configurations are geometric (i.e. Regge like) one has two kind of contributions to the amplitude assymptotics: those coming from degenerate and non-degenerate configurations. If one (by hand) restricts to the non-degenerate configurations then one has
\be
W_{\Delta^{\star}}^{\gamma}(j_f)\sim \frac{c}{\lambda^{(33 n_e-6n_v-4n_f)}} \exp(i\lambda S^{\va E}_{\va Regge}(\Delta^{\star},j_f)),
\ee
where $n_e$, $n_v$, and $n_f$ denote the number of edges, vertices, and faces in the two complex $\Delta^{\star}$ respectively.
There are recent works by M. Han where asymptotics of general simplicial geometry amplitudes are studied in the context of the EPRL model \cite{Han:2011rf, Han:2011re}.


The problem of computing the two point function and higher correlation functions in the context of spin foam 
has received lots of attention recently. The framework for the definition of the correlation functions in the background independent 
setting has been generally discussed by Rovelli in \cite{Rovelli:2005yj} and correspods to a special application of a more general 
proposal investigated by Oeckl \cite{Oeckl:2011yi, Oeckl:2011qd, Oeckl:2010ra, Colosi:2009cp, Colosi:2008jf, Colosi:2008fv, Colosi:2007bj, Oeckl:2006rs}. It was then applied to the Barrett-Crane model in \cite{Alesci:2007tg, Alesci:2007tx, Bianchi:2006uf}, where it was discovered that certain components of the two point function could not yield the expected result 
compatible with Regge gravity in the semiclassical limit. This was used as the main motivation of the weakening of the
imposition of the Plebanski constraints leading to the new models. Soon thereafter it was argued that the difficulties of the Barrett-Crane model
where indeed absent in the EPRL model \cite{Alesci:2008ff}. The two point function for the EPRL model was calculated in \cite{Bianchi:2009ri} and it was shown to produce
a result in agreement with that of Regge calculus\cite{Bianchi:2008ae,Magliaro:2008zz}
in the limit $\gamma\to 0$. 

The fact that, for the new model,  the double scaling limit  $\gamma\to 0$ and $j\to \infty$ with $\gamma j$=constant  defines the appropriate regime where the fluctuation  behave as in Regge gravity (in the leading order) has been further clarified in \cite{Magliaro:2011dz}. This indicates that the quantum fluctuations in the new models are more general than simply metric fluctuations. The fact the the new models are not metric at all scales should not be surprising as we know that the Plebanski constraints that produce metric general relativity out of BF theory has been implemented only semiclassically (in the large spin limit). At the deep Planckian regime fluctuations are more general than metric. {However, it not clear at this stage why this is controlled by the Immirzi parameter. }

All the previous calculations involve a complex with a single four-simplex. The first computation involving more than one simplex was performed in 
\cite{Mamone:2009pw,Bianchi:2006uf} for the case of the Barrett-Crane model. 
Certain peculiar properties were found and it is not clear at this stage whether these issues remain in the EPRL model. 
Higher order correlation functions have been computed in \cite{Rovelli:2011kf}, the results are in agreement with Regge gravity in the $\gamma\to 0$ limit.

\section{Acknowledgements}
\label{section:acknowledgements}

I would like to thank the help of many people in the field that have helped me in various ways. 
I am grateful to  Eugenio Bianchi, Carlo Rovelli and Simone Speziale for the many for the many discussions on 
aspects and details of the recent literature. Many detailed calculations that contributed to the presentation of the new models in this review where
done in collaboration with Mercedes Vel\'azquez to whom I would like to express my gratitude. 
I would also like to thank You Ding, Florian Conrady, Laurent Freidel, Muxin Han, Merced Montesinos for help and  valuable interaction.






\end{document}